\documentclass[aps,prb,twocolumn,preprintnumbers,superscriptaddress,amsmath]{revtex4-2}
\usepackage{graphicx} % Required for inserting images

\usepackage{amsmath}
\usepackage{amssymb}
\usepackage{enumitem}
\usepackage{multirow}
\usepackage{physics}
\usepackage[dvipsnames]{xcolor}
\usepackage[T1]{fontenc}
\usepackage{graphicx}% Include figure files
\usepackage{dcolumn}% Align table columns on decimal point
\usepackage{bm}% bold math
\usepackage[retainorgcmds]{IEEEtrantools}
\usepackage{xcolor}
\usepackage[colorlinks=true,linkcolor=blue,citecolor=blue,urlcolor=blue]{hyperref}
\usepackage{bbold}
\usepackage{mathtools}
\usepackage{accents}
\usepackage{verbatim} % comments
\usepackage{dutchcal}
\newcommand\crule[3][black]{\textcolor{#1}{\rule{#2}{#3}}}

\definecolor{light_blue}{HTML}{17BECF}
\definecolor{blue}{HTML}{1F77B4}
\definecolor{green}{HTML}{2CA02C}
\definecolor{yellow}{HTML}{FAC205}
\definecolor{orange}{HTML}{FF7F0E}
\definecolor{pink}{HTML}{E377C2}
\definecolor{red}{HTML}{D62728}
\definecolor{brown}{HTML}{8C564B}
\definecolor{light_red}{HTML}{EF4026}
\definecolor{violet}{HTML}{7E1E9C}

\renewcommand{\selectlanguage}[1]{}

\renewcommand{\vec}[1]{\boldsymbol{#1}}
\newcommand{\R}{\vec{R}}
\renewcommand{\t}{\vec{\tau}}
\renewcommand{\k}{\vec{k}}

\renewcommand{\r}{\vec{r}}

\newcommand{\eig}{\epsilon}
\newcommand{\eigo}{\epsilon^{0}}

\newcommand{\kF}{k_{\mathrm{F}}}

\newcommand{\qph}{\vec{Q}}

\newcommand{\qscr}{\vec{q}}

\newcommand{\inveps}{\varepsilon^{-1}}
\newcommand{\eps}{\varepsilon}

\newcommand{\w}{\omega}

\newcommand{\vcoul}{v}

\newcommand{\GW}{\mathrm{GW}}
\newcommand{\GoWo}{\mathrm{G^0W^0}}
\newcommand{\SX}{\mathrm{SX}}
\newcommand{\SXo}{\mathrm{SX^0}}
\renewcommand{\H}{\mathrm{H}}
\newcommand{\ZD}{\mathrm{2D}}
\newcommand{\eh}{\mathrm{eh}}
\newcommand{\F}{\mathrm{F}}
\newcommand{\emb}{\mathrm{emb}}
\newcommand{\dl}{\mathrm{dl}}
\newcommand{\dm}{\mathrm{dm}}

\renewcommand{\dh}{\mathrm{dh}}

\newcommand{\DFT}{\mathrm{DFT}}

\newcommand{\G}{\mathbf{G}}
\newcommand{\orb}{\varphi}

\newcommand{\editor}[2]{%
  \expandafter\newcommand\csname #1note\endcsname[1]{%
    \textcolor{#2}{(\textbf{#1:} \textit{##1})}}%
  \expandafter\newcommand\csname #1\endcsname[1]{%
    \textcolor{#2}{##1}}%
  \expandafter\newcommand\csname #1cancel\endcsname[1]{%
    \textcolor{#2}{\sout{##1}}}%
  \expandafter\newcommand\csname #1change\endcsname[2]{%
    \textcolor{#2}{\sout{##1} ##2}}%
  \newenvironment{#1text}{\color{#2}}{\color{black}}
}
\definecolor{Blu}{rgb}{0.00,0.00,1.00}
\definecolor{Red}{rgb}{1.00,0.00,0.00}
\definecolor{Orange}{rgb}{1.00,0.45,0.00}
\definecolor{Green}{rgb}{0.360,0.6,0.36}
\definecolor{tangerine}{rgb}{0.944,0.522,0}
\definecolor{brown}{rgb}{0.633,0.156,0.156}
\definecolor{lime}{rgb}{0.5,1.0,0.0313}
\definecolor{limedark}{rgb}{0.333, 0.666, 0.020}
\definecolor{applegreen}{rgb}{0.55, 0.71, 0.0}
\definecolor{green1}{rgb}{0.0, 0.5, 0.0}
\definecolor{green2}{rgb}{0.25, 0.5, 0.016}
\definecolor{BluBondi}{rgb}{0.00,0.58,0.71}
\definecolor{Cyan}{rgb}{0.00,0.50,0.50}

\editor{AG}{Red}
\editor{FM}{limedark}
\editor{GC}{BluBondi}
\editor{FMau}{Red}

\begin{document}
\title{High- and low-energy many-body effects of graphene in a unified approach}

\author{Alberto Guandalini}
\email{alberto.guandalini@uniroma1.it}
\affiliation{Dipartimento di Fisica, Universit\`a di Roma La Sapienza, Piazzale Aldo Moro 5, I-00185 Roma, Italy}

\author{Giovanni Caldarelli}
\affiliation{Dipartimento di Fisica, Universit\`a di Roma La Sapienza, Piazzale Aldo Moro 5, I-00185 Roma, Italy}

\author{Francesco Macheda}
\affiliation{Dipartimento di Fisica, Universit\`a di Roma La Sapienza, Piazzale Aldo Moro 5, I-00185 Roma, Italy}

\author{Francesco Mauri}
\affiliation{Dipartimento di Fisica, Universit\`a di Roma La Sapienza, Piazzale Aldo Moro 5, I-00185 Roma, Italy}

\begin{abstract}
    We show that the many-body features of graphene band structure and electronic response can be accurately evaluated by applying many-body perturbation theory to a tight-binding (TB) model.
    In particular, we compare TB results for the optical conductivity with previous ab-initio calculations, showing a nearly perfect agreement both in the low energy region near the Dirac cone ($\sim 100$ meV), and at the higher energies of the $\pi$ plasmon ($\sim 5$ eV).
    A reasonable agreement is reached also for the density-density response at the Brillouin zone corner.
    With the help of the reduced computational cost of the TB model, we study the effect of self-consistency on the screened interaction ($W$) and on the quasi-particle corrections, a task that is not yet achievable in ab-initio frameworks.
    We find that self-consistency is important to reproduce the experimental results on the divergence of the Fermi velocity, as confirmed by previous studies, while it marginally affects the optical conductivity.
    Finally, we study the robustness of our results against doping or the introduction of a uniform dielectric environment.
\end{abstract}

\maketitle

%%%%%%%%%%%%%%%%%%%%%%
\section{Introduction}
%%%%%%%%%%%%%%%%%%%%%%
Since its discovery in $2004$~\cite{novoselov_two-dimensional_2005,Novoselov_2016}, graphene excitations have been thoroughly studied in the literature due to its remarkable transport, plasmonic and opto-electronic applications~\cite{Zhang_2005,Bonaccorso_2010,Grigorenko_2012, Garcia_2014}.
Initially, theoretical studies within the framework of the random-phase approximation (RPA) have been used to describe some electronic graphene properties, such as the Dirac~\cite{Hwang_2007,Sarma_13} and $\pi$ plasmon dispersions~\cite{Taft1965,Zeppenfeld1967,Kinyanjui2012,Wachsmuth2013,Wachsmuth2014,Kramberger2008}.
Nevertheless, the inclusion of the Coulomb interaction between electrons, beyond the RPA, is relevant to describe several spectroscopic features, such as the logarithmic divergence of the Fermi velocity near the neutrality point~\cite{siegel2011many} and the shape and position of the $\pi$ plasmon peak both in optics~\cite{Yang2009,Yang2011} and low momentum-transfer electron-energy loss (EEL) spectroscopy~\cite{guandalini_2023}.
Further, as suggested in previous studies, excitonic effects may be relevant in determining the optical phonon dispersion near the K point.
In particular, theoretical arguments based on  the renormalization group approach\cite{Basko_2008} and experimental Raman spectra~\cite{Venanzi_2023,PhysRevB.109.075420} suggest a steepening of the Kohn-anomaly spectrum guided by an enhancement of the electron-phonon interactions.
The same observations seem to be consistent with Raman spectra of bilayer graphene \cite{Graziotto2024-zk}.
These effects cannot be reproduced by a single particle framework, e.g. a TB model parametrized with experimental data or density-functional theory (DFT) in a local or semi-local approximation.
Thus, we refer to them as many-body effects.

Many-body effects in graphene have been studied with ab-initio methods in the framework of many-body perturbation theory.
Within the G$^0$W$^0$~\cite{Hedin_1965,Strinati_1982,Hybertsen_1986,Godby_1988} plus Bethe-Salpeter~\cite{Hedin_1965,Strinati_1988,Onida_2002} (BSE) formalism, the high-energy $\pi$ plasmon peak at $\sim 5$ eV have been successfully reproduced.~\cite{Yang2009,Yang2011,guandalini_2023}
G$^0$W$^0$ studies~\cite{Trevisanutto_2008,Attaccalite_2009,Guandalini_2024} have been also able to reproduce the low-energy ($\sim 100$ meV) Fermi velocity logarithmic behaviour, despite the renormalization is underestimated with respect to experimental data~\cite{Elias_11,Sonntag_2018}.
The reason of this discrepancy may be attributed to the necessity to include more classes of Feynman diagrams, e.g. by calculating self-consistently the screened interaction and quasi-particle properties. Also, ab-initio frameworks require a too high computational cost to converge graphene calculations at such low-energy scale, both for the freestanding and low-doping cases, probably affecting the robustness of the results.

Dirac cone models~\cite{CastroNeto_rev2009} with an Hartree-Fock self-energy~\cite{Sonntag_2018} or with renormalization group approaches~\cite{Gonzales_1994,Gonzales_1999} have shown to be able to reproduce the experimental Fermi velocity increase near the neutrality point.
However, the parameter choice is not unique between different studies, and there is no possibility to describe with the same framework higher-energies $\pi$-plasmon features.
A TB model plus many-body perturbation theory approach, as the one studied in Ref.~\onlinecite{Stauber_17}, successfully describes the Fermi velocity renormalization, containing also the physics to describe many-body effects above the range of validity of the Dirac cone model.
However, in Ref.~\onlinecite{Stauber_17}, the effect on doping on the velocity renormalization has not been studied and graphene is considered as a 2D sheet without thickness through the definition of the Coulomb potential.
In addition, up to now a BSE approach has never been used on top of a graphene TB model to describe both low- and high-energy many body effects in a unified framework.

In this work, we develop a computational scheme based on a TB model plus many-body perturbation theory corrections that is able to describe both low- and high- energy many-body effects in graphene with high accuracy, thus avoiding the need to resort to costly ab-initio methods.
The quasi-particle corrections are calculated with the static screened exchange (SX) approximation, while the electron-hole interaction is included by solving the BSE with the same static screened interaction used for the quasi-particle properties.
We focus on the effects of self-consistency between the quasi-particle corrections and the screened interaction, which we find to be important to compare with experiments.
Finally, we discuss the effect of both doping and of a static dielectric environment on both the band structure and the electronic response.

The paper is organized as follows: in Sec.~\ref{sec:MBPT} we summarize the main equations of many-body perturbation theory concerning the quasi-particle properties (Sec.~\ref{sec:QP}) and the evaluation of density response functions (Sec.~\ref{Sec_BSE}).
In Sec.~\ref{sec:dielectric_2D}, we focus on two-dimensional (2D) materials and describe how we approximate the electronic orbitals in the non-periodic direction.
The application of many-body perturbation theory to the graphene TB model is described in Sec.~\ref{sec:TB_MBPT} for what concerns quasi-particle corrections (Sec.~\ref{eq:TB_QP}), screened interaction (Sec.~\ref{sec:scr_int}), the BSE (Sec.~\ref{sec:TB_BSE}) and how to evaluate doping and environmental effects (Sec.~\ref{sec:dop_emb}).
Numerical details of the calculations are provided in Sec.~\ref{sec:numdet}.
In Sec.~\ref{sec:results}, we show the results. We firstly study the band structure, Fermi velocity renormalization and optical conductivity and compare with experiments in Sec.~\ref{sec:bands_vFermi_opt_cond}. 
Next, in Sec.~\ref{sec:dop_env_eff} we analyze doping and environmental effects.
The conclusions are drawn in Sec.~\ref{sec:conclusions}.
Details about the TB model are included in Appx.~\ref{appx:TB_model} and in Appx.~\ref{sec:sigma_TB} we derive the expression of the SX self-energy within the the TB model.
Next, in Appx.~\ref{appx:validation} we compare the TB results with ab-initio results from the literature.
Finally, in Appx.~\ref{appx:TB_all} we provide additional data about doping and environment effects in the group velocity renormalization and dielectric response functions.

%%%%%%%%%%%%%%%%%%%%%%%%%%%%%%%%%%%%%%%%%%%%%%%%%%%%%%%%%%%%%%%%%%%
\section{Summary of many-body perturbation theory}\label{sec:MBPT}
%%%%%%%%%%%%%%%%%%%%%%%%%%%%%%%%%%%%%%%%%%%%%%%%%%%%%%%%%%%%%%%%%%%

\begin{figure}
\includegraphics[width=\linewidth]{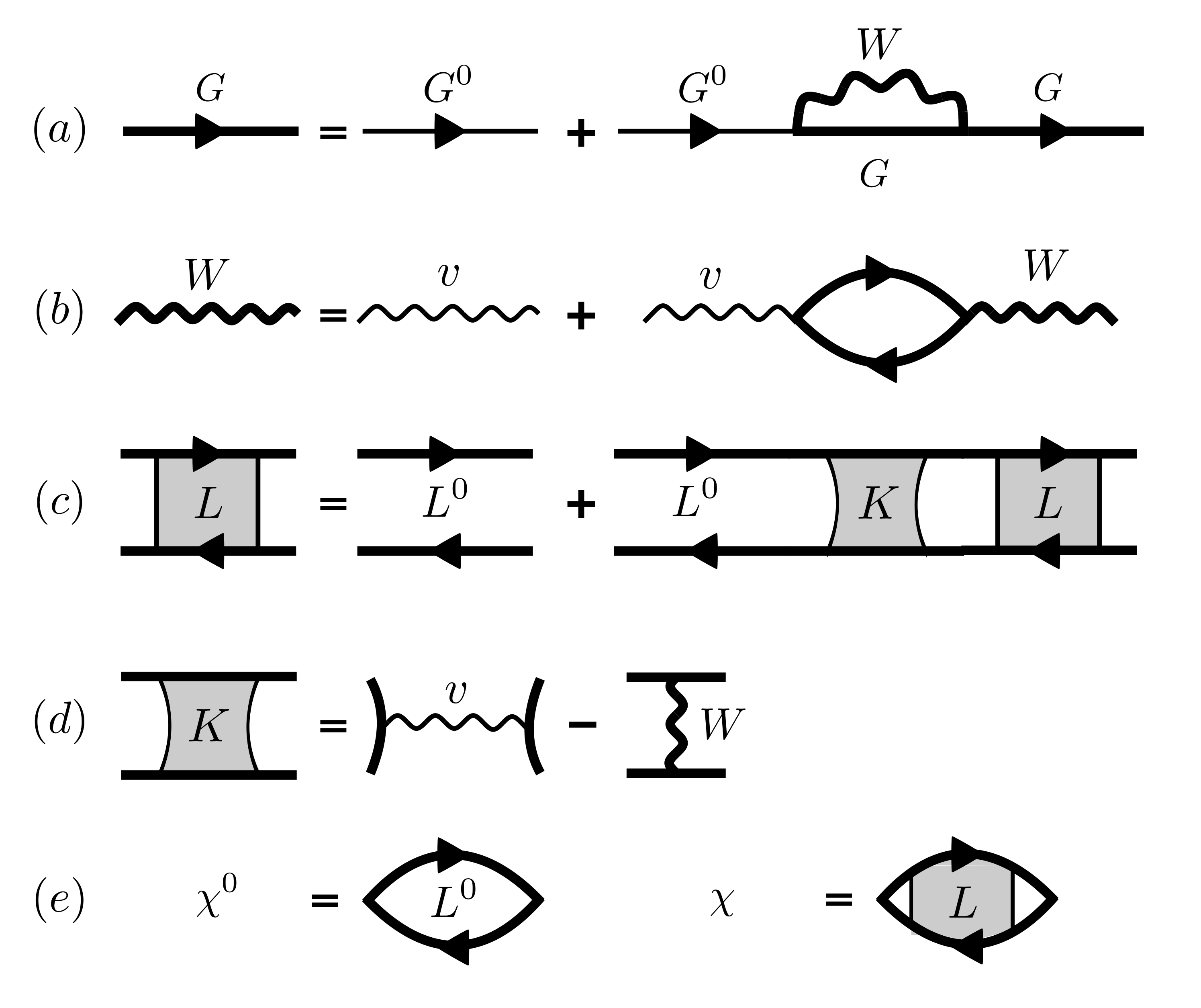}
\caption{Sketch of Feynman diagrams included in our calculations.
In particular, (a) the quasi-particle Dyson equation with a SX self-energy [see Eq.~\eqref{eq:SX_def}],
(b) the Dyson equation of the screened interaction $W$ in the RPA, (c) the Bethe-Salpeter equation [see Eq.~\eqref{eq:BSE_def}] with the kernel sketched in (d) [see Eq.~\eqref{eq:K_def}].
In (e), the definitions of $\chi^0$ and $\chi$ as contractions of the non-interacting $L^0$ and interacting $L$ electron-hole propagators.
(a) and (b) are self-consistently solved to obtain the band structure within the SX approximation, while all equations (a)-(e) are self-consistently solved to obtain the dielectric response properties.}
\label{fig:fey_diag}
\end{figure}

In this section, we review the main concepts of many body perturbation theory. In particular, we summarize the key ingredients for the calculation of quasi-particle and dielectric properties in Sec.~\ref{sec:QP} and ~\ref{Sec_BSE} respectively~\cite{Reining_16}.
The set of Feynman diagrams included in our calculations are sketched in Fig.~\ref{fig:fey_diag}

%%%%%%%%%%%%%%%%%%%%%%%%%%%%%%%%%%%%%%%%%%%%%%%%%%%%%%
\subsection{Quasi-particle excitations}\label{sec:QP}
%%%%%%%%%%%%%%%%%%%%%%%%%%%%%%%%%%%%%%%%%%%%%%%%%%%%%%

We consider a periodic crystal with $N$ electrons per unit cell interacting via the Coulomb interaction. In the following, we discretize reciprocal space momenta by considering a Born-Von Karman supercell made of $N_k$ unit cells. The Hamiltonian of this system may be written as
\begin{equation}
    \hat{H} = \hat{H}^0+\hat{H}^1 \ , 
\end{equation}
where $\hat{H}^0 = \hat{T}+\hat{V}$ is a single-particle Hamiltonian, $\hat{T}$ being the kinetic energy and $\hat{V}$ the potential one, and $\hat{H}^1$ is the Coulomb repulsion between electrons.
$\hat{V}$ includes the interaction between electrons and nuclei, and usually includes a mean field self-consistent potential, e.g. the Hartree or Kohn-Sham one, so to improve the starting point before applying perturbation theory.
If this is the case, the mean field potential must be subtracted from $\hat{H}^1$ so to avoid double counting.
In this work, $\hat{H^0}$ is described by a TB Hamiltonian which parameters are determined via ab-initio calculations, and therefore implicitly contain the effect of a self-consistent mean field potential.
The double counting is avoided by considering only the irreducible vertex in the expression for the electronic self-energy, as described later.

The single particle problem for the periodic system is solved by determining the eigenvalues $\eig^0_{m\k}$ (band energies) and eigenvectors $\orb^0_{m\k}$ (Bloch functions) of $\hat{H}^0$ as
\begin{equation}\label{eq:H_0}
    \hat{H}^0\orb^0_{m\k} = \eig^0_{m\k}\orb^0_{m\k} \ .
\end{equation}
From Eq.~\eqref{eq:H_0}, we define the non-interacting Green's function as
\begin{multline}
    G^0(\r,\r',\w) = 
    \lim\limits_{\eta\to 0^+}
    \sum\limits_{m\k}\orb^0_{m\k}(\r)\orb^{0*}_{m\k}(\r')\\
    \times\left[\frac{f^0_{m\k}}{\w-\eig^0_{m\k}-i\eta}
    +\frac{(1-f^0_{m\k})}{\w-\eig^0_{m\k}+i\eta} \right] \ ,
\end{multline}
where $f^0_{m\k}$ are the Fermi-Dirac occupation numbers corresponding to $\eig^0_{m\k}$.

In the framework of many body perturbation theory, it can be shown that interaction effects are effectively included in a similar fashion as Eq.~\eqref{eq:H_0} by including a dynamical self-energy operator $\Sigma$
\begin{equation}\label{eq:Ham_QP}
    \left[\hat{H}^0 + \Sigma(\eig_{m\k})\right]\orb_{m\k} = \eig_{m\k} \orb_{m\k} \ .
\end{equation}
$\eig_{m\k}$ and $\orb_{m\k}$ are the quasi-particle energies and orbitals that enter in the determination of the interacting single-particle Green's function
\begin{multline}
    G(\r,\r',\w) = \lim\limits_{\eta\to 0^+}\sum\limits_{m\k}\orb_{m\k}(\r)\orb^*_{m\k}(\r')\\
    \times\left[\frac{f_{m\k}}{\w-\eig_{m\k}-i\eta}
    +\frac{(1-f_{m\k})}{\w-\eig_{m\k}+i\eta} \right] \ ,
\end{multline}
where $f_{m\k}$ are the quasi-particle occupation numbers corresponding to $\eig_{m\k}$.

Eq.~\eqref{eq:Ham_QP} may be reformulated in terms of a Dyson equation that relate $G$ and $G^0$ via $\Sigma$ as 
\begin{equation}\label{eq:Dyson_G}
    G(\w) = G^0(\w) +G^0(\w)\Sigma(\w) G(\w)\ ,
\end{equation}
where space coordinates and integrals have been omitted to ease of notation.
In practice, Eq.~\eqref{eq:Dyson_G} is often simplified via the the quasi-particle orbital approximation, which consists in taking $\orb \approx \orb^0$.
With this approximation and assuming non-degeneracy for simplicity, quasi-particle energies may be written as
\begin{equation}
    \eig_{n\k} =  \eig^0_{n\k}+ \mel{n\k}{\Sigma(\eig_{n\k})}{n\k},
\label{Eq_QP_full}
\end{equation}
where we used the Dirac notation to identify the Bloch functions. By Taylor-expanding the self-energy $\Sigma$ around the bare energy $\eigo$, the above equation may also be rewritten as
\begin{equation}
    \eig_{n\k} =  \eig^0_{n\k}+ Z_{n\k}\mel{n\k}{\Sigma(\eig^0_{n\k})}{n\k},
\label{Eq_QP_full}
\end{equation}
where
$Z_{n\mathbf{k}} = \left[\left. 1-\langle n {\bf k}|\partial\Sigma(\omega)/\partial \omega| n {\bf k}  \rangle\right|_{\,\omega=\eigo_{n {\bf k}}} \right]^{-1}$
is the renormalization factor.

%%%%%%%%%%%%%%%%%%%%%%%%%%%%%%%%%%%%%%%%%%%%%%%%%%%
\subsubsection{Approximations of the self-energy}
%%%%%%%%%%%%%%%%%%%%%%%%%%%%%%%%%%%%%%%%%%%%%%%%%%%

The exact expression of the self-energy as a function of the screened interaction $W$ is given by the Hedin's equations.
A detailed explanation can be found in Ref.~\onlinecite{Reining_16}.
In this paragraph we remind the expressions for the self-energy that are relevant for this work.

Within the GW approximation\cite{Hedin_65,Strinati_82,Hybertsen_86,Godby_88}, the self-energy is given by first-order perturbation theory with respect to the screened interaction $W$ as
\begin{multline}\label{eq:def_GW}
    \Sigma^{\GW}(\r,\r',t-t') \\
    = iG(\r,\r',t-t')W(\r,\r',t-t'+0^+)\ .
\end{multline}
In the screened-exchange (SX) approximation, the dynamical screened interaction $W$ in Eq.~\eqref{eq:def_GW} is approximated by its static/instantaneous counterpart, thus
\begin{multline}\label{eq:SX_def}
    \Sigma^{\SX}(\r,\r',t-t') \\
    = iG(\r,\r',t-t')W(\r,\r')\delta(t-t') \ ,
\end{multline}
where $W(\r,\r')=W(\r,\r',\w = 0)$ is the static screened interaction.
The Fourier transform of $\Sigma^{\SX}$ over time $\Sigma^{\SX}(\r,\r',\w) = \Sigma^{\SX}(\r,\r')$ is frequency independent and reads
\begin{align}\label{eq_SX_def}
    \Sigma^{\SX}(\r,\r')
    = -n(\r,\r')W(\r,\r') \ , \\
    n(\r,\r') = \sum_{m\k} f_{m\k} \orb_{m\k}(\r)\orb^*_{m\k}(\r')\ ,
\end{align}
where $n(\r,\r')$ is the quasi-particle density matrix.
The quasi-particle Dyson equation in Eq.~\eqref{eq:Dyson_G} with a SX self-energy is sketched in terms of Feynman diagrams in Fig.~\ref{fig:fey_diag} (a).

Finally, in the G$^0$W$^0$ approximation, both the one-particle Green function $G$ and screened interaction $W$ are approximated by their bare counterparts
\begin{multline}
    \Sigma^{\GoWo}(\r,\r',t-t') \\
    = iG^0(\r,\r',t-t')W^0(\r,\r',t-t'+0^+) \ ,
\end{multline}
where $W^0$ is the screened interaction of the bare system.

The static approximation of the screened interaction may be applied also to the G$^0$W$^0$ self-energy, obtaining the non self-consistent screened-exchange (SX$^0$) approximation
\begin{equation}
        \Sigma^{\SXo}(\r,\r') = -n^0(\r,\r')W^0(\r,\r') \ .
\end{equation}

%%%%%%%%%%%%%%%%%%%%%%%%%%%%%%%%%%%%%%%%%%%%%%%%%%%%%%%%%%%%%%%%%%%%%%%%%%%%%%%%%
\subsection{Density response functions and dielectric properties}\label{Sec_BSE}
%%%%%%%%%%%%%%%%%%%%%%%%%%%%%%%%%%%%%%%%%%%%%%%%%%%%%%%%%%%%%%%%%%%%%%%%%%%%%%%%%

The electron-hole (e-h) propagator $L$ is the main quantity of interest in the calculation of density-density response and dielectric functions with many-body perturbation theory~\cite{Reining_16}. $L$ is determined via the BSE, which expresses the linear response of the single-particle Green's function $G$ to an external non-local source\cite{Hedin_65,Strinati_88,Onida_02} and reads
\begin{equation}\label{eq:BSE_def}
    L(\qph,\w) = L^0(\qph,\w) + L^0(\qph,\w)K(\qph)L(\qph,\w) \ .
\end{equation}
$\qph$ is the center-of-mass momentum of the e-h pair, 
$K$ is a kernel including the effect of interaction and $L^0$ is the bare e-h propagator, describing the non-interacting evolution of two quasi-particles, with the following real space representation:
\begin{multline}
    L^0(\r_1,\r_2,\r_3,\r_4,\w) = 2\sum\limits_{nm}\sum\limits_{\k\k'} (f_{m\k'}-f_{n\k})\\\times\lim\limits_{\eta \to 0^+}\frac{\orb_{n\k}(\r_1)\orb^*_{m\k'}(\r_2)\orb^*_{n\k}(\r_3)\orb_{m\k'}(\r_4)}{\w-(\eig_{n\k}-\eig_{m\k'})+i\eta} \ .
\end{multline}
If the single-particle Green's function $G$ is obtained from a SX self-energy, the kernel is composed by two terms
\begin{align}\label{eq:K_def}
K = K^{\H} + K^{\SX},
\end{align}
where $K^{\H}$ is the response due to the Hartree potential and $K^{\SX}$ contains the attractive e-h interaction.
Their matrix elements are conveniently expressed in the Hilbert space of e-h single particle transitions $\lbrace{\ket{t}\rbrace} = \lbrace |n\k \to m\k+\qph\rangle\rbrace$, where $n$, $m$ are band indexes and $\k$ is a wave-vector of the Brillouin zone (BZ).
In particular we consider $\bra{t_1} = \langle n_1\k_1 \to m_1\k_1+\qph|$ and $\ket{t_2} = |n_2\k_2 \to m_2\k_2+\qph\rangle$, where $\qph$ can assume values outside the first Brillouin zone, and obtain the matrix elements
\begin{multline}\label{eq:Xi_x}
    \mel{t_1}{K^{\H}(\qph)}{t_2} = 
    2\int d\r \int d\r'
    \orb^*_{m_1\k_1+\qph}(\r)
    \orb^*_{n_2\k_2}(\r')\\
    \times
    v(\r-\r')
    \orb_{n_1\k_1}(\r)
    \orb_{m_2\k_2+\qph}(\r'),
\end{multline}
\begin{multline}\label{eq:Xi_c}
    \mel{t_1}{K^{\SX}(\qph)}{t_2} =    
    -\int d\r \int d\r'
    \orb^*_{m_1\k_1+\qph}(\r)
    \orb^*_{n_2\k_2}(\r')\\
    \times
    W(\r,\r')
    \orb_{n_1\k_1}(\r')
    \orb_{m_2\k_2+\qph}(\r) \ .
\end{multline}
$v$ is the Coulomb interaction
\begin{align}
v(\r-\r') = \frac{1}{\varepsilon_r}\frac{1}{|\r-\r'|},
\label{eq:Coulkern}
\end{align}
where $\varepsilon_r$ is an effective dielectric constant taking into account the possible dielectric screening of an embedding isotropic medium. Notice that $\bra{t_1}$ and $\ket{t_2}$ contain electron-hole transitions mediated by the same wave-vector $\qph$, due to momentum conservation as made explicit in the following by Eqs. (\ref{eq:Xix_2D}) and (\ref{eq:Xic_2D}).
The BSE given by Eq.~\eqref{eq:BSE_def} along with the kernel in Eq.~\eqref{eq:K_def} are represented in terms of Feynman diagrams in Fig.~\ref{fig:fey_diag} (c) and (d).

In the SX approximation, the BSE may be recast as an eigenvalue problem of the two-body Hamiltonian 
\begin{equation}\label{eq:Ham_eh}
    \mel{t_1}{\hat{H}^{\eh}}{t_2} =
    E_{t_1}(\qph) \delta_{t_1t_2} +
    f_{t_1}\mel{t_1}{K(\qph)}{t_2} \ ,
\end{equation}
where $E_{t_1}(\qph) = (\eig_{m\k+\qph}-\eig_{n\k})$ are the e-h excitation energies and $f_{t_1} = f_{n\k}-f_{m\k+\qph}$ the occupation factors.
We note in Eq.~\eqref{eq:Ham_eh} we are including both resonant and anti-resonant transitions, thus we are not employing the Tamm–Dancoff approximation~\cite{Reining_16}.
The eigenvalues and eigenvectors of the e-h Hamiltonian in Eq.~\eqref{eq:Ham_eh} are excitonic energies and wavefunctions\cite{Gatti_13}.
The diagonal component of the density-density response function may be related to the inverse of the e-h Hamiltonian as
\begin{multline}\label{eq:chi_def}
    \chi(\qph,\w) \\
    = \sum\limits_{t_1t_2}\rho_{t_1}^{*}(\qph)\left[\mel{t_1}{\w\mathbb{1}-\hat{H}^{\eh}(\qph)}{t_2}\right]^{-1}\rho_{t_2}(\qph),    
\end{multline}
\begin{equation}
    \rho_{t}(\qph) = \mel{n\k}{e^{-i\qph\cdot r}}{m\k+\qph}.
\end{equation}
$\rho_{t}(\qph)$ are the oscillator strengths; if one is interested in the non-diagonal components of Eq. (\ref{eq:chi_def}), then only the oscillator strengths have to be changed in order to contain the dependence on local fields. 
The inverse dielectric function is expressed, in terms of $\chi$, as
\begin{equation}
    \inveps(\qph,\w) = 1 + v(\qph)\chi(\qph,\w) \ ,
\end{equation}
where $v(\qph)$ is the Fourier transform of the Coulomb interaction in Eq. (\ref{eq:Coulkern}).

For the remainder of this section, we suppose $\qph$ is a small momentum transfer near $\Gamma$.
As mentioned in the introduction, in this work we are interested in the optical absorption, i.e. the current response computed under the condition of null macroscopic electric field. The Coulomb kernel without the macroscopic component, $\overline{v}$ is obtained by setting $\overline{v}_{\G=0}(\qph) = 0$, where $v_{\G}(\qph)$ is the Fourier transform of the Coulomb potential. The corresponding kernel, given by Eq.~\eqref{eq:Xi_x} by substituting $v$ with $\bar{v}$, is labelled as $\overline{K}^{\H}$.

The correspondent density-density response function is
\begin{multline}\label{eq:chi_bar_def}
    \overline{\chi}(\qph,\w) \\=\sum\limits_{t_1t_2}\rho_{t_1}^{*}(\qph)\left[\mel{t_1}{\w\mathbb{1}-\hat{\bar{H}}^{\eh}(\qph)}{t_2}\right]^{-1}\rho_{t_2}(\qph) \ .
\end{multline}
In the above equation, $\hat{\bar{H}}^{\eh}$ is the same of Eq.~\eqref{eq:Ham_eh} but with $K^{\H}$ replaced by $\overline{K}^{\H}$.
The response function $\overline{\chi}$ is related to the macroscopic dielectric matrix via
\begin{equation}
    \eps(\qph,\w) =
    1 - v(\qph)\overline{\chi}(\qph,\w) \ .
\end{equation}
The formulation presented in this section is valid for three-dimensional crystals. For systems that are periodic in two directions but finite sized the third, e.g. graphene, more care as to be taken, as shown in the next section.

%%%%%%%%%%%%%%%%%%%%%%%%%%%%%%%%%%%%%%%%%%%%%%%%%%%%%%%%%%%%%%%%%%%%%%%%%%%%%%%%%%%%%%%%%%%%%%%
\section{Dielectric properties of a 2D crystal with finite thickness}\label{sec:dielectric_2D}
%%%%%%%%%%%%%%%%%%%%%%%%%%%%%%%%%%%%%%%%%%%%%%%%%%%%%%%%%%%%%%%%%%%%%%%%%%%%%%%%%%%%%%%%%%%%%%%

Graphene is a low-dimensional system.
In the last years, several studies have shown the importance to take into account the 2D nature of monolayer materials in the description of electronic screening~\cite{Huser_2013,Latini_15,qiu2016screening,Guandalini2023npjCM}. 
In ab-initio calculations based on a plane-wave representation of the orbitals, one possibility is to cutoff the Coulomb interaction so to remove spurious interaction between replicas~\cite{Beigi_2006, Rozzi_2006}.
In our model, we instead take into account the nanosized nature of graphene by hypothesizing an analytic form of the orbitals in the non-periodic directions, as discussed in Ref.~\onlinecite{PhysRevB.107.094308} and summarized in the following. 

We consider a 2D crystal, thus a system periodic in two dimensions $(x,y)$ and nanosized in the $z$ direction with a typical thickness $d$.
In the long wavelength limit ($|\qph|d \ll 1$) and within the RPA, the screening properties of the layer are the same of a dielectric sheet located at $z=0$ \cite{Sohier2015}. 
Outside this limit, finite thickness effects may become relevant, and they may depend on the nature of the electronic orbitals that determine the out-of-plane thickness of the material.
In this work, we are mainly interested in long-rage effects due to the Coulomb interaction.
For this reason, the accurate description of the $z$ dependence of the electronic orbitals is not critical, and we therefore approximate it by a rectangular profile \cite{PhysRevB.107.094308}.
Thus, we write the electronic orbitals as
\begin{equation}\label{eq:fin_thick}
    \orb_{i\k}(\r) \approx \orb_{i\k}(\r_{\parallel})\frac{\tilde{\Theta}(z)}{\sqrt{d}} \ ,
\end{equation}
where $\r_\parallel = (x,y)$ and  $\tilde{\Theta}(z)\equiv \Theta(|z|-d/2)$ is the Heaviside theta.
We emphasize that, since the Bloch theorem is valid only in the in-plane coordinates, the crystallographic momentum $\k$ is oriented along the 2D periodic directions.

Thanks to Eq. (\ref{eq:fin_thick}), $z$ integrals in the electronic self energy $\Sigma$ and kernels $K$ may be performed analytically to find effective equations of 2D sheets with a modified Coulomb interaction.

In fact, by substituting Eq.~\eqref{eq:fin_thick} into Eq.~\eqref{eq_SX_def}, the matrix elements of $\Sigma^{\SX}$ between two generic states $\ket{a}$ and $\ket{b}$ are
\begin{multline}\label{eq:SX_2D}
    \mel{a}{\Sigma^{\SX}}{b}
    = \sum\limits_{m\k}\sum\limits_{\G\G'} f_{m\k'}
    \mel{a}{e^{i(\qscr+\G)\cdot\r_{\parallel}}}{m\k}\\
    \times W^{\ZD}_{\G\G'}(\qscr)
    \mel{m\k}{e^{-i(\qscr+\G')\cdot\r_{\parallel}}}{b}
\end{multline}
where
\begin{equation}
    W^{\ZD}_{\G\G'}(\qscr) =
    \frac{1}{d^2}\int\limits_{-d/2}^{d/2} dz
    \int\limits_{-d/2}^{d/2} dz'
    W_{\G\G'}(\qscr,z,z').\label{eq:W2d}
\end{equation}
Eq. \ref{eq:W2d} is the Fourier transform of the screened interaction along the periodic directions and averaged along the non-periodic direction, where $\mathbf{q}$ is reduced to the first Brillouin zone and $\lbrace\G\rbrace$ are 2D reciprocal lattice vectors.

In the same way, by substituting Eq.~\eqref{eq:fin_thick} into  Eqs.~\eqref{eq:Xi_x}-\eqref{eq:Xi_c}, and Fourier transforming $v$ and $W$ along the periodic directions, we find
\begin{widetext}
    \begin{equation}\label{eq:Xix_2D}
    \mel{t_1}{K^{\H}(\qph)}{t_2} = 
    \frac{2}{A}\sum\limits_{\G}
    \mel{m_1\k_1+\qph}{e^{i(\qph+\G)\cdot\r_{\parallel}}}{n_1\k_1}
    v^{\ZD}_{\G}(\qph)
    \mel{n_2\k_2}{e^{-i(\qph+\G)\cdot\r'_{\parallel}}}{m_2\k_2+\qph} ,
\end{equation}

\begin{equation}\label{eq:Xic_2D}
    \mel{t_1}{K^{\SX}(\qph)}{t_2} =
    \frac{1}{A}\sum\limits_{\G\G'}
    \mel{m_1\k_1+\qph}{e^{i(\k_1-\k_2+\G)\cdot\r_{\parallel}}}{m_2\k_2+\qph}
    W_{\G\G'}^{\ZD}(\k_1-\k_2)
    \mel{n_2\k_2}{e^{-i(\k_1-\k_2+\G')\cdot\r'_{\parallel}}}{n_1\k_1} ,
\end{equation}
where $A$ is the graphene unit-cell area and 
\begin{equation}\label{eq:v_2D_fin_thick}
    v^{\ZD}_{\G}(\qscr) = 
    \frac{1}{d^2}\int\limits_{-d/2}^{d/2} dz
    \int\limits_{-d/2}^{d/2} dz' v_{\G}(\qscr,z,z') = 
    \frac{2\pi}{\varepsilon_r|\qscr+\G|}F(|\qscr+\G|d)
\end{equation}
\end{widetext}
is an effective 2D Coulomb interaction that includes finite (orbital-independent) thickness effects through the form factor
\begin{multline}
    F(x) = \frac{1}{d^2} \int\limits_{-d/2}^{d/2} dz \int\limits_{-d/2}^{d/2} dz' e^{-|x/d||z-z'|}\\ = \frac{2}{x}\left(1+\frac{e^{-x}-1}{x}\right) \ .
\end{multline}

The Fourier transform of the density-density response function, assuming the $z$ profile of Eq.~\eqref{eq:fin_thick}, shows the same rectangular $z$ profile as the electronic orbitals
\begin{equation}
    \chi(\qph,z,z',\w) = \chi^{\ZD}(\qph,\w)\frac{\tilde{\Theta}(z)\tilde{\Theta}(z')}{d^2} \ .
\end{equation}
This is inherited by the out-of-plane form of $\chi^0$, which is directly expressed as a function of orbitals, via the Dyson equation. The 2D effective density-density response functions $\chi^{\ZD}$ and $\overline{\chi}^{\ZD}$ are evaluated from Eqs.~\eqref{eq:chi_def} and \eqref{eq:chi_bar_def} with the kernels defined in Eqs.~\eqref{eq:Xix_2D}-\eqref{eq:Xic_2D}.

The 3D Fourier transforms of the inverse and macroscopic dielectric functions are not  defined for a 2D crystal, as described in more detail in Ref.~\onlinecite{Nazarov_15}.
Still, we can define effective 2D dielectric functions averaged along the $z$ direction as
\begin{multline}\label{eq:inv_eps_2D}
    \varepsilon^{-1,\ZD}(\qph,\w) = \frac{1}{d^2}
    \int\limits_{-d/2}^{d/2} dz
    \int\limits_{-d/2}^{d/2} dz'
    \inveps(\qph, z, z', \w) \\
    =1+v^{\ZD}(\qph)\chi^{\ZD}(\qph,\w),
\end{multline}
\begin{multline}\label{eq:eps_2D}
    \eps^{\ZD}(\qph,\w) = \frac{1}{d^2}
    \int\limits_{d/2}^{d/2} dz
    \int\limits_{d/2}^{d/2} dz'
    \eps_{M}(\qph, z, z', \w) \\
    = 1-v^{\ZD}(\qph)\overline{\chi}^{\ZD}(\qph,\w).
\end{multline}

Finally, the long wavelength limit of the 2D effective conductivity is related to the 2D dielectric function via the following relation~\cite{Cudazzo_2011}:
\begin{equation}\label{eq:cudazzo_formula}
    \eps^{\ZD}(\qph\to 0,\omega)-1 = \lim\limits_{|\qph| \to 0} 2\pi i\frac{\sigma^{\ZD}(\omega)Q}{\omega+i\eta } \ .
\end{equation}
By combining Eq.~(\ref{eq:cudazzo_formula}) with Eq.~(\ref{eq:inv_eps_2D}), we find the 2D optical conductivity as
\begin{equation}
    \sigma^{\ZD}(\w) = \lim\limits_{|\qph| \to 0} \frac{i(\w+i\eta)\overline{\chi}^{\ZD}(\qph,\w)}{|\qph|^2} \ .
\end{equation}
We note that $\overline{\chi}^{\ZD} \to 0$ as $|\qph| \to 0$ with a quadratic power, thus the optical conductivity is finite.

%%%%%%%%%%%%%%%%%%%%%%%%%%%%%%%%%%%%%%%%%%%%%%%%%%%%%%%%%%%%%%%%%%%%%%%%%%%%%%%%%%%%%%%%
\section{Many-body perturbation theory applied to graphene TB model}\label{sec:TB_MBPT}
%%%%%%%%%%%%%%%%%%%%%%%%%%%%%%%%%%%%%%%%%%%%%%%%%%%%%%%%%%%%%%%%%%%%%%%%%%%%%%%%%%%%%%%%

In this section, we describe how we evaluate the quasi-particle corrections and response functions of graphene starting from the five nearest neighbour TB model for graphene described in Appx.~\ref{appx:TB_model}.
In addition we show how we model the screened interaction $W$ for graphene.

%%%%%%%%%%%%%%%%%%%%%%%%%%%%%%%%%%%%%%%%%%%%%%%%%%%%%%%%%%%
\subsection{Screened-exchange self-energy}\label{eq:TB_QP}
%%%%%%%%%%%%%%%%%%%%%%%%%%%%%%%%%%%%%%%%%%%%%%%%%%%%%%%%%%%

Our aim is to evaluate the SX self-energy corrections within the TB model for graphene.
In this case, the quasi-particle Hamiltonian, given by Eq.~\eqref{eq:Ham_QP}, may be written in the basis of the Bloch functions as
\begin{align}\label{eq:H_TB_SX}
H^{\SX}_{\k} = H^{0}_{\k}+ \mqty(
        0 & \Sigma_{\k}^{\SX} \\
        \Sigma_{\k}^{\SX*} & 0
    )\ ,
    H^{0}_{\k} = 
    \mqty(
        \mathcal{g}_{\k} & \mathcal{f}_{\k} \\
        \mathcal{f}^*_{\k} & \mathcal{g}_{\k}
    )\ .
\end{align}
$H^{0}_{\k}$ is the TB Hamiltonian, obtained as described in App. \ref{appx:TB_model}; its matrix elements are defined in Eq.~\eqref{eq:gk} and \eqref{eq:fk}.
For freestanding graphene (i.e. at zero doping), the diagonal terms of the self-energy are $\k$- and atom-independent, thus they can be neglected.
For simplicity, we neglect the diagonal terms of the self-energy even in the doped case, as for low doping these terms are not expected to change the physical conclusions.

The eigenvalues of the Hamiltonian in Eq.~\eqref{eq:H_TB_SX} give the quasi-particle energies
\begin{equation}\label{eq:eig_SX}
    \eig_{\pi^*/\pi\k} = \mathcal{g}_{\k}\pm|\mathcal{f}_{\k}+\Sigma_{\k}^{\SX}| \ .
\end{equation}
Self-energy corrections do not change the phase of the spin components $\phi$ in a first nearest neighbour tight-binding model, as demonstrated in the supporting information of Ref.~\onlinecite{Stauber_17}.
We numerically verify that also in our five-nearest neighour tight-binding model the phase $\phi$ [defined in Eq.~\eqref{eq:phi_TB}] do not appreciably change.
Thus, we can safely use the quasi-particle orbital approximation $\orb_{n\k} \approx \orb^0_{n\k}$ already described in Sec.~\ref{sec:QP}.

We now consider Eq.~\eqref{eq:SX_2D}, and disregard non-diagonal local field effects. As we derive in Appx.~\ref{sec:sigma_TB}, $\Sigma_{\k}^{\SX}$ assumes the form
\begin{multline}\label{eq:SX_TB}
    \Sigma^{\SX}_{\k} = \frac{-1}{N_kA}\sum\limits_{\k'\G'}\Delta f_{\k'}|\mathcal{F}_{\k-\k'}|^2
    W^{\ZD}_{\G'}(\k-\k') \phi^*_{\k'-\G'}\ ,
\end{multline}
where $\Delta f_{\k'} = f_{\pi^*\k'}-f_{\pi\k'}$ , $\phi$ is the spinorial phase defined in Eq.~\eqref{eq:phi_TB} and $\mathcal{F}_{\k-\k'}$ is the atomic form factor defined in Eq.~\eqref{eq:pz_momentum}.
We note $W$ depends only on one $\G$ vector as we neglect some local field effects in screening.
This approximation is described in Sec.~\ref{sec:scr_int}.

The screened Coulomb of Eq.~\eqref{eq:SX_TB} implicitly depends on the set of energies $\lbrace \eig_{\pi^*/\pi\k} \rbrace$ determined via Eq. \ref{eq:SX_TB}. This means that a self-consistent procedure is needed in order to evaluate $W$, in the SX approximation, which we implement in our framework. Beside this, we also study the case where the screened interaction is evaluated once and for all, in a non self-consistent way, using the TB energies; we refer to this screened interaction as $W^0$, and to the approximation as SX$^0$. In this approximation, Eqs.~\eqref{eq:H_TB_SX} and \eqref{eq:eig_SX} are still valid, but with a self-energy containing $W^0$ as
\begin{multline}
    \Sigma^{\SXo}_{\k} = \frac{-1}{N_kA}\sum\limits_{\k'\G'}\Delta f^0_{\k'}|\mathcal{F}_{\k-\k'}|^2
    W^{0,\ZD}_{\G'}(\k-\k') \phi^*_{\k'-\G'} \ ,
\end{multline}
where $\Delta f^0_{\k'} = f^0_{\pi^*\k'}-f^0_{\pi\k'}$.
As it regards the electronic populations, we find that for the electronic temperature studied in this work ($T = 0$ or $T=4$ K), we can safely set $\Delta f_{\k'} \approx \Delta f^0_{\k'}$.
In the following subsection, we model the screened interaction in graphene for both SX and SX$^0$ approximations.

%%%%%%%%%%%%%%%%%%%%%%%%%%%%%%%%%%%%%%%%%%%%%%%%%%%%%
\subsection{Screened interaction}\label{sec:scr_int}
%%%%%%%%%%%%%%%%%%%%%%%%%%%%%%%%%%%%%%%%%%%%%%%%%%%%%
In this section we describe our model for the screened interaction, within the RPA, that is used both in the evaluation of the quasi-particle self-energy Eq. (\ref{eq:SX_2D}) and correlation kernel of the BSE equation Eq. (\ref{eq:Xic_2D}).
The effective screened interaction of a 2D crystal within the orbital approximation given by Eq.~\eqref{eq:fin_thick} may be written as
\begin{equation}\label{eq:W_Q}
    W^{\ZD}_{\G\G'}(\qph,\w) = \tilde{\varepsilon}^{-1,\ZD}_{\G\G'}(\qph,\w)\vcoul^{\ZD}_{\G'}(\qph) \ ,
\end{equation}
where $\tilde{\varepsilon}^{-1,\ZD}$ is the inverse dielectric function, obtained from Eq.~\eqref{eq:inv_eps_2D} with a density-density response $\tilde{\chi}$ obtained within the RPA
\begin{multline}\label{eq:RPA_2D}
    \tilde{\chi}^{\ZD}_{\G\G'}(\qph,\w)\\ = \left[\mathbb{1}-\chi^{0,\ZD}_{\G\G''}(\qph,\w)v^{\ZD}_{\G''}(\qph)\right]^{-1}\chi^{0,\ZD}_{\G''\G'}(\qph,\w) \ .
\end{multline}
A sketch of the Feyman diagrams included in the screened interaction $W$ can be found in Fig.~\ref{fig:fey_diag} (b).

$\chi^{0,\ZD}$ is the Fourier transform of the irreducible polarizability  
\begin{widetext}
\begin{equation}\label{eq:chi_irr}
    \chi^{0,\ZD}_{\G\G'}(\qscr,\w) = 2
    \lim\limits_{\eta \to 0^+}\sum\limits_{m,n=\pi}^{\pi^*}\sum\limits_{\k}(f_{m\k}-f_{n\k+\qscr})\frac{\mel{n\k+\qscr}{e^{i(\qscr+\G)\cdot\r}}{m\k}
    \mel{m\k}{e^{-i(\qscr+\G')\cdot\r}}{n\k+\qscr}}{\w-(\eig_{n\k+\qscr}-\eig_{m\k})+i\eta} \ .
\end{equation}
\end{widetext}
For simplicity, we neglect off-diagonal matrix elements of $\chi^{0,\ZD}_{\G\G'}$, which has been shown to be a good approximation for graphene \cite{Sohier2015}.
Thus, we write $\chi^{0,\ZD}_{\G\G'} = \chi^{0,\ZD}_{\G}\delta_{\G\G'}$, $W^{\ZD}_{\G\G'} = W^{\ZD}_{\G}\delta_{\G\G'}$ and we define the short-hand notations $\chi^{0,\ZD}_{\G} \equiv \chi^{0,\ZD}_{\G\G}$ and $W_{\G} = W_{\G\G}$.
Within the proposed approximations, the screened interaction can be written as
\begin{equation}\label{eq:W_2D_TB}
    W^{\ZD}_{\G}(\qscr) = \frac{v^{\ZD}_{\G}(\qscr)}{1-v^{\ZD}_{\G}(\qscr)\chi^{0,\ZD}_{\G}(\qscr)} \ .
\end{equation}
The electronic screening is encoded in the irreducible polarizability.

In the self-consistent SX, Eqs.~\eqref{eq:eig_SX}, \eqref{eq:SX_TB}, \eqref{eq:chi_irr} and \eqref{eq:W_2D_TB} must be solved self-consistently, as the irreducible polarizability depend on the quasi-particle energies.
In Fig.~\ref{fig:fey_diag} (a) and (b) we sketch the Feynman diagrams included in the self-consistent SX.
This is analogous to self-consistent quasi-particle  GW theory~\cite{Schilfgaarde_2006} with the static approroximation of the screened interaction.
In the non-self consistent case (SX$^0$) we simply use $W^0$.

Practically, we approximate the static screened interaction $W$ or $W^0$ of graphene with screening obtained from low energy models. 
In fact, it is known from the literature that low-energy excitations are the most important ones in the description of long-wavelength screening~\cite{Guandalini_2024}, that is the responsible of graphene many-body features.
This procedure speeds up the calculations, as the evaluation of the screened interaction is usually the most time-consuming part of a many-body perturbation theory simulation.
We now list the explicit forms of $W$ and $W^0 $ used in this work for different doping levels. The effect of the embedding in the screened interaction is treated in Sec.~\ref{sec:dop_emb}.

%%%%%%%%%%%%%%%%%%%%%%%%%%%%%%%%%%%%%%%%%%%%%%%%%%%%%%%%%%%%%%%%
\subsubsection{$W^0$ of freestanding graphene}\label{sec:Wo_free}
%%%%%%%%%%%%%%%%%%%%%%%%%%%%%%%%%%%%%%%%%%%%%%%%%%%%%%%%%%%%%%%

For this zero doping case, we approximate the static irreducible polarizability, given by Eq.~\eqref{eq:chi_irr}, with the one of the Dirac cone model at $T=0$ K
\begin{equation}\label{eq:chirr_Dirac}
    \chi^{0,\ZD}_{\G}(\qscr) = -\frac{|\qscr+\G|}{4v_{\F}} \ ,
\end{equation}
where $v_{\F}$ is the Fermi velocity of the TB model.

By inserting Eqs.~\eqref{eq:chirr_Dirac} and \eqref{eq:v_2D_fin_thick} into Eq.~\eqref{eq:W_2D_TB}, the screened interaction can be evaluated analytically.

%%%%%%%%%%%%%%%%%%%%%%%%%%%%%%%%%%%%%%%%%%%%%%%%%%%%%%%
\subsubsection{$W^0$ of doped graphene}\label{sec:Wo_dop}
%%%%%%%%%%%%%%%%%%%%%%%%%%%%%%%%%%%%%%%%%%%%%%%%%%%%%%%

Also in this case we approximate the static irreducible polarizability of the TB model with the one of the Dirac cone model at $T = 0$ K.
In case of finite doping, the irreducible polarizability has been already derived in Ref.~\onlinecite{Hwang_07}, and it reads
\begin{widetext}
\begin{equation}\label{eq:chirr_Dirac_dop}
    \chi_{\G}^{0,\ZD}(\qscr) = -\frac{2\kF}{\pi v_{\F}}
    \begin{dcases}
    \begin{aligned}
    1 &, \  \  \ |\qscr+\G| \le 2\kF\\
    1+\frac{\pi |\qscr+\G|}{8\kF}-\frac{1}{2}\sqrt{1-\frac{4\kF^2}{|\qscr+\G|^2}}-
    \frac{|\qscr+\G|}{4\kF}\sin^{-1}\frac{2\kF}{|\qscr+\G|} &, \  \  \ |\qscr+\G| > 2\kF \ 
    \end{aligned}
    \end{dcases},
\end{equation}
\end{widetext}
where $k_{\F} = \sqrt{\pi n}$ is the Fermi wavevector, $n$ being the carrier (electrons or holes) density.
In the limit $k_{\F} \to 0$, Eq.~\eqref{eq:chirr_Dirac_dop} tends to Eq.~\eqref{eq:chirr_Dirac}.

As in the previous case, by inserting Eqs.~\eqref{eq:chirr_Dirac} and \eqref{eq:v_2D_fin_thick} into Eq. \eqref{eq:W_2D_TB}, the screened interaction can be evaluated analytically.

%%%%%%%%%%%%%%%%%%%%%%%%%%%%%%%%%%%%%%%%%%%%%%%%%%%%%%%%%%%%
\subsubsection{$W$ of freestanding graphene}\label{sec:W_free}
%%%%%%%%%%%%%%%%%%%%%%%%%%%%%%%%%%%%%%%%%%%%%%%%%%%%%%%%%%%%

In this case, we want to calculate the irreducible polarizability obtained from the quasi-particle band structure in Eq.~\eqref{eq:eig_SX}.
The Dirac cone approximation can no more be applied, due to the Fermi velocity renormalization with a logarithmic divergence at the neutrality point, as shown later in detail.

However, in Ref.~\onlinecite{Guandalini_2024} a proper low-energy model of the quasi-particle band structure is derived.
We suppose the band structure of the quasi-particle cone has the following functional form
\begin{equation}\label{eq:E_sm_model}
\begin{split}
    \eig_{\pi^*/\pi}(k) &=\pm v_{\F} k \pm f v_{\F} \frac{k}{2} \left[\cosh^{-1}{\left(\frac{2 k_c}{k}\right)} +\frac{1}{2}\right]  \ ,
\end{split}
\end{equation}
where $f$ and $k_c$ are numerical parameters fitted with numerical data obtained by sampling the surroundings of the Dirac point.

Next, the k sum of the irreducible polarizability in Eq.~\eqref{eq:chi_irr} is evaluated numerically in elliptic coordinates to exploit the  symmetries of the sum.
The resulting screened interaction is evaluated on a numerical logarithmic grid around $q = 0$, then evaluated when needed by the code with a linear interpolation scheme.

%%%%%%%%%%%%%%%%%%%%%%%%%%%%%%%%%%%%%%%%%%%%%%%%%%%%%
\subsubsection{$W$ of doped graphene}\label{sec:W_dop}
%%%%%%%%%%%%%%%%%%%%%%%%%%%%%%%%%%%%%%%%%%%%%%%%%%%%

The procedure we used to calculate the screened interaction is the same as Sec.~\ref{sec:W_free}, but the low-energy quasi-particle band structure is interpolated with Chebyshev polynomials instead of fitted with Eq.~\eqref{eq:E_sm_model}. In fact, to the best of our knowledge, an analytical low-energy model that can be exploit to speed up the calculation is not present in the literature.

%%%%%%%%%%%%%%%%%%%%%%%%%%%%%%%%%%%%%%%%%%%%%%%%%%%%%%%
\subsection{Bethe-Salpeter equation}\label{sec:TB_BSE}
%%%%%%%%%%%%%%%%%%%%%%%%%%%%%%%%%%%%%%%%%%%%%%%%%%%%%%%

The kernels in Eqs.~\eqref{eq:Xix_2D} and \eqref{eq:Xic_2D} are evaluated by using the matrix elements in Eq.~\eqref{eq:mel_TB}.
We model screened interaction $W^{\ZD}$ as explained in Sec.~\ref{sec:scr_int} and use it in the kernel of Eq.~\eqref{eq:Xic_2D}.
Non interacting excitation energies $E_{t_1}(\qph)$ are evaluated or from Eq.~\eqref{eq:eig_SX} with SX quasi-particle energies or from G$^0$W$^0$ TB parameters (see Appx.~\ref{appx:validation}). % We will consider both these cases in the results section.
From a diagrammatic perspective, the solution of the BSE requires the self-consistent solution of all the diagrammatic equations sketched in Fig.~\ref{fig:fey_diag}.

%%%%%%%%%%%%%%%%%%%%%%%%%%%%%%%%%%%%%%%%%%%%%%%%%%%%%%%%%%%%%%%%%
\subsection{Doping and environmental effects}\label{sec:dop_emb}
%%%%%%%%%%%%%%%%%%%%%%%%%%%%%%%%%%%%%%%%%%%%%%%%%%%%%%%%%%%%%%%%%

We model doping effects by changing the chemical potential $\mu$ in the occupation numbers $f_{m\k}$ and using a proper screened interaction as explained in Sec.~\ref{sec:scr_int}.
In principle, as doping changes the Kohn-Sham potential, the DFT TB parametrization is doping dependent.
However, we neglect the TB fit dependence on doping, as this effect is negligible~\cite{Sohier2015}.

Environmental effects are taken into account with a static uniform dielectric constant $\varepsilon_r \neq 1$ in the Coulomb interaction $v^{\ZD}$ given by Eq.~\eqref{eq:v_2D_fin_thick}.
For what concerns the screened interaction, the environment reduces its intensity through a reduction of $v^{\ZD}$ in Eq.~\eqref{eq:W_Q}.
However, there is also a screening reduction, enchancing $W$, given by an increase of the inverse dielectric function, as can be seen from Eq.~\eqref{eq:inv_eps_2D}.
The net effect is a reduction of $W$ with respect to the freestanding case, but with a lower value than $\varepsilon_r$ due to the reduced screening caused by the environment.

%%%%%%%%%%%%%%%%%%%%%%%%%%%%
\section{Numerical details}
%%%%%%%%%%%%%%%%%%%%%%%%%%%

\label{sec:numdet}
\textit{Ab-initio}---DFT calculations have been done with the QUANTUM ESPRESSO package~\citep{QE_2020} within the local-density approximation exchange-correlation functional.~\citep{LDA} 
We adopted norm-conserving pseudopotentials to model the electron-ion interaction and the kinetic energy cutoff for the wavefunctions is set to $90$ Ry.

G$^0$W$^0$ quasi-particle band structure and the BSE excitation spectra have been calculated with the YAMBO package~\cite{yambo_2009,yambo_2019}.
The bare Coulomb interaction has been truncated with a slab cutoff in order to remove spurious interactions between replicas.~\cite{Beigi_2006, Rozzi_2006, PhysRevB.96.075448}
The screened interaction $W$ has been considered within the plasmon pole approximation~\cite{Godby_1989} in G$^0$W$^0$ calculations and with the static approximation in BSE calculations. 
Following Ref.~\onlinecite{guandalini_2023}, we used plasmon pole instead of multi-pole G$^0$W$^0$ band structure as an input to static BSE, as the detailed description of the frequency-dependence of $W$ in the band structure requires the same frequency treatment in the BSE kernel.

A cutoff energy of $10$ Ry has been applied to its Fourier transform.
We used $100$ states both in the sum-over-states of the irreducible polarizability and of the self-energy.
We numerically verified that the above numerical parameters converge the gap at M within $10$ meV.
The optical conductivity and density-density response function have been calculated by selecting $3$ valence and $12$ conduction bands nearer to the Fermi level and a finite damping $\eta = 0.1$ eV.
The optical conductivity spectrum is additionally smeared out with a gaussian smearing to smooth the spectrum with a standard deviation of $0.17$ eV.
We used a Mohnkhorst-Pack grid~\cite{Monkhorst_1976} of $60\times 60$ in G$^0$W$^0$ and of $90\times 90$ in the BSE calculations.

\textit{TB}---All TB calculations have been done with a python3 code designed by the authors. The parameters of the five nearest neighbour TB are fitted to the DFT or G$^0$W$^0$ calculations. When we perform SX$^0$ or SX calculations, the TB parameters are fitted to DFT. The electronic temperature is set to $4$ K. 
In the Hartree and SX kernels, given by Eq.~\eqref{eq:Xix_2D}, the sum over $\G$ was performed up to the first three shells of $\G$ vectors.
%Instead, in the SX kernel [see Eq.~\eqref{eq:Xic_2D}] we cutoff the screened interaction $W$ for $q > 4\pi/3a$, where $a$ is the graphene lattice constant.
We stress than in our 2D formulation the $\G$ vectors corresponds to those oriented along the periodic directions.
The local fields oriented along the non-periodic direction are implicitly taken into account via the real space formulation provided in Sec.~\ref{sec:dielectric_2D}.
We used a Monkhorst-Pack grid of $181\times 181$ with $\eta = 0.2$ eV in the calculations  of \ref{fig:Zeff_check} and Fig.~\ref{fig:el_resp_vs_yambo}, while a grid of $361\times 361$ in the response functions in Fig.~\ref{fig:TB_vs_exp_opt}, right panel of Fig.~\ref{fig:TB_dop_env} and Fig.~\ref{fig:el_resp_TB}, with a damping $\eta = 0.1$ eV.
All other calculations have been performed with telescopic grids~\cite{Binci_2021} with $p=3$, $\mathcal{N}=25$, $l=8$ and $L=12$, with the exception of the doped response functions of doped graphene obtained with $l=8$ and $L=10$.
We checked numerically that all the calculations are converged.

\textit{Bethe-Salpeter}---The BSE, both in ab-initio and TB calculations, is solved with a pseudo-Hermitian Lanczos algorithm, that is particularly efficient to calculate the propagators [see Eqs.~\eqref{eq:chi_def} and \eqref{eq:chi_bar_def}] at different $\omega$ in just one calculation, due to an iterative procedure able to find a basis representation where the BSE matrix is tridiagonal.~\cite{Benedict_1998,Benedict_1999,Gruning_2011}

%%%%%%%%%%%%%%%%%%%%%%%%%%%%%%%%%%%%%
\section{Results}\label{sec:results}
%%%%%%%%%%%%%%%%%%%%%%%%%%%%%%%%%%%%%

In this section, we summarize the results obtained with the computational scheme proposed in the theoretical section.
Firstly, we discuss the quasi-particle band structure, and compare the Fermi velocity renormalization with experiments.
Then, we study the optical conductivity.
In both cases, we focus on the role of self-consistency in the quasi-particle correction.
Finally, we study the effects of doping and dielectric embedding on both the group velocity near the K point and the optical conductivity.
A validation between our TB and ab-initio calculations can be found in Appx.~\ref{appx:validation}. In Appx.~\ref{appx:TB_all}, we show a wider range of data sets for completeness, along with the density-density response function at $q=$ K.

In all of the figures in the results section, we highlight two region intervals in the low-energy sector of the graphene spectrum. 
The first region ($|\eig_{n\k}| \lesssim 0.1$ eV) is typically relevant for low-doping electron-phonon mediated transport properties or for the hydrodynamic transport regime dominated by e-e interaction~\cite{Elias_11,Crossno_2016}. In such a region, the band structure is strongly affected by the many-body renormalization of Fermi velocities, with an expected large impact on the evaluation of transport properties.
The second region is determined by the phonon energy range ($|\eig_{n\k}| \lesssim 0.2$ eV), where electron excitations are in resonance with optical phonons, originating non-adiabatic effects~\cite{Lazzeri_2006}.

\subsection{Band structure, Fermi velocity renormalization and optical conductivity in the SX approximation}\label{sec:bands_vFermi_opt_cond}

\begin{figure}
\includegraphics[width=\linewidth]{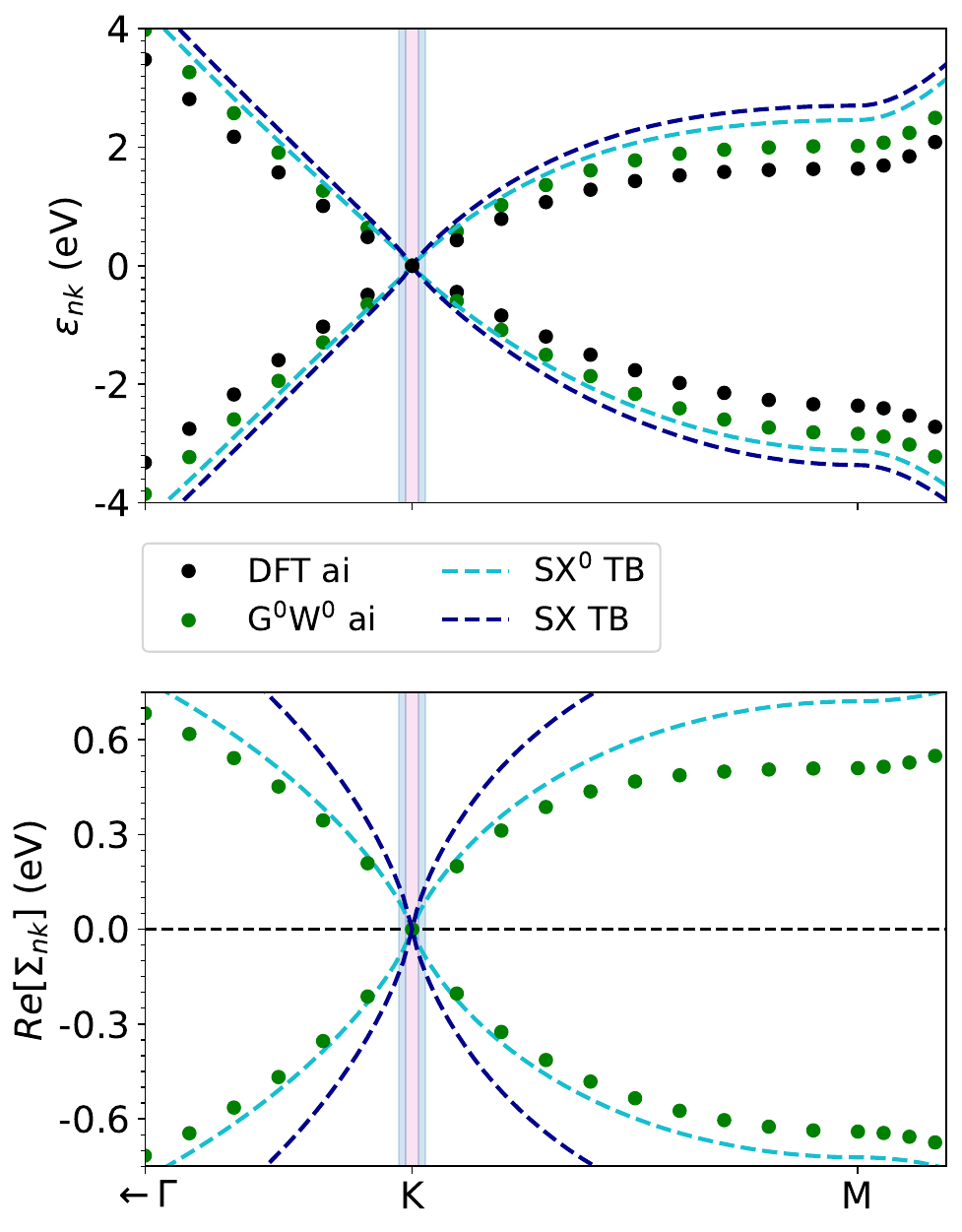}
\caption{Electronic energies (top) and real part of the self-energies (bottom) of the $\pi/\pi^*$ bands of freestanding
graphene obtained with ab-initio calculations (dots) and our work (dashed lines). Black is used for DFT calculations, green for G$^0$W$^0$ and (light) blue for (SX$^0$) SX. 
Ab-initio (tight-binding) calculations are labeled with `ai' (`TB').
The pink shaded region indicates the energy range relevant for Fermi energy renormalization, while the blue-gray region the energy range of optical phonons.}
\label{fig:bands_vs_yambo}
\end{figure}

In Fig.~\ref{fig:bands_vs_yambo}, we compare the quasi-particle band structure and self-energy of freestanding graphene in the vicinity of the Fermi level.
The ab-initio and G$^0$W$^0$ results are taken from Ref.~\onlinecite{Guandalini_2024}.
We note the k-point sampling of the G$^0$W$^0$ band structure is represented by a coarse grid of points, while the TB band structure with continuous line.
This trend reflects the possibility to converge the TB band structure very near to the Dirac point, as opposite to the ab-initio G$^0$W$^0$ calculation.
SX$^0$ bands are slightly more steep than the G$^0$W$^0$ band structure, mainly due to the missing of the renormalization factor $Z_{n\mathbf{k}}$, that is $\approx 0.75$ in the G$^0$W$^0$ calculation.
The $Z_{n\mathbf{k}} \neq 1$ trend is a signature of the energy-dependence of the self-energy.
In fact, the SX$^0$ TB and G$^0$W$^0$ ab-initio self-energies are very similar in the vicinity of the K point, in particular in the low-energy range relevant for transport and phonon properties.
Near the M point, the SX$^0$ self-energy slightly deviate from the G$^0$W$^0$ probably due to the missing of the dynamical nature of the screened interaction $W$. SX calculations are qualitatively similar to the SX$^0$ ones, with strongly quantitative differences near the K point that we will highlight in the following.

% \subsection{Optical conductivity}

\begin{figure}[h!]
\includegraphics[width=\linewidth]{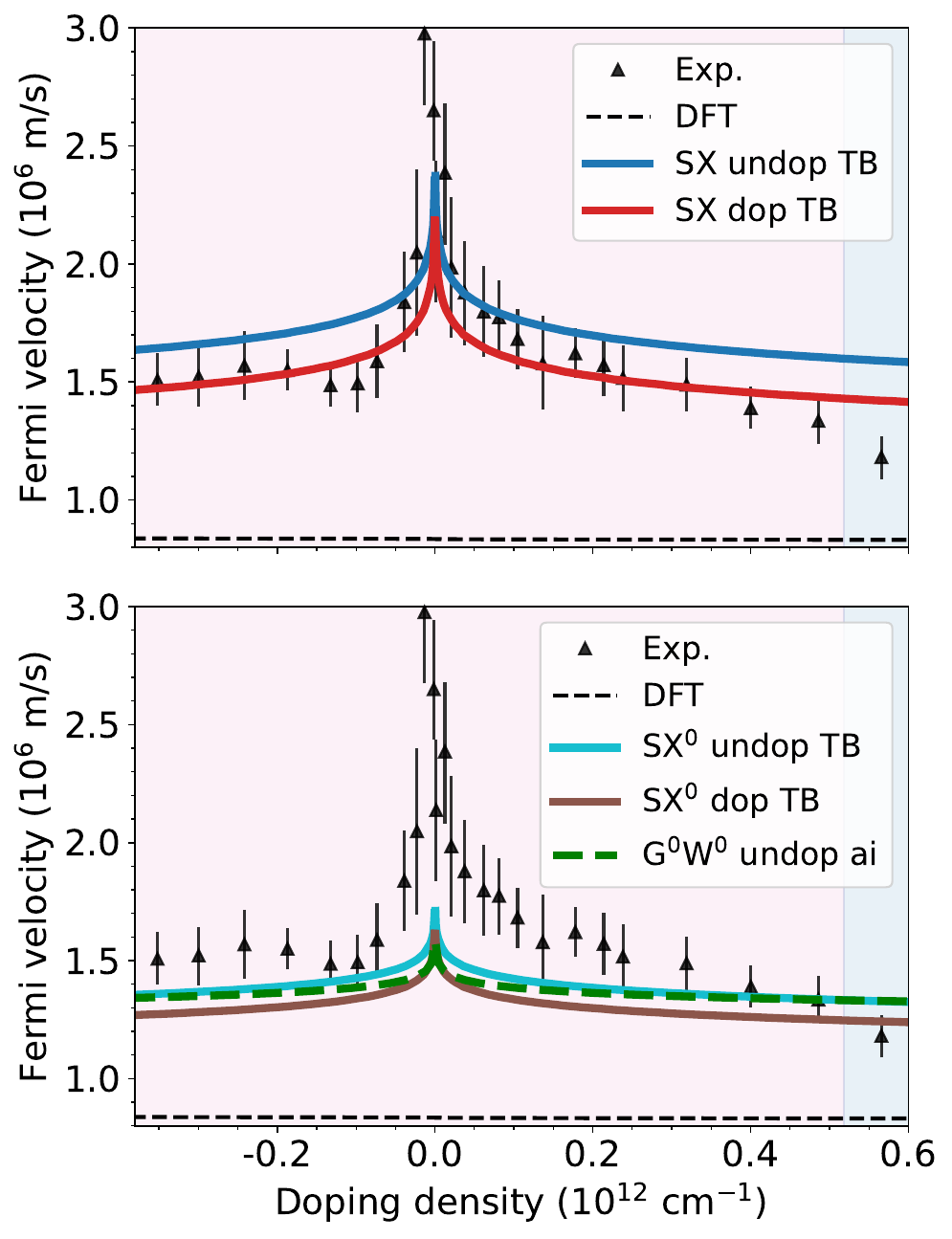}
\caption{Calculated Fermi velocities of doped graphene as a function of doping concentration compared with experiments.
Black triangles are experimental data obtained from Ref.~\onlinecite{Elias_11} and DFT data are marked with dashed lines.
Non self-consistent G$^0$W$^0$ results obtained from Ref.~\onlinecite{Guandalini_2024} are plotted in green.
We label with ``dop'' calculations where we included doping effects also in the quasi-particle corrections and screening, while with ``undop'' the group velocity of freestanding graphene correspondent to the doping level. Positive (negative) doping densities correspond to electron (hole) doping.
}
%Light blue (blue) curves are theoretical results obtained within the (non self consistent) static SX approximation.
%Positive (negative) doping densities correspond to electron (hole) doping.}
\label{fig:TB_vs_exp_vf}
\end{figure}

In Fig.~\ref{fig:TB_vs_exp_vf}, we compare the Fermi velocity renormalization calculated with SX$^0$ and SX starting from a TB band structure, G$^0$W$^0$ obtained ab-initio as in Ref.~\onlinecite{Guandalini_2024}, and the experimental results.
%In results labeled with ``dop'', we calculate at each point the screened interaction and quasi-particle band structure correspondent with the doping, while in results labeled with ``dop'' we used 
While the DFT results provide a constant and underestimated Fermi velocity, the SX$^0$ Fermi velocity show the qualitative correct behavior, even if the renormalization effect is underestimated in the same fashion of previous G$^0$W$^0$ calculations~\cite{Guandalini_2024}.
Only in the SX case we reach a renormalization comparable with the experiments, in agreement with what found in Ref.~\onlinecite{Stauber_17}.
The inclusion of a proper band structure and screened interaction with doping effects included is important to describe the decay of the renormalization at higher dopings.
We note these effects are difficult to include with an ab-initio calculation, due to the large amounts of screened interactions (one for each doping level) are necessary to compute.
In the limit of vanishing doping density ($n \to 0$), the Fermi velocities with and without doping effects included would coincide.
In order to verify this statement numerically, lower doping would be considered.
However, lower doping levels are out of convergence with respect to our grid sampling, thus have not been included.

\begin{figure}[h!]
\includegraphics[width=\linewidth]{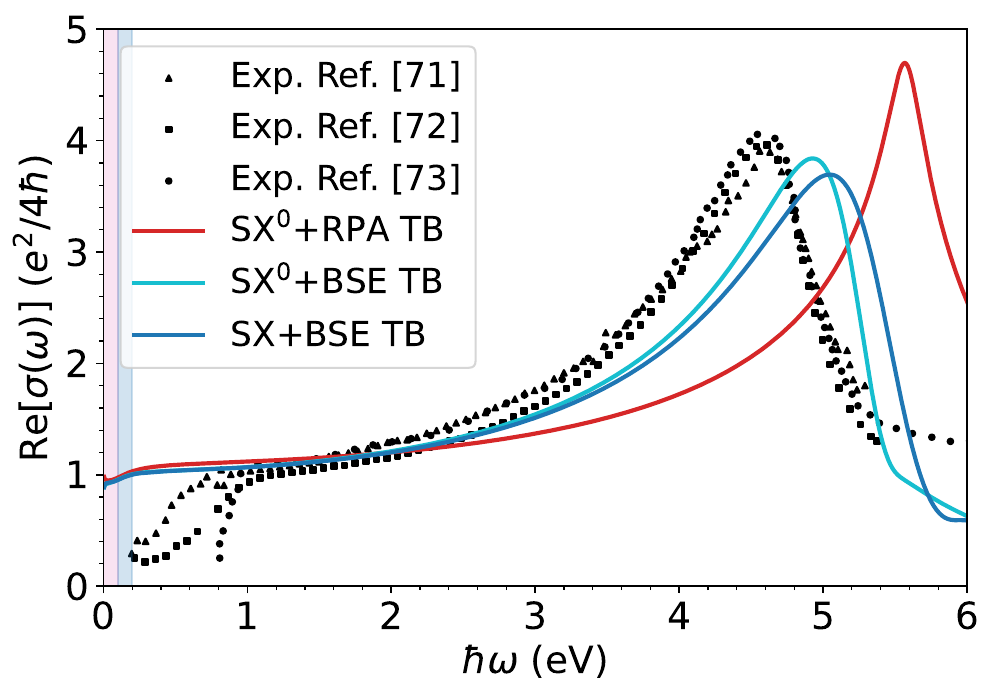}
\caption{Optical conductivity of freestanding graphene obtained with our tight-binding model compared with experiments.
Black triangles, squares and dots are experimental data obtained from Refs.~\onlinecite{mak2012optical}, \onlinecite{chang2014extracting} and \onlinecite{li2016broadband} respectively.
Light blue (blue) curves are theoretical results obtained within the (non self consistent) static SX approximation by solving the BSE.
The red curve is obtained using a SX$^0$ band structure within the RPA approximation, thus by neglecting $K^{\SX}$ in Eq.~\eqref{eq:K_def} (i.e. removing the $W$ diagram in the sketch (d) of Fig.~\ref{fig:fey_diag}).}
\label{fig:TB_vs_exp_opt}
\end{figure}

In Fig.~\ref{fig:TB_vs_exp_opt}, we compare instead the optical conductivity, calculated with the SX$^0$ or SX band structure solving the BSE or within the RPA, with experiments.
As already shown in the literature~\cite{Yang2009,guandalini_2023}, the inclusion of excitonic effects through the BSE is mandatory in order to have an optical conductivity in agreement with experiments.
Contrary to the band structure, the optical conductivity is not sensitive to the self-consistent calculation of the screened interaction.
In fact, there is a cancellation between the quasi-particle corrections, that enhance the gap thus blue shifting the $\pi$ plasmon position, and the excitonic effects, which red shift the spectrum~\cite{guandalini_2023}.
Overall, SX results are in excellent agreement with experimental data.
%We note the $\pi$ plasmon position of the ab-initio calculation is red-shifted with respect to the experimental one. This is probably due to the inclusion of dynamical screening effects in the quasi-particle band structure which are not included in the BSE, thus cancellation does not occur properly.

In summary, we have shown in this section that the use of a self-consistent screened interaction $W$ is relevant to understand low-energy many-body effects, while at higher energies it seems less crucial, as evidenced by previous studied with a non self-consistent $W$~\cite{Yang2009,guandalini_2023}.

\subsection{Doping and environmental effects}\label{sec:dop_env_eff}

Having validated our model for freestanding graphene in the previous section, we now deal with the effects of doping and environment on the group velocity renormalization and optical conductivity, which are easily treated within our TB approach. %To understand the importance of these effects, we analyze the long wavelength inverse dielectric function and the low energy band structure, which are interdependent as discussed below.

\begin{figure*}[t]
\includegraphics[width=0.9\linewidth,keepaspectratio]{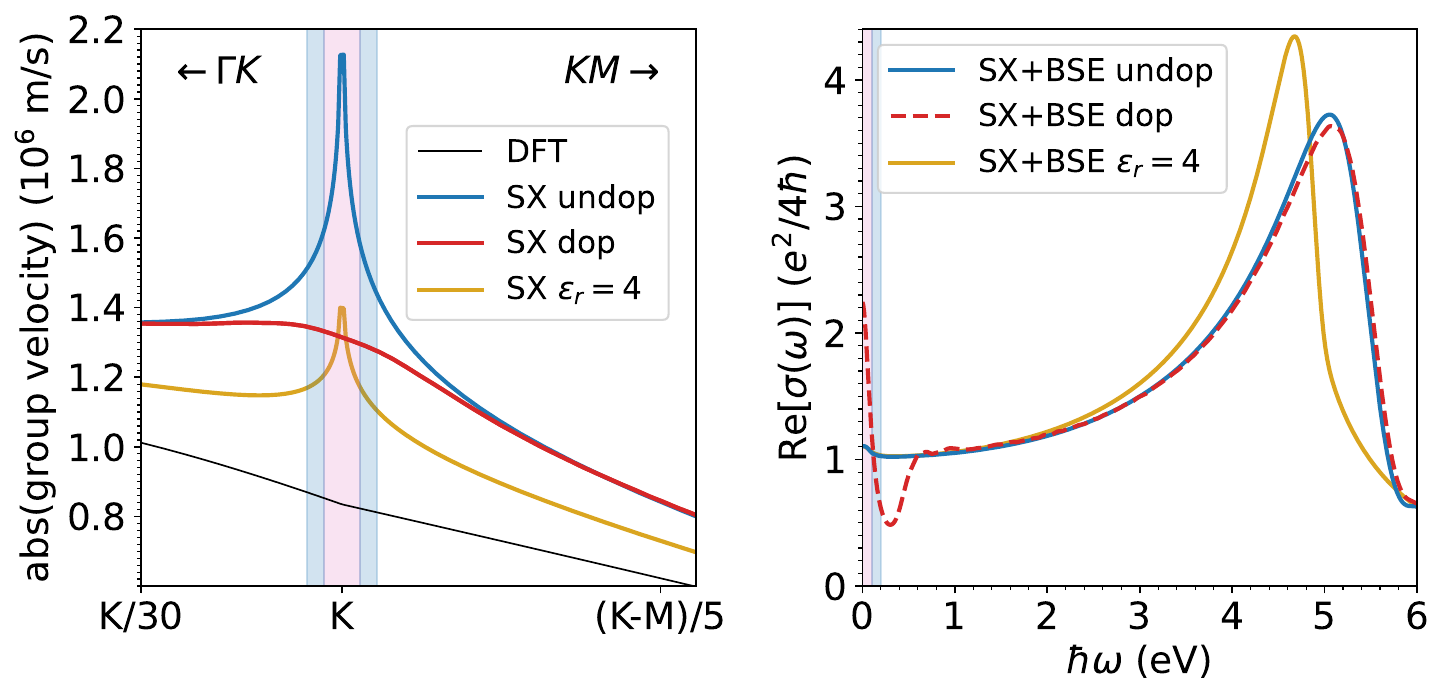}
\caption{Group velocities around the K point (left) and real part of the optical conductivity (right) of graphene freestanding (blue), doped (red) and embedded (yellow).
The DFT group velocity is shown in black.
The group velocities are obtained with the SX approximation, while optical conductivities with the SX band structure by solving the BSE.
The electron doping level is $n= 2.075 \times 10^{12}$ cm$^{-2}$.
The doped case displays the Dirac charge plasmon peak in the low energy energy region.
The embedding is modeled with a static uniform dielectric constant of $\varepsilon_r = 4$ as explained in Sec.~\ref{sec:dop_emb}.
The pink shaded region indicates the energy range relevant for Fermi energy renormalization, while the blue-gray region the energy range of optical phonons.}
\label{fig:TB_dop_env}
\end{figure*}

In Fig.~\ref{fig:TB_dop_env}, we show the group velocity of the $\pi$ band in the neighborhood of the Dirac point with the SX approximation for the case of freestanding, doped ($n = 2.075 \times 10^{12}$ cm$^{-2}$) and embedded ($\varepsilon_r = 4$) graphene,  and optical conductivity obtained with the same band structure by solving the BSE.
$\varepsilon_r \approx 4$ is the in-plane dielectric constant of hBN, that is often used to encapsulate graphene.
The group velocity of doped graphene does not diverge at the Dirac point, due to the quenching of the long wavelength component of screened interaction due to metallic screening (see Appx.~\ref{appx:TB_all} for a discussion about doping effects in the inverse dielectric function).
Instead, the dielectric environment reduces the renormalization of the group velocity, but without removing it. This is because the screened interaction is reduced by the embedding, but still not quenched in the long-wavelength limit contrary to the doped case.

Doping and environment affect less the dielectric response with respect than the band structure.
This is due to the partial cancellation between quasi-particle blue-shift and excitonic red shift.
This cancellation is nearly perfect in the doped case, where the $\pi$ plasmon is indistinguishable from the freestanding one.
The embedding instead causes a small red-shift of the $\pi$ plasmon with an enhancement of the peak height.

%%%%%%%%%%%%%%%%%%%%%%%%%%%%%%%%%%%%%%%%%%%%%
\section{Conclusions}\label{sec:conclusions}
%%%%%%%%%%%%%%%%%%%%%%%%%%%%%%%%%%%%%%%%%%%%%

We proposed a TB plus many-body perturbation theory model to describe low- and high-energy  many-body effects in graphene with a unified framework.
Within the static screening approximation, the renormalization of the Fermi velocity is in agreement with experiments, provided that the quasi-particle corrections are calculated self-consistently with the screened interaction, in accordance with Ref.~\onlinecite{Stauber_17}.
The frequency dependence of the screened interaction may be relevant for doped graphene, due to the presence of the low-energy Drude plasmon, as emphasized in Ref.~\onlinecite{Liang_2015} for the fundamental gap of doped transition-metal-dichalcogenide monolayers. However, the inclusion of finite frequency effects in the screened interaction is by itself a major topic on its own, and will be the addressed in a future work.
The optical conductivity is in accordance both with ab-initio calculations and experiments. In this case, the self-consistent solution of the quasi-particle corrections and the screened interaction does not play a critical role due to the partial cancellation between quasi-particle corrections and excitonic effects.
Also the density-density response function at $q=\mathrm{K}$ (shown in Appx.~\ref{appx:TB_all}) and low energy is in accordance with ab-initio calculations.
Instead, at higher energies some deviations occur, probably due to the approximate description of local field effects in the TB model, that does not accurately catch the microscopic description of the electronic orbitals.

We further studied the effect of doping and environment on the quasi-particle band structure and electronic response.
We found that doping quenches the logarithmic divergence of the group velocity, while the environment reduces the group velocity due to the screening of electron-electron interaction, but without changing qualitatively the results.
In the optical conductivity, the environmental effects slightly red shift the $\pi$ plasmon peak.
Instead, the presence of doping do not change the plasmon position.
This counter-intuitive behaviours, since in both cases the electron-electron interaction is more screened than in the freestanding case, are due to the partial cancellation between quasi-particle corrections and excitonic effects.
The same results are valid also for the density-density response function at $q=\mathrm{K}$.

In principle, this framework can be extended to take into account also the lattice response and electron-phonon coupling matrix elements. The proposed unified framework will allow also to study whether the many-body effects described in this paper may influence the lattice response, as we will do in a subsequent work.

\section{Acknowlegements}
This project has received funding from the European Research Council (ERC) under the European Union’s Horizon 2020 research and innovation programe (MORE-TEM ERC-SYN project, grant agreement No 951215).

\appendix

%%%%%%%%%%%%%%%%%%%%%%%%%%%%%%%%%%%%%%%%
\section{TB model}\label{appx:TB_model}
%%%%%%%%%%%%%%%%%%%%%%%%%%%%%%%%%%%%%%%%

We summarize in this section the graphene electronic structure obtained from a $5$-th nearest neighbour TB model as derived in the appendix of Ref.~\onlinecite{Venezuela_11}.
Here, we summarize the model for the sake of completeness.
The TB Bloch wavefunctions are defined as
\begin{equation}\label{eq:TB_bloch}
    \ket{s\k} = \sum\limits_{l}e^{i\k\cdot(\R_l+\t_s)}\ket{sl} \ ,
\end{equation}
where $\lbrace\R_l\rbrace$ are the lattice vectors and $\lbrace\t_s\rbrace$ carbon positions within the unit cell.
$\ket{sl}$ is the $p_z$ orbital of the $s$ atom in the $l$ unit cell.
The TB Hamiltonian $H^{0}_{ss'\k} = \mel{s\k}{\hat{H}^{0}}{s'\k}/N$, where $N$ the number of unit cells, can be written as
\begin{equation}\label{eq:TB_H}
    H^{0}_{\k} = 
    \mqty(
        \mathcal{g}_{\k} & \mathcal{f}_{\k} \\
        \mathcal{f}^*_{\k} & \mathcal{g}_{\k}
    )\ ,
\end{equation}
where 
\begin{equation}\label{eq:fk}
    \mathcal{f}_{\k} = -t_1\sum\limits_{i=1}^{3} e^{i\k\cdot\mathbf{C}^{1}_i}
    -t_3\sum\limits_{i=1}^{3} e^{i\k\cdot\mathbf{C}^{3}_i}
    -t_4\sum\limits_{i=1}^{6} e^{i\k\cdot\mathbf{C}^{4}_i}\ ,
\end{equation}
\begin{equation}\label{eq:gk}
    \mathcal{g}_{\k} = -t_2\sum\limits_{i=1}^{6} e^{i\k\cdot\mathbf{C}^{2}_i}
    -t_5\sum\limits_{i=1}^{6} e^{i\k\cdot\mathbf{C}^{5}_i} = g^*_{\k}\ ,
\end{equation}
and $t_n$ are the hopping parameters.
$\mathbf{C}^{n}_i$ is the $i$-th vectors connecting the reference atom to a $n$-th nearest neighbor.\\
By diagonalizing the Hamiltonian in Eq.~\eqref{eq:TB_H}, we have the $\pi$ and $\pi^*$ bands
\begin{equation}\label{eq:eig_TB}
    \eig_{\pi^*/\pi\k} = \mathcal{g}_{\k}\pm|\mathcal{f}_{\k}|
\end{equation}
and the corresponding eigenvectors
\begin{equation}\label{eq:eigv_TB}
    a_{\pi^*/\pi\k} = \frac{1}{\sqrt{2}}\mqty(
        1 \\
        \pm\phi_{\k}
    )\ ,
\end{equation}
where 
\begin{equation}\label{eq:phi_TB}
    \phi_{\k} = f^*_{\k}/|f_{\k}|
\end{equation}
are the spinorial phases.

The TB parameters $\mathbf{t}$, used in Eqs.~\eqref{eq:fk} and \eqref{eq:gk}, have been fitted over the DFT band structure.
After the fit, the optimal parameters found are $\mathbf{t}^{\DFT} = \mqty(-2.8810 & 0.2797 & -0.2034  & 0.1017 & 0.0763 )$ (eV), in accordance with Ref.~\onlinecite{Venezuela_11}.
To mimic the G$^0$W$^0$ band structure, again in accordance with Ref.~\onlinecite{Venezuela_11}, we use the following TB parameters:  $\mathbf{t}^{G^0W^0} = \alpha_{\GW}\mathbf{t}^{0}$, where $\alpha_{\GW} = 1.18$.

\begin{figure}[h!]
\includegraphics[width=\linewidth]{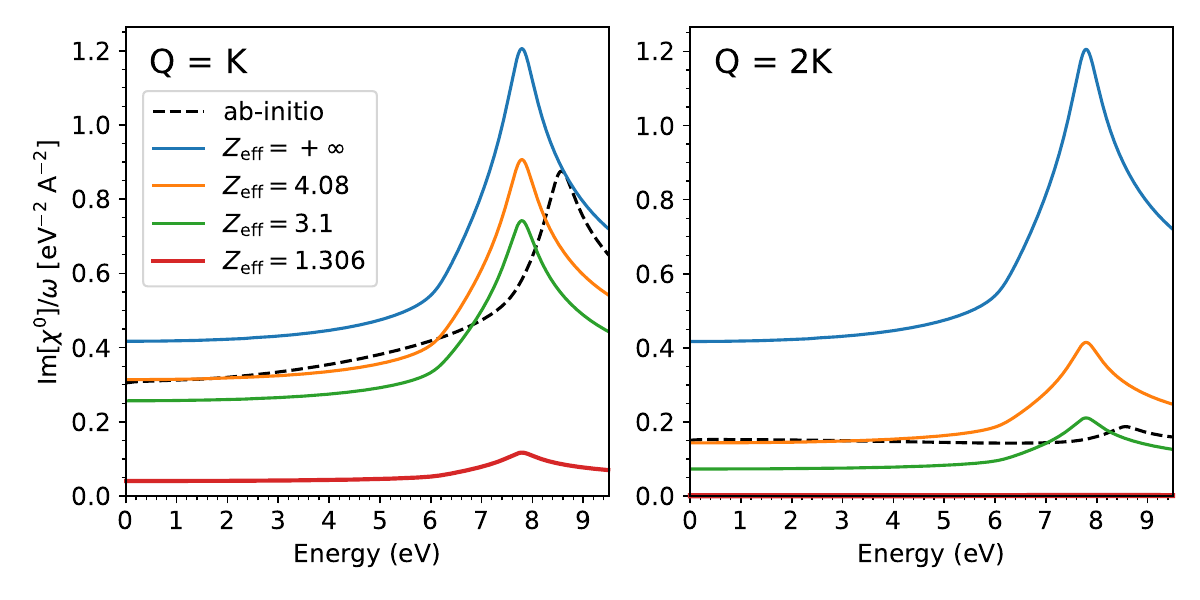}
\caption{Irreducible polarizability $\chi^0$ (see Eq.\eqref{eq:chi_irr}) over energy $\omega$ obtained with ab-initio method (black dashed line) and the tight-binding model with different $Z_{\mathrm{eff}}$ [see Eq.~\eqref{eq:pz_momentum}]. The irreducible polarizabilities are obtained at $q = K$ (left panel) and $q = 2K$ (right panel). }
\label{fig:Zeff_check}
\end{figure}

Both in the calculation of the SX self-energy and BSE kernels [see Eqs.~\eqref{eq:SX_2D}, \eqref{eq:Xix_2D} and \eqref{eq:Xic_2D}], we have matrix elements that can be written within this model as
\begin{equation}\label{eq:mel_TB}
    \mel{m\k}{e^{i\qscr\cdot \r_{\parallel}}}{n\k'} =
    \mathcal{F}_{\k-\k'}
    \frac{1}{2} [1-n m\phi^*_{\k}\phi_{\k'}]\delta_{\qscr,\k-\k'} \ ,
\end{equation}
where
\begin{equation}
\mathcal{F}_{\k-\k'} = \int d\r_{\parallel} |\orb(\r_{\parallel})|^2 e^{i(\k-\k')\cdot\r_{\parallel}}
\end{equation}
is the 2D atomic form factor .
We already approximated, via Eq.~\eqref{eq:fin_thick}, the $z$ profile of the $p_z$ orbitals with a rectangular shape.
The $x$ and $y$ profile is instead modeled with a hydrogenic $p$ wavefunction integrated along the $z$ direction
\begin{equation}\label{eq:pz_real}
    |\orb(\r_{\parallel})|^2 = \frac{Z_{\mathrm{eff}}}{32\pi}\int dz  \ r^2 e^{-Z_{\mathrm{eff}}r} \cos(\theta),
\end{equation}
where $r^2 = r_{\parallel}^2+z^2$, $\cos(\theta) = z/r$ and $Z_{\mathrm{eff}}$ is an adjustable parameter which determines the wavefunction extension in real space.
By substituting Eq.~\eqref{eq:pz_real} in the definition of $\mathcal{F}_{\k-\k'}$, we find
\begin{equation}\label{eq:pz_momentum}
    \mathcal{F}_{\k-\k'} = \frac{1}{(1+|\k-\k'|^2/Z_{\mathrm{eff}}^2)^3}.
\end{equation}
For an explicit derivation of Eq.~\eqref{eq:pz_momentum}, we refer to the supplemental material of Ref.~\onlinecite{Stauber_17}.
%For simplicity, we neglect the $\r_{\parallel}$ orbital shape by setting $|\orb(\r_{\parallel})|^2 = \delta(\r_{\parallel})$, which correspond to $\mathcal{F}_{\k-\k'} = 1$.
We tested several values of $Z_{\mathrm{eff}} = 4.08$ and $1.306$ as proposed in Ref.~\onlinecite{Stauber_17}, and $Z_{\mathrm{eff}} = 3.1$ obtained by fitting the $2p_z$ wavefunction of carbon obtained with an all-electron density-functional theory code.
We tested also the limiting case $Z_{\mathrm{eff}} \to +\infty$, which corresponds to $|\orb(\r_{\parallel})|^2 \to \delta(\r_{\parallel})$. 
In Fig.~\ref{fig:Zeff_check} we plot the irreducible polarizability $\chi^0$ divided by $\omega$ obtained with our tight-binding model along with ab-initio results, considered as the benchmark.
The momentum transfers $q = K$ and $q = 2K$ are an ideal test case to select the best $Z_{\mathrm{eff}}$, as transitions near the Dirac cones (which are better reproduced with our TB model) are selected, while the form factors at high $\k-\k'$ are employed at the same time.
As $\chi^0$ at low energies is linear in $\omega$ at these momentum transfers, $\chi^0/\omega$ is proportional to the square modulus of the matrix elements.
From our results in Fig.~\ref{fig:Zeff_check}, it is clear that $Z_{\mathrm{eff}} = 4.08$ is the best choice, as it better reproduce the ab-initio calculations both at $q = K$ and $q = 2K$.

%%%%%%%%%%%%%%%%%%%%%%%%%%%%%%%%%%%%%%%%%%%%%%%%%%%%
\section{Screened exchange self-energy in the tight-binding model}\label{sec:sigma_TB}
%%%%%%%%%%%%%%%%%%%%%%%%%%%%%%%%%%%%%%%%%%%%%%%%%%%
In this appendix we derive the expression of the off-diagonal SX self energy term in the tight-binding basis as given by Eq.~\eqref{eq:SX_TB}.
Starting from Eq.~\eqref{eq:SX_2D}, the matrix elements of $\Sigma^{\SX}$ are given by
\begin{multline}\label{eq:S_SX_bloch}
    \mel{s_1\k}{\Sigma^{\SX}}{s_2\k} = \frac{-1}{N_{\k}A}\sum\limits_{m\k'}
    \sum\limits_{\G} f_{mk'}\mel{s_1\k}{e^{i(\k-\k'+\G)}}{m\k'}\\
    \times W^{\ZD}_{\G}(\k-\k')\mel{m\k'}{e^{-i(\k-\k'+G )}}{s_2\k},
\end{multline}
where $\ket{s_1\k}$, $\ket{s_2\k}$ are given by Eq.~\eqref{eq:TB_bloch}.
The matrix elements between atomic Bloch wavefunctions and band states in Eq.~\eqref{eq:S_SX_bloch} can be written as
\begin{equation}\label{eq:mel_TB_S}
\mel{s_1\k}{e^{i(\k-\k')}}{m\k'} = a^{s_1}_{m\k'-\G}\mathcal{F}_{\k-\k'},
\end{equation}
where $a^{s_1}_{m\k'-\G}$ are the eigenvectors of the TB Hamiltonian given by Eq.~\eqref{eq:eigv_TB} and formally extended over the first BZ in order to include local-field effects and $\mathcal{F_{\k-\k'}}$ are the atomic form factors.
By substituting Eq.~\eqref{eq:mel_TB_S} into Eq.~\eqref{eq:S_SX_bloch}, we find
\begin{multline}
    \mel{s_1\k}{\Sigma^{\SX}}{s_2\k} = \frac{-1}{N_{\k}A}\sum\limits_{m\k'}
    \sum\limits_{\G} f_{mk'} |\mathcal{F}_{\k-\k'}|^2\\
    \times a^{s_1}_{m\k'-\G}a^{s_2*}_{m\k'-\G}
    W^{\ZD}_{\G}(\k-\k').
\end{multline}
We are interested in the off-diagonal term, thus $\Sigma^{\SX}_{\k} = \mel{1\k}{\Sigma^{\SX}}{2\k}$.
By inserting the expression of the eigenvector of the TB Hamiltonian given by Eq.~\eqref{eq:eigv_TB}, $\Sigma^{\SX}_{\k}$ can be written as
\begin{multline}\label{eq:S_TB_intermediate}
    \Sigma^{\SX}_{\k} = \frac{-1}{N_{\k}A}\sum\limits_{m\k'}
    \sum\limits_{\G} f_{mk'} |\mathcal{F}_{\k-\k'}|^2\\
    \times sign(m)\phi^*_{\k'-\G}
    W^{\ZD}_{\G}(\k-\k'),
\end{multline}
where $sign(\pi) = -1$ and $sign(\pi^*) = 1$.
We note that only $sign(m)$ and $f_{mk'}$ depend on the band index. Thus, we can define the difference of the $\pi$ and $\pi^*$ occupations as
\begin{equation}
    \Delta f_{\k'} = \sum\limits_m sign(m)f_{m\k'} = f_{\pi^*\k'}-f_{\pi\k'}
\end{equation}
and substitute this expression in Eq.~\eqref{eq:S_TB_intermediate} to find Eq.~\eqref{eq:SX_TB}.

%%%%%%%%%%%%%%%%%%%%%%%%%%%%%%%%%%%%%%%%%%%%%%%%%%%%
\section{Validation of the dielectric response with ab-initio calculations}\label{appx:validation}
\begin{figure}[h!]
\includegraphics[width=\linewidth]{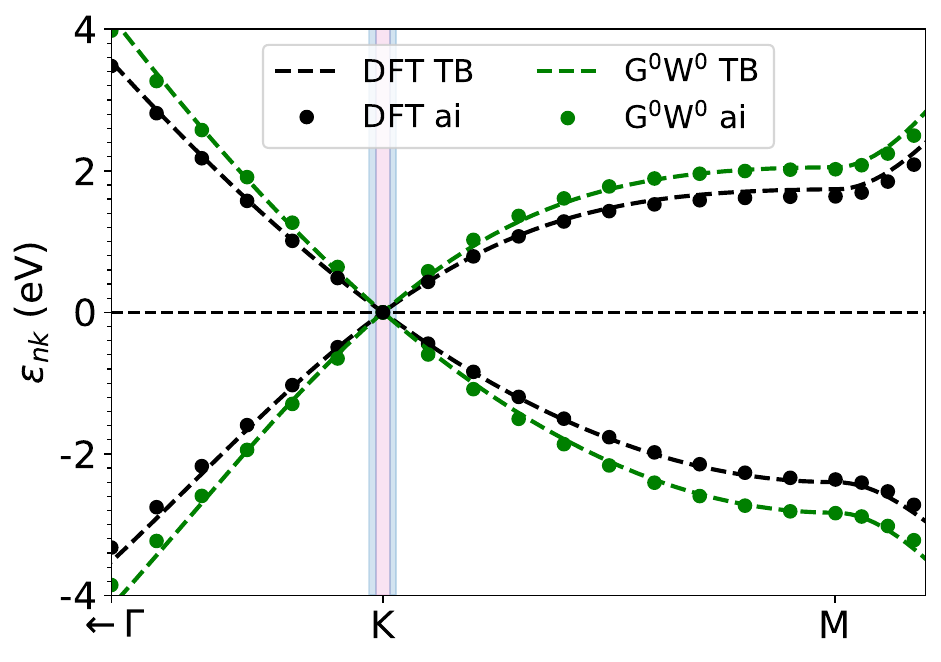}
\caption{Band structure of graphene near the Fermi level obtained from ab-initio calculations with DFT (black dots) and G$^0$W$^0$ (green dots). Dashed lines represents TB calculations with parameters fitted over DFT (black) and G$^0$W$^0$.}
\label{fig:bands_fit}
\end{figure}

%%%%%%%%%%%%%%%%%%%%%%%%%%%%%%%%%%%%%%%%%%%%%%%%%%%
As described in Sec. \ref{sec:numdet}, the TB parameters for DFT and G$^0$W$^0$ band-structures are fitted on the respective ab-initio calculations, while for the SX$^0$ and SX calculations we use the parameters fitted on DFT as the bare system.
The TB and ab-initio results for both DFT and G$^0$W$^0$ show an excellent agreement, as can be seen from Fig.~\ref{fig:bands_fit}.

\begin{figure}[h!]
\includegraphics[width=\linewidth]{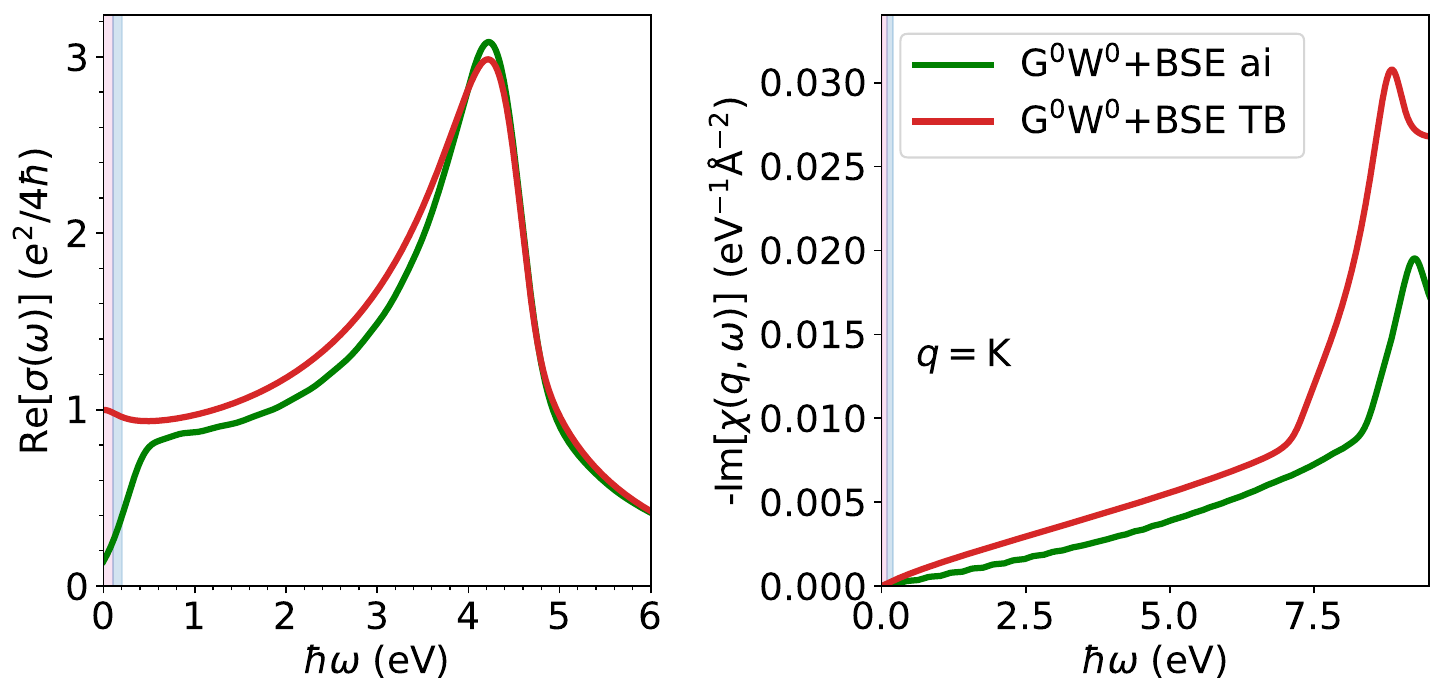}
\caption{Real part of the optical conductivity (left) and imaginary part of
the density-density response function evaluated at $q=\mathrm{K}$ (right) for
freestanding graphene. We indicate with green lines the ab-initio results
and with red lines those obtained starting from our TB model.}
\label{fig:el_resp_vs_yambo}
\end{figure}

In Fig.~\ref{fig:el_resp_vs_yambo}, we plot the optical conductivity and density-density response function at $q=\mathrm{K}$ of freestanding graphene obtained with the G$^0$W$^0$+BSE method, and compare between ab-initio and TB calculations.
We stress that with TB G$^0$W$^0$ we mean a TB model with parameters fitted over the ab-initio G$^0$W$^0$ band structure.
For what concerns the optical conductivity, there is a perfect agreement between TB and ab-initio calculations about the position and shape of the $\pi$ plasmon peak.
The TB calculation is converged at small $\w$ due to the denser TB k-grid, allowed by the lower computational effort required by our calculations, showing the Drude plasmon peak originated from the finite electronic temperature, while the ab-initio calculation do not show the Drude plasmon due to the lower k-grid and the removal of the $\w = 0$ transition at the Dirac point due to numerical instabilities.
The TB calculation of the density-density response function at $q=\mathrm{K}$ is in satisfactory agreement with the ab-initio calculation. At high momentum transfer, the accuracy of the band structure in a higher energy range is more important, as well as the accurate description of the microscopic shape of the electrons wavefunctions. Further, in this regime an accurate inclusion of local field effects is required, which is not achieved by a TB model. Still, we find remarkable that the qualitative behavior of the ab-initio spectrum is well reproduced by the TB model with a computational cost which is two orders of magnitude lower.

%%%%%%%%%%%%%%%%%%%%%%%%%%%%%%%%%%%%%%%%%%%%%%%%%%%%%%%%%%%%%%%%%%%%
\section{Additional studies on doping and environmental effects}\label{appx:TB_all}
%%%%%%%%%%%%%%%%%%%%%%%%%%%%%%%%%%%%%%%%%%%%%%%%%%%%%%%%%%%%%%%%%%%%
In this appendix, we show a more complete set of data with respect to the results section for completeness.
The color code of all the figures, along with detailed information about the system considered, are listed  in Tab.~\ref{Tab_CC}.

\begin{table}
\centering
\begin{ruledtabular}
\begin{tabular}{llccll}
\\[-3pt]
System                & n ($10^{12}$ cm$^{-2}$)  & $\varepsilon_r$& $W$  & label & color\\
\hline
Freestanding & None    & 1 & \ref{sec:Wo_free} & SX$^0$          &  \crule[light_blue]{3mm}{3mm}\\
Freestanding & None    & 1 & \ref{sec:W_free}  & SX              &  \crule[blue]{3mm}{3mm}\\
Embedded     & None    & 4 & \ref{sec:Wo_free} & SX$^{0}_{\emb}$ &  \crule[green]{3mm}{3mm}\\
Embedded     & None    & 4 & \ref{sec:W_free} & SX$_{\emb}$      &  \crule[yellow]{3mm}{3mm}\\
Low-doped    & $0.129$ & 1 & \ref{sec:Wo_dop} & SX$^{0}_{\dl}$   &  \crule[light_red]{3mm}{3mm}\\
Low-doped    & $0.129$ & 1 & \ref{sec:W_dop}  & SX$_{\dl}$       &  \crule[orange]{3mm}{3mm}\\
Medium-doped & $0.519$ & 1 & \ref{sec:Wo_dop} & SX$^{0}_{\dm}$   &  \crule[violet]{3mm}{3mm}\\
Medium-doped & $0.519$ & 1 & \ref{sec:W_dop}  & SX$_{\dm}$       &  \crule[pink]{3mm}{3mm}\\
High-doped   & $2.075$ & 1 & \ref{sec:Wo_dop} & SX$^{0}_{\dh}$   &  \crule[brown]{3mm}{3mm}\\
High-doped   & $2.075$ & 1 & \ref{sec:W_dop}  & SX$_{\dh}$       &  \crule[red]{3mm}{3mm}\\
\end{tabular}
\end{ruledtabular}
\caption{Summary of the graphene systems considered in Figs.~\ref{fig:epsm1_TB}-\ref{fig:el_resp_TB}  with doping density, static dielectric background, section where we model $W$, figure label and line color.}
\label{Tab_CC}
\end{table}

\begin{figure}[h!]
\includegraphics[width=\linewidth]{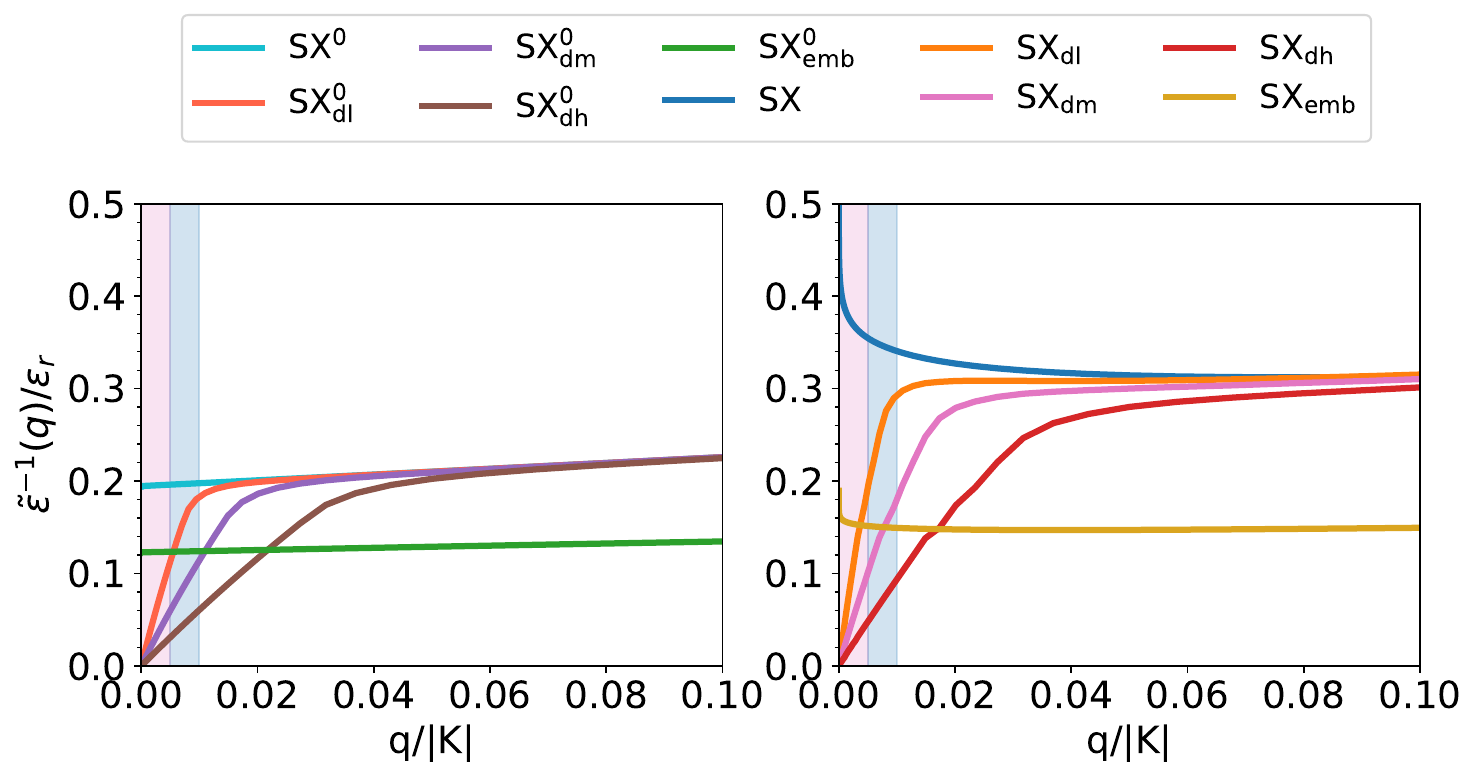}
\caption{Inverse static dielectric function of graphene obtained with the RPA starting from a DFT (left) or SX (right) irreducible polarizability [see Eq.~\eqref{eq:chi_irr}]. 
The color-code of different doping levels, embedding and theory-levels in given in Tab. \eqref{Tab_CC}.}
\label{fig:epsm1_TB}
\end{figure}

In Fig.~\ref{fig:epsm1_TB}, we show the inverse static dielectric function of freestanding, doped and embedded ($\varepsilon_r = 4$) graphene for both SX$^0$ and SX.
For freestanding SX$^0$ calculations, the inverse dielectric function is nearly constant at low momentum transfers, as already shown in Ref.~\onlinecite{Sohier2015}.
Doping instead produces a metallic screening that is reflected in the drop of the inverse dielectric function in the long wavelength limit. The higher the doping, the faster is the decay.
The effect of the static environment is to reduce the inverse dielectric function by $\approx 1.7$ with respect to the freestanding case.
The inverse dielectric constant is not reduced by simply $\varepsilon_r = 4$ because the embedding reduces also the Coulomb interaction entering the RPA resummation [see Eq.~\eqref{eq:RPA_2D}], thus reducing the intensity of the RPA correction.
The SX inverse dielectric function is larger than the SX$^0$ case, meaning that screening effects are weaker as a result of the self-consistent calculation.
This is due to the quantitative difference of band structures between the two calculations near the K point.
%, as anticipated in the previous section and discussed later in more detail.
As a consequence, in the SX case the quasi-particle corrections reduce screening effects.
In general, freestanding and doped SX inverse dielectric functions are $\approx 1.5$ larger than their SX$^0$ counterpart, thus enhancing the electron interaction and the corresponding many-body effects.
The embedded inverse dielectric function is instead nearly independent on the self-consistent procedure as the quasi-particle corrections are quenched by the environmental screening.
Most importantly, both the freestanding and embedded inverse dielectric functions tends logarithmically to $1$ in the long-wavelength limit, due to the logarithmic divergence of the group velocity in the quasi-particle bands, which is quenched in the doped case.

\begin{figure}[t]
\includegraphics[width=\linewidth]{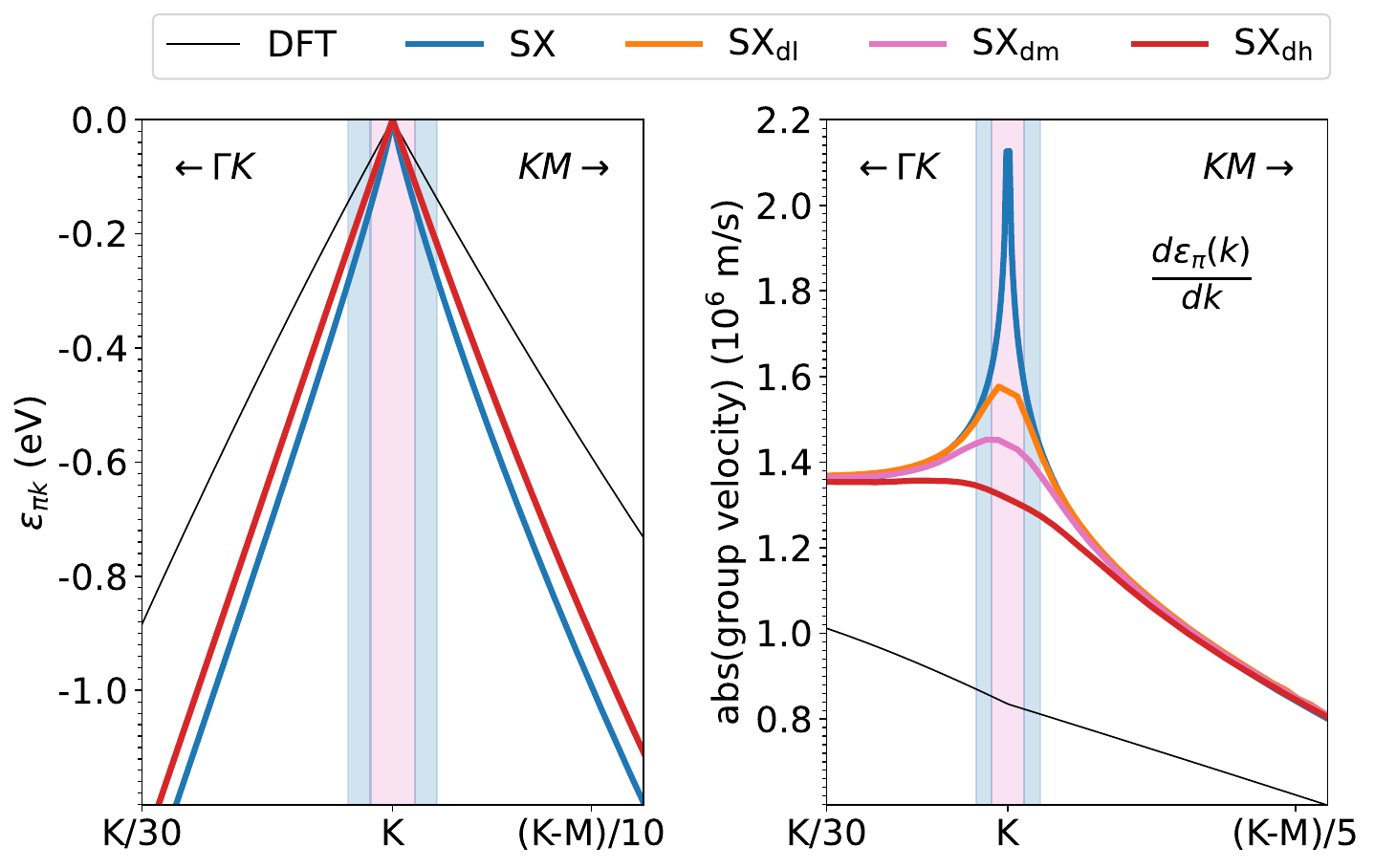}
\caption{Electronic energies of the $\pi$ band (left) and  group velocity (right) of
graphene near the K point. 
Beside the color-code given in Tab. \eqref{Tab_CC}, we use black for the DFT results of freestanding graphene.}
\label{fig:bands_TB}
\end{figure}

In Fig.~\ref{fig:bands_TB}, we show the electronic band structure and group velocity of the $\pi$ band in the neighborhood of the Dirac point for freestanding and doped graphene with three different levels of doping (see Tab.~\ref{Tab_CC} for details) with te SX approximation.
%As anticipated, the SX quasi-particle corrections are larger than the SX$^0$ ones. In particular, in the range of 0.2 eV around K it shows a larger logarithmically enhanced group velocity. This is due to the weaker screening effects obtained in the self-consistent calculations, as discussed above. 
The doped quasi-particle corrections are lower than the freestanding case due to metallic screening, as previously discussed.
Thus, the doped $\pi$ band is nearer to the DFT band with respect to the freestanding case.
While the band structures of the three different dopings are indistinguishable, it is clear from the group velocity that increasing the doping increases the quenching of the group velocity renormalization, due to the enhanced metallic screening.

\begin{figure}[h]
\includegraphics[width=\linewidth]{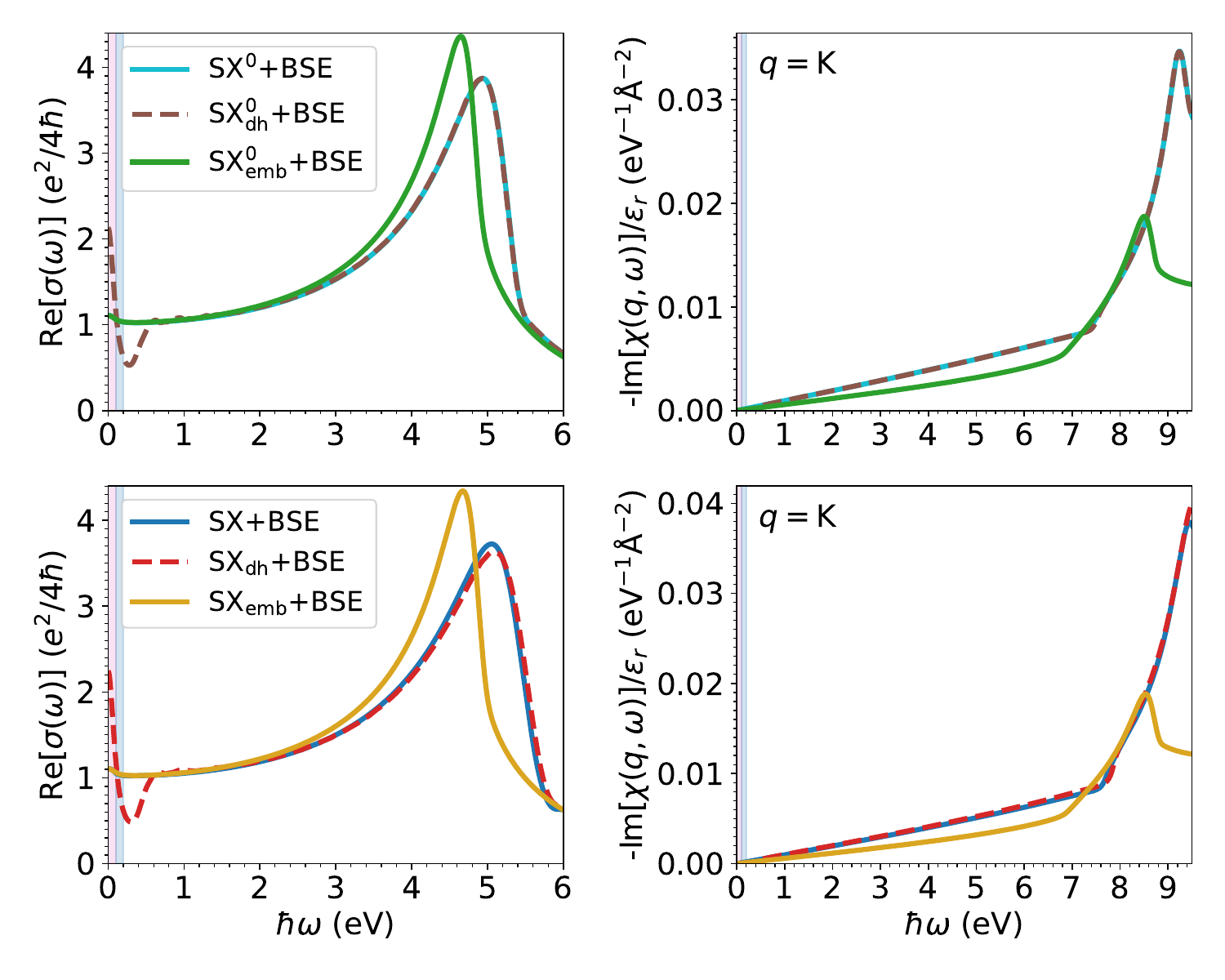}
\caption{Real part of the optical conductivity (left panels) and imaginary part of the density-density response function evaluated at $q=\mathrm{K}$ (right panels) of graphene obtained with the SX+BSE method.
The color code is given in Tab.~\ref{Tab_CC}.}
\label{fig:el_resp_TB}
\end{figure}

In Fig.~\ref{fig:el_resp_TB}, we show the optical conductivity and density-density response functions at $q=\mathrm{K}$ as a function of energy for freestanding, doped and embedded graphene.
As shown in the main text for freestanding graphene, we note that the electronic response is less affected by self-consistency also in the doping and embedded cases due to the partial cancellation between quasi-particle corrections and excitonic effects.
This partial cancellation result in a non trivial relation between the amount of screening and the peak shift.
In fact, the self-consistent procedure, that enhance the excitonic effect, causes instead a blue-shift, due to the enhance of the quasi-particle gap. For embedded graphene instead the self-consistent solution of the screened interaction does not affect the spectra, as noted for the inverse dielectric function.
The density-density response function at $q=\mathrm{K}$ at low energy is nearly insensitive of doping and self-consistency.
Instead, the high energy $\pi$ plasmon peak is red-shifted by the embedding, and nearly invariant with respect to doping and self-consistency.


\begin{thebibliography}{75}%
\makeatletter
\providecommand \@ifxundefined [1]{%
 \@ifx{#1\undefined}
}%
\providecommand \@ifnum [1]{%
 \ifnum #1\expandafter \@firstoftwo
 \else \expandafter \@secondoftwo
 \fi
}%
\providecommand \@ifx [1]{%
 \ifx #1\expandafter \@firstoftwo
 \else \expandafter \@secondoftwo
 \fi
}%
\providecommand \natexlab [1]{#1}%
\providecommand \enquote  [1]{``#1''}%
\providecommand \bibnamefont  [1]{#1}%
\providecommand \bibfnamefont [1]{#1}%
\providecommand \citenamefont [1]{#1}%
\providecommand \href@noop [0]{\@secondoftwo}%
\providecommand \href [0]{\begingroup \@sanitize@url \@href}%
\providecommand \@href[1]{\@@startlink{#1}\@@href}%
\providecommand \@@href[1]{\endgroup#1\@@endlink}%
\providecommand \@sanitize@url [0]{\catcode `\\12\catcode `\$12\catcode
  `\&12\catcode `\#12\catcode `\^12\catcode `\_12\catcode `\%12\relax}%
\providecommand \@@startlink[1]{}%
\providecommand \@@endlink[0]{}%
\providecommand \url  [0]{\begingroup\@sanitize@url \@url }%
\providecommand \@url [1]{\endgroup\@href {#1}{\urlprefix }}%
\providecommand \urlprefix  [0]{URL }%
\providecommand \Eprint [0]{\href }%
\providecommand \doibase [0]{https://doi.org/}%
\providecommand \selectlanguage [0]{\@gobble}%
\providecommand \bibinfo  [0]{\@secondoftwo}%
\providecommand \bibfield  [0]{\@secondoftwo}%
\providecommand \translation [1]{[#1]}%
\providecommand \BibitemOpen [0]{}%
\providecommand \bibitemStop [0]{}%
\providecommand \bibitemNoStop [0]{.\EOS\space}%
\providecommand \EOS [0]{\spacefactor3000\relax}%
\providecommand \BibitemShut  [1]{\csname bibitem#1\endcsname}%
\let\auto@bib@innerbib\@empty
%</preamble>
\bibitem [{\citenamefont {Novoselov}\ \emph {et~al.}(2005)\citenamefont
  {Novoselov}, \citenamefont {Geim}, \citenamefont {Morozov}, \citenamefont
  {Jiang}, \citenamefont {Katsnelson}, \citenamefont {Grigorieva},
  \citenamefont {Dubonos},\ and\ \citenamefont
  {Firsov}}]{novoselov_two-dimensional_2005}%
  \BibitemOpen
  \bibfield  {author} {\bibinfo {author} {\bibfnamefont {K.~S.}\ \bibnamefont
  {Novoselov}}, \bibinfo {author} {\bibfnamefont {A.~K.}\ \bibnamefont {Geim}},
  \bibinfo {author} {\bibfnamefont {S.~V.}\ \bibnamefont {Morozov}}, \bibinfo
  {author} {\bibfnamefont {D.}~\bibnamefont {Jiang}}, \bibinfo {author}
  {\bibfnamefont {M.~I.}\ \bibnamefont {Katsnelson}}, \bibinfo {author}
  {\bibfnamefont {I.~V.}\ \bibnamefont {Grigorieva}}, \bibinfo {author}
  {\bibfnamefont {S.~V.}\ \bibnamefont {Dubonos}},\ and\ \bibinfo {author}
  {\bibfnamefont {A.~A.}\ \bibnamefont {Firsov}},\ }\bibfield  {title}
  {{\selectlanguage {en}\bibinfo {title} {Two-dimensional gas of massless dirac
  fermions in graphene}},\ }\href {https://doi.org/10.1038/nature04233}
  {\bibfield  {journal} {\bibinfo  {journal} {Nature}\ }\textbf {\bibinfo
  {volume} {438}},\ \bibinfo {pages} {197} (\bibinfo {year}
  {2005})}\BibitemShut {NoStop}%
\bibitem [{\citenamefont {Novoselov}\ \emph {et~al.}(2004)\citenamefont
  {Novoselov}, \citenamefont {Geim}, \citenamefont {Morozov}, \citenamefont
  {Jiang}, \citenamefont {Zhang}, \citenamefont {Dubonos}, \citenamefont
  {Grigorieva},\ and\ \citenamefont {Firsov}}]{Novoselov_2016}%
  \BibitemOpen
  \bibfield  {author} {\bibinfo {author} {\bibfnamefont {K.~S.}\ \bibnamefont
  {Novoselov}}, \bibinfo {author} {\bibfnamefont {A.~K.}\ \bibnamefont {Geim}},
  \bibinfo {author} {\bibfnamefont {S.~V.}\ \bibnamefont {Morozov}}, \bibinfo
  {author} {\bibfnamefont {D.}~\bibnamefont {Jiang}}, \bibinfo {author}
  {\bibfnamefont {Y.}~\bibnamefont {Zhang}}, \bibinfo {author} {\bibfnamefont
  {S.~V.}\ \bibnamefont {Dubonos}}, \bibinfo {author} {\bibfnamefont {I.~V.}\
  \bibnamefont {Grigorieva}},\ and\ \bibinfo {author} {\bibfnamefont {A.~A.}\
  \bibnamefont {Firsov}},\ }\bibfield  {title} {\bibinfo {title} {Electric
  field effect in atomically thin carbon films},\ }\href
  {https://doi.org/10.1126/science.1102896} {\bibfield  {journal} {\bibinfo
  {journal} {Science}\ }\textbf {\bibinfo {volume} {306}},\ \bibinfo {pages}
  {666} (\bibinfo {year} {2004})}\BibitemShut {NoStop}%
\bibitem [{\citenamefont {Zhang}\ \emph {et~al.}(2005)\citenamefont {Zhang},
  \citenamefont {Tan}, \citenamefont {Stormer},\ and\ \citenamefont
  {Kim}}]{Zhang_2005}%
  \BibitemOpen
  \bibfield  {author} {\bibinfo {author} {\bibfnamefont {Y.}~\bibnamefont
  {Zhang}}, \bibinfo {author} {\bibfnamefont {Y.-W.}\ \bibnamefont {Tan}},
  \bibinfo {author} {\bibfnamefont {H.~L.}\ \bibnamefont {Stormer}},\ and\
  \bibinfo {author} {\bibfnamefont {P.}~\bibnamefont {Kim}},\ }\bibfield
  {title} {\bibinfo {title} {Experimental observation of the quantum hall
  effect and berry's phase in graphene},\ }\href
  {https://doi.org/10.1038/nature04235} {\bibfield  {journal} {\bibinfo
  {journal} {Nature}\ }\textbf {\bibinfo {volume} {438}},\ \bibinfo {pages}
  {201} (\bibinfo {year} {2005})}\BibitemShut {NoStop}%
\bibitem [{\citenamefont {Bonaccorso}\ \emph {et~al.}(2010)\citenamefont
  {Bonaccorso}, \citenamefont {Sun}, \citenamefont {Hasan},\ and\ \citenamefont
  {Ferrari}}]{Bonaccorso_2010}%
  \BibitemOpen
  \bibfield  {author} {\bibinfo {author} {\bibfnamefont {F.}~\bibnamefont
  {Bonaccorso}}, \bibinfo {author} {\bibfnamefont {Z.}~\bibnamefont {Sun}},
  \bibinfo {author} {\bibfnamefont {T.}~\bibnamefont {Hasan}},\ and\ \bibinfo
  {author} {\bibfnamefont {A.~C.}\ \bibnamefont {Ferrari}},\ }\bibfield
  {title} {\bibinfo {title} {Graphene photonics and optoelectronics},\ }\href
  {https://doi.org/10.1038/nphoton.2010.186} {\bibfield  {journal} {\bibinfo
  {journal} {Nature Photonics}\ }\textbf {\bibinfo {volume} {4}},\ \bibinfo
  {pages} {611} (\bibinfo {year} {2010})}\BibitemShut {NoStop}%
\bibitem [{\citenamefont {Grigorenko}\ \emph {et~al.}(2012)\citenamefont
  {Grigorenko}, \citenamefont {Polini},\ and\ \citenamefont
  {Novoselov}}]{Grigorenko_2012}%
  \BibitemOpen
  \bibfield  {author} {\bibinfo {author} {\bibfnamefont {A.}~\bibnamefont
  {Grigorenko}}, \bibinfo {author} {\bibfnamefont {M.}~\bibnamefont {Polini}},\
  and\ \bibinfo {author} {\bibfnamefont {K.}~\bibnamefont {Novoselov}},\
  }\bibfield  {title} {\bibinfo {title} {Graphene plasmonics},\ }\href
  {https://doi.org/10.1038/nphoton.2012.262} {\bibfield  {journal} {\bibinfo
  {journal} {Nature Photon.}\ }\textbf {\bibinfo {volume} {6}},\ \bibinfo
  {pages} {749} (\bibinfo {year} {2012})}\BibitemShut {NoStop}%
\bibitem [{\citenamefont {García~de Abajo}(2014)}]{Garcia_2014}%
  \BibitemOpen
  \bibfield  {author} {\bibinfo {author} {\bibfnamefont {F.~J.}\ \bibnamefont
  {García~de Abajo}},\ }\bibfield  {title} {\bibinfo {title} {Graphene
  plasmonics: Challenges and opportunities},\ }\href
  {https://doi.org/10.1021/ph400147y} {\bibfield  {journal} {\bibinfo
  {journal} {ACS Photonics}\ }\textbf {\bibinfo {volume} {1}},\ \bibinfo
  {pages} {135} (\bibinfo {year} {2014})}\BibitemShut {NoStop}%
\bibitem [{\citenamefont {Hwang}\ and\ \citenamefont
  {Das~Sarma}(2007{\natexlab{a}})}]{Hwang_2007}%
  \BibitemOpen
  \bibfield  {author} {\bibinfo {author} {\bibfnamefont {E.~H.}\ \bibnamefont
  {Hwang}}\ and\ \bibinfo {author} {\bibfnamefont {S.}~\bibnamefont
  {Das~Sarma}},\ }\bibfield  {title} {\bibinfo {title} {Dielectric function,
  screening, and plasmons in two-dimensional graphene},\ }\href
  {https://doi.org/10.1103/PhysRevB.75.205418} {\bibfield  {journal} {\bibinfo
  {journal} {Phys. Rev. B}\ }\textbf {\bibinfo {volume} {75}},\ \bibinfo
  {pages} {205418} (\bibinfo {year} {2007}{\natexlab{a}})}\BibitemShut
  {NoStop}%
\bibitem [{\citenamefont {Das~Sarma}\ and\ \citenamefont
  {Li}(2013)}]{Sarma_13}%
  \BibitemOpen
  \bibfield  {author} {\bibinfo {author} {\bibfnamefont {S.}~\bibnamefont
  {Das~Sarma}}\ and\ \bibinfo {author} {\bibfnamefont {Q.}~\bibnamefont {Li}},\
  }\bibfield  {title} {\bibinfo {title} {Intrinsic plasmons in two-dimensional
  dirac materials},\ }\href {https://doi.org/10.1103/PhysRevB.87.235418}
  {\bibfield  {journal} {\bibinfo  {journal} {Phys. Rev. B}\ }\textbf {\bibinfo
  {volume} {87}},\ \bibinfo {pages} {235418} (\bibinfo {year}
  {2013})}\BibitemShut {NoStop}%
\bibitem [{\citenamefont {Taft}\ and\ \citenamefont
  {Philipp}(1965)}]{Taft1965}%
  \BibitemOpen
  \bibfield  {author} {\bibinfo {author} {\bibfnamefont {E.~A.}\ \bibnamefont
  {Taft}}\ and\ \bibinfo {author} {\bibfnamefont {H.~R.}\ \bibnamefont
  {Philipp}},\ }\bibfield  {title} {\bibinfo {title} {{Optical Properties of
  Graphite}},\ }\href {https://doi.org/10.1103/PhysRev.138.A197} {\bibfield
  {journal} {\bibinfo  {journal} {Phys. Rev.}\ }\textbf {\bibinfo {volume}
  {138}},\ \bibinfo {pages} {A197} (\bibinfo {year} {1965})}\BibitemShut
  {NoStop}%
\bibitem [{\citenamefont {Zeppenfeld}(1967)}]{Zeppenfeld1967}%
  \BibitemOpen
  \bibfield  {author} {\bibinfo {author} {\bibfnamefont {K.}~\bibnamefont
  {Zeppenfeld}},\ }\bibfield  {title} {\bibinfo {title} {{Anisotropic Plasmon
  Behaviour in Graphite}},\ }\href
  {https://doi.org/10.1016/0375-9601(67)90683-4} {\bibfield  {journal}
  {\bibinfo  {journal} {Phys. Lett. A}\ }\textbf {\bibinfo {volume} {25}},\
  \bibinfo {pages} {335} (\bibinfo {year} {1967})}\BibitemShut {NoStop}%
\bibitem [{\citenamefont {Kinyanjui}\ \emph {et~al.}(2012)\citenamefont
  {Kinyanjui}, \citenamefont {Kramberger}, \citenamefont {Pichler},
  \citenamefont {Meyer}, \citenamefont {Wachsmuth}, \citenamefont {Benner},\
  and\ \citenamefont {Kaiser}}]{Kinyanjui2012}%
  \BibitemOpen
  \bibfield  {author} {\bibinfo {author} {\bibfnamefont {M.~K.}\ \bibnamefont
  {Kinyanjui}}, \bibinfo {author} {\bibfnamefont {C.}~\bibnamefont
  {Kramberger}}, \bibinfo {author} {\bibfnamefont {T.}~\bibnamefont {Pichler}},
  \bibinfo {author} {\bibfnamefont {J.~C.}\ \bibnamefont {Meyer}}, \bibinfo
  {author} {\bibfnamefont {P.}~\bibnamefont {Wachsmuth}}, \bibinfo {author}
  {\bibfnamefont {G.}~\bibnamefont {Benner}},\ and\ \bibinfo {author}
  {\bibfnamefont {U.}~\bibnamefont {Kaiser}},\ }\bibfield  {title} {\bibinfo
  {title} {Direct probe of linearly dispersing 2d interband plasmons in a
  free-standing graphene monolayer},\ }\href
  {https://doi.org/10.1209/0295-5075/97/57005} {\bibfield  {journal} {\bibinfo
  {journal} {Europhys. Lett.}\ }\textbf {\bibinfo {volume} {97}},\ \bibinfo
  {pages} {57005} (\bibinfo {year} {2012})}\BibitemShut {NoStop}%
\bibitem [{\citenamefont {Wachsmuth}\ \emph {et~al.}(2013)\citenamefont
  {Wachsmuth}, \citenamefont {Hambach}, \citenamefont {Kinyanjui},
  \citenamefont {Guzzo}, \citenamefont {Benner},\ and\ \citenamefont
  {Kaiser}}]{Wachsmuth2013}%
  \BibitemOpen
  \bibfield  {author} {\bibinfo {author} {\bibfnamefont {P.}~\bibnamefont
  {Wachsmuth}}, \bibinfo {author} {\bibfnamefont {R.}~\bibnamefont {Hambach}},
  \bibinfo {author} {\bibfnamefont {M.~K.}\ \bibnamefont {Kinyanjui}}, \bibinfo
  {author} {\bibfnamefont {M.}~\bibnamefont {Guzzo}}, \bibinfo {author}
  {\bibfnamefont {G.}~\bibnamefont {Benner}},\ and\ \bibinfo {author}
  {\bibfnamefont {U.}~\bibnamefont {Kaiser}},\ }\bibfield  {title} {\bibinfo
  {title} {{High-energy collective electronic excitations in free-standing
  single-layer graphene}},\ }\href {https://doi.org/10.1103/PhysRevB.88.075433}
  {\bibfield  {journal} {\bibinfo  {journal} {Phys. Rev. B}\ }\textbf {\bibinfo
  {volume} {88}},\ \bibinfo {pages} {075433} (\bibinfo {year}
  {2013})}\BibitemShut {NoStop}%
\bibitem [{\citenamefont {Wachsmuth}\ \emph {et~al.}(2014)\citenamefont
  {Wachsmuth}, \citenamefont {Hambach}, \citenamefont {Benner},\ and\
  \citenamefont {Kaiser}}]{Wachsmuth2014}%
  \BibitemOpen
  \bibfield  {author} {\bibinfo {author} {\bibfnamefont {P.}~\bibnamefont
  {Wachsmuth}}, \bibinfo {author} {\bibfnamefont {R.}~\bibnamefont {Hambach}},
  \bibinfo {author} {\bibfnamefont {G.}~\bibnamefont {Benner}},\ and\ \bibinfo
  {author} {\bibfnamefont {U.}~\bibnamefont {Kaiser}},\ }\bibfield  {title}
  {\bibinfo {title} {{Plasmon bands in multilayer graphene}},\ }\href
  {https://doi.org/10.1103/PhysRevB.90.235434} {\bibfield  {journal} {\bibinfo
  {journal} {Phys. Rev. B}\ }\textbf {\bibinfo {volume} {90}},\ \bibinfo
  {pages} {235434} (\bibinfo {year} {2014})}\BibitemShut {NoStop}%
\bibitem [{\citenamefont {Kramberger}\ \emph {et~al.}(2008)\citenamefont
  {Kramberger}, \citenamefont {Hambach}, \citenamefont {Giorgetti},
  \citenamefont {Rummeli}, \citenamefont {Knupfer}, \citenamefont {Fink},
  \citenamefont {Buchner}, \citenamefont {Reining}, \citenamefont {Einarsson},
  \citenamefont {Maruyama}, \citenamefont {Sottile}, \citenamefont {Hannewald},
  \citenamefont {Olevano}, \citenamefont {Marinopoulos},\ and\ \citenamefont
  {Pichler}}]{Kramberger2008}%
  \BibitemOpen
  \bibfield  {author} {\bibinfo {author} {\bibfnamefont {C.}~\bibnamefont
  {Kramberger}}, \bibinfo {author} {\bibfnamefont {R.}~\bibnamefont {Hambach}},
  \bibinfo {author} {\bibfnamefont {C.}~\bibnamefont {Giorgetti}}, \bibinfo
  {author} {\bibfnamefont {M.~H.}\ \bibnamefont {Rummeli}}, \bibinfo {author}
  {\bibfnamefont {M.}~\bibnamefont {Knupfer}}, \bibinfo {author} {\bibfnamefont
  {J.}~\bibnamefont {Fink}}, \bibinfo {author} {\bibfnamefont {B.}~\bibnamefont
  {Buchner}}, \bibinfo {author} {\bibfnamefont {L.}~\bibnamefont {Reining}},
  \bibinfo {author} {\bibfnamefont {E.}~\bibnamefont {Einarsson}}, \bibinfo
  {author} {\bibfnamefont {S.}~\bibnamefont {Maruyama}}, \bibinfo {author}
  {\bibfnamefont {F.}~\bibnamefont {Sottile}}, \bibinfo {author} {\bibfnamefont
  {K.}~\bibnamefont {Hannewald}}, \bibinfo {author} {\bibfnamefont
  {V.}~\bibnamefont {Olevano}}, \bibinfo {author} {\bibfnamefont {A.~G.}\
  \bibnamefont {Marinopoulos}},\ and\ \bibinfo {author} {\bibfnamefont
  {T.}~\bibnamefont {Pichler}},\ }\bibfield  {title} {\bibinfo {title} {Linear
  plasmon dispersion in single-wall carbon nanotubes and the collective
  excitation spectrum of graphene},\ }\href
  {https://doi.org/10.1103/PhysRevLett.100.196803} {\bibfield  {journal}
  {\bibinfo  {journal} {Phys. Rev. Lett.}\ }\textbf {\bibinfo {volume} {100}},\
  \bibinfo {pages} {196803} (\bibinfo {year} {2008})}\BibitemShut {NoStop}%
\bibitem [{\citenamefont {Siegel}\ \emph {et~al.}(2011)\citenamefont {Siegel},
  \citenamefont {Park}, \citenamefont {Hwang}, \citenamefont {Deslippe},
  \citenamefont {Fedorov}, \citenamefont {Louie},\ and\ \citenamefont
  {Lanzara}}]{siegel2011many}%
  \BibitemOpen
  \bibfield  {author} {\bibinfo {author} {\bibfnamefont {D.~A.}\ \bibnamefont
  {Siegel}}, \bibinfo {author} {\bibfnamefont {C.-H.}\ \bibnamefont {Park}},
  \bibinfo {author} {\bibfnamefont {C.}~\bibnamefont {Hwang}}, \bibinfo
  {author} {\bibfnamefont {J.}~\bibnamefont {Deslippe}}, \bibinfo {author}
  {\bibfnamefont {A.~V.}\ \bibnamefont {Fedorov}}, \bibinfo {author}
  {\bibfnamefont {S.~G.}\ \bibnamefont {Louie}},\ and\ \bibinfo {author}
  {\bibfnamefont {A.}~\bibnamefont {Lanzara}},\ }\bibfield  {title} {\bibinfo
  {title} {Many-body interactions in quasi-freestanding graphene},\ }\href
  {https://doi.org/10.1073/pnas.1100242108} {\bibfield  {journal} {\bibinfo
  {journal} {Proceedings of the National Academy of Sciences}\ }\textbf
  {\bibinfo {volume} {108}},\ \bibinfo {pages} {11365} (\bibinfo {year}
  {2011})}\BibitemShut {NoStop}%
\bibitem [{\citenamefont {Yang}\ \emph {et~al.}(2009)\citenamefont {Yang},
  \citenamefont {Deslippe}, \citenamefont {Park}, \citenamefont {Cohen},\ and\
  \citenamefont {Louie}}]{Yang2009}%
  \BibitemOpen
  \bibfield  {author} {\bibinfo {author} {\bibfnamefont {L.}~\bibnamefont
  {Yang}}, \bibinfo {author} {\bibfnamefont {J.}~\bibnamefont {Deslippe}},
  \bibinfo {author} {\bibfnamefont {C.-H.}\ \bibnamefont {Park}}, \bibinfo
  {author} {\bibfnamefont {M.~L.}\ \bibnamefont {Cohen}},\ and\ \bibinfo
  {author} {\bibfnamefont {S.~G.}\ \bibnamefont {Louie}},\ }\bibfield  {title}
  {\bibinfo {title} {Excitonic effects on the optical response of graphene and
  bilayer graphene},\ }\href {https://doi.org/10.1103/PhysRevLett.103.186802}
  {\bibfield  {journal} {\bibinfo  {journal} {Phys. Rev. Lett.}\ }\textbf
  {\bibinfo {volume} {103}},\ \bibinfo {pages} {186802} (\bibinfo {year}
  {2009})}\BibitemShut {NoStop}%
\bibitem [{\citenamefont {Yang}(2011)}]{Yang2011}%
  \BibitemOpen
  \bibfield  {author} {\bibinfo {author} {\bibfnamefont {L.}~\bibnamefont
  {Yang}},\ }\bibfield  {title} {\bibinfo {title} {Excitons in intrinsic and
  bilayer graphene},\ }\href {https://doi.org/10.1103/PhysRevB.83.085405}
  {\bibfield  {journal} {\bibinfo  {journal} {Phys. Rev. B}\ }\textbf {\bibinfo
  {volume} {83}},\ \bibinfo {pages} {085405} (\bibinfo {year}
  {2011})}\BibitemShut {NoStop}%
\bibitem [{\citenamefont {Guandalini}\ \emph
  {et~al.}(2023{\natexlab{a}})\citenamefont {Guandalini}, \citenamefont
  {Senga}, \citenamefont {Lin}, \citenamefont {Suenaga}, \citenamefont
  {Ferretti}, \citenamefont {Varsano}, \citenamefont {Recchia}, \citenamefont
  {Barone}, \citenamefont {Mauri}, \citenamefont {Pichler},\ and\ \citenamefont
  {Kramberger}}]{guandalini_2023}%
  \BibitemOpen
  \bibfield  {author} {\bibinfo {author} {\bibfnamefont {A.}~\bibnamefont
  {Guandalini}}, \bibinfo {author} {\bibfnamefont {R.}~\bibnamefont {Senga}},
  \bibinfo {author} {\bibfnamefont {Y.-C.}\ \bibnamefont {Lin}}, \bibinfo
  {author} {\bibfnamefont {K.}~\bibnamefont {Suenaga}}, \bibinfo {author}
  {\bibfnamefont {A.}~\bibnamefont {Ferretti}}, \bibinfo {author}
  {\bibfnamefont {D.}~\bibnamefont {Varsano}}, \bibinfo {author} {\bibfnamefont
  {A.}~\bibnamefont {Recchia}}, \bibinfo {author} {\bibfnamefont
  {P.}~\bibnamefont {Barone}}, \bibinfo {author} {\bibfnamefont
  {F.}~\bibnamefont {Mauri}}, \bibinfo {author} {\bibfnamefont
  {T.}~\bibnamefont {Pichler}},\ and\ \bibinfo {author} {\bibfnamefont
  {C.}~\bibnamefont {Kramberger}},\ }\bibfield  {title} {\bibinfo {title}
  {Excitonic effects in energy-loss spectra of freestanding graphene},\ }\href
  {https://doi.org/10.1021/acs.nanolett.3c03863} {\bibfield  {journal}
  {\bibinfo  {journal} {Nano Letters}\ }\textbf {\bibinfo {volume} {23}},\
  \bibinfo {pages} {11835} (\bibinfo {year} {2023}{\natexlab{a}})}\BibitemShut
  {NoStop}%
\bibitem [{\citenamefont {Basko}\ and\ \citenamefont
  {Aleiner}(2008)}]{Basko_2008}%
  \BibitemOpen
  \bibfield  {author} {\bibinfo {author} {\bibfnamefont {D.~M.}\ \bibnamefont
  {Basko}}\ and\ \bibinfo {author} {\bibfnamefont {I.~L.}\ \bibnamefont
  {Aleiner}},\ }\bibfield  {title} {\bibinfo {title} {Interplay of coulomb and
  electron-phonon interactions in graphene},\ }\href
  {https://doi.org/10.1103/PhysRevB.77.041409} {\bibfield  {journal} {\bibinfo
  {journal} {Phys. Rev. B}\ }\textbf {\bibinfo {volume} {77}},\ \bibinfo
  {pages} {041409} (\bibinfo {year} {2008})}\BibitemShut {NoStop}%
\bibitem [{\citenamefont {Venanzi}\ \emph {et~al.}(2023)\citenamefont
  {Venanzi}, \citenamefont {Graziotto}, \citenamefont {Macheda}, \citenamefont
  {Sotgiu}, \citenamefont {Ouaj}, \citenamefont {Stellino}, \citenamefont
  {Fasolato}, \citenamefont {Postorino}, \citenamefont
  {Mi\ifmmode~\check{s}\else \v{s}\fi{}eikis}, \citenamefont {Metzelaars},
  \citenamefont {K\"ogerler}, \citenamefont {Beschoten}, \citenamefont
  {Coletti}, \citenamefont {Roddaro}, \citenamefont {Calandra}, \citenamefont
  {Ortolani}, \citenamefont {Stampfer}, \citenamefont {Mauri},\ and\
  \citenamefont {Baldassarre}}]{Venanzi_2023}%
  \BibitemOpen
  \bibfield  {author} {\bibinfo {author} {\bibfnamefont {T.}~\bibnamefont
  {Venanzi}}, \bibinfo {author} {\bibfnamefont {L.}~\bibnamefont {Graziotto}},
  \bibinfo {author} {\bibfnamefont {F.}~\bibnamefont {Macheda}}, \bibinfo
  {author} {\bibfnamefont {S.}~\bibnamefont {Sotgiu}}, \bibinfo {author}
  {\bibfnamefont {T.}~\bibnamefont {Ouaj}}, \bibinfo {author} {\bibfnamefont
  {E.}~\bibnamefont {Stellino}}, \bibinfo {author} {\bibfnamefont
  {C.}~\bibnamefont {Fasolato}}, \bibinfo {author} {\bibfnamefont
  {P.}~\bibnamefont {Postorino}}, \bibinfo {author} {\bibfnamefont
  {V.}~\bibnamefont {Mi\ifmmode~\check{s}\else \v{s}\fi{}eikis}}, \bibinfo
  {author} {\bibfnamefont {M.}~\bibnamefont {Metzelaars}}, \bibinfo {author}
  {\bibfnamefont {P.}~\bibnamefont {K\"ogerler}}, \bibinfo {author}
  {\bibfnamefont {B.}~\bibnamefont {Beschoten}}, \bibinfo {author}
  {\bibfnamefont {C.}~\bibnamefont {Coletti}}, \bibinfo {author} {\bibfnamefont
  {S.}~\bibnamefont {Roddaro}}, \bibinfo {author} {\bibfnamefont
  {M.}~\bibnamefont {Calandra}}, \bibinfo {author} {\bibfnamefont
  {M.}~\bibnamefont {Ortolani}}, \bibinfo {author} {\bibfnamefont
  {C.}~\bibnamefont {Stampfer}}, \bibinfo {author} {\bibfnamefont
  {F.}~\bibnamefont {Mauri}},\ and\ \bibinfo {author} {\bibfnamefont
  {L.}~\bibnamefont {Baldassarre}},\ }\bibfield  {title} {\bibinfo {title}
  {Probing enhanced electron-phonon coupling in graphene by infrared resonance
  raman spectroscopy},\ }\href {https://doi.org/10.1103/PhysRevLett.130.256901}
  {\bibfield  {journal} {\bibinfo  {journal} {Phys. Rev. Lett.}\ }\textbf
  {\bibinfo {volume} {130}},\ \bibinfo {pages} {256901} (\bibinfo {year}
  {2023})}\BibitemShut {NoStop}%
\bibitem [{\citenamefont {Graziotto}\ \emph
  {et~al.}(2024{\natexlab{a}})\citenamefont {Graziotto}, \citenamefont
  {Macheda}, \citenamefont {Sohier}, \citenamefont {Calandra},\ and\
  \citenamefont {Mauri}}]{PhysRevB.109.075420}%
  \BibitemOpen
  \bibfield  {author} {\bibinfo {author} {\bibfnamefont {L.}~\bibnamefont
  {Graziotto}}, \bibinfo {author} {\bibfnamefont {F.}~\bibnamefont {Macheda}},
  \bibinfo {author} {\bibfnamefont {T.}~\bibnamefont {Sohier}}, \bibinfo
  {author} {\bibfnamefont {M.}~\bibnamefont {Calandra}},\ and\ \bibinfo
  {author} {\bibfnamefont {F.}~\bibnamefont {Mauri}},\ }\bibfield  {title}
  {\bibinfo {title} {Theory of infrared double-resonance raman spectrum in
  graphene: The role of the zone-boundary electron-phonon enhancement},\ }\href
  {https://doi.org/10.1103/PhysRevB.109.075420} {\bibfield  {journal} {\bibinfo
   {journal} {Phys. Rev. B}\ }\textbf {\bibinfo {volume} {109}},\ \bibinfo
  {pages} {075420} (\bibinfo {year} {2024}{\natexlab{a}})}\BibitemShut
  {NoStop}%
\bibitem [{\citenamefont {Graziotto}\ \emph
  {et~al.}(2024{\natexlab{b}})\citenamefont {Graziotto}, \citenamefont
  {Macheda}, \citenamefont {Venanzi}, \citenamefont {Marchese}, \citenamefont
  {Sotgiu}, \citenamefont {Ouaj}, \citenamefont {Stellino}, \citenamefont
  {Fasolato}, \citenamefont {Postorino}, \citenamefont {Metzelaars},
  \citenamefont {K{\"o}gerler}, \citenamefont {Beschoten}, \citenamefont
  {Calandra}, \citenamefont {Ortolani}, \citenamefont {Stampfer}, \citenamefont
  {Mauri},\ and\ \citenamefont {Baldassarre}}]{Graziotto2024-zk}%
  \BibitemOpen
  \bibfield  {author} {\bibinfo {author} {\bibfnamefont {L.}~\bibnamefont
  {Graziotto}}, \bibinfo {author} {\bibfnamefont {F.}~\bibnamefont {Macheda}},
  \bibinfo {author} {\bibfnamefont {T.}~\bibnamefont {Venanzi}}, \bibinfo
  {author} {\bibfnamefont {G.}~\bibnamefont {Marchese}}, \bibinfo {author}
  {\bibfnamefont {S.}~\bibnamefont {Sotgiu}}, \bibinfo {author} {\bibfnamefont
  {T.}~\bibnamefont {Ouaj}}, \bibinfo {author} {\bibfnamefont {E.}~\bibnamefont
  {Stellino}}, \bibinfo {author} {\bibfnamefont {C.}~\bibnamefont {Fasolato}},
  \bibinfo {author} {\bibfnamefont {P.}~\bibnamefont {Postorino}}, \bibinfo
  {author} {\bibfnamefont {M.}~\bibnamefont {Metzelaars}}, \bibinfo {author}
  {\bibfnamefont {P.}~\bibnamefont {K{\"o}gerler}}, \bibinfo {author}
  {\bibfnamefont {B.}~\bibnamefont {Beschoten}}, \bibinfo {author}
  {\bibfnamefont {M.}~\bibnamefont {Calandra}}, \bibinfo {author}
  {\bibfnamefont {M.}~\bibnamefont {Ortolani}}, \bibinfo {author}
  {\bibfnamefont {C.}~\bibnamefont {Stampfer}}, \bibinfo {author}
  {\bibfnamefont {F.}~\bibnamefont {Mauri}},\ and\ \bibinfo {author}
  {\bibfnamefont {L.}~\bibnamefont {Baldassarre}},\ }\bibfield  {title}
  {\bibinfo {title} {Infrared resonance raman of bilayer graphene: Signatures
  of massive fermions and band structure on the {2D} peak},\ }\href
  {https://doi.org/10.1021/acs.nanolett.3c03502} {\bibfield  {journal}
  {\bibinfo  {journal} {Nano Lett.}\ }\textbf {\bibinfo {volume} {24}},\
  \bibinfo {pages} {1867} (\bibinfo {year} {2024}{\natexlab{b}})}\BibitemShut
  {NoStop}%
\bibitem [{\citenamefont {Hedin}(1965{\natexlab{a}})}]{Hedin_1965}%
  \BibitemOpen
  \bibfield  {author} {\bibinfo {author} {\bibfnamefont {L.}~\bibnamefont
  {Hedin}},\ }\bibfield  {title} {\bibinfo {title} {New method for calculating
  the one-particle green's function with application to the electron-gas
  problem},\ }\href {https://doi.org/10.1103/PhysRev.139.A796} {\bibfield
  {journal} {\bibinfo  {journal} {Phys. Rev.}\ }\textbf {\bibinfo {volume}
  {139}},\ \bibinfo {pages} {A796} (\bibinfo {year}
  {1965}{\natexlab{a}})}\BibitemShut {NoStop}%
\bibitem [{\citenamefont {Strinati}\ \emph
  {et~al.}(1982{\natexlab{a}})\citenamefont {Strinati}, \citenamefont
  {Mattausch},\ and\ \citenamefont {Hanke}}]{Strinati_1982}%
  \BibitemOpen
  \bibfield  {author} {\bibinfo {author} {\bibfnamefont {G.}~\bibnamefont
  {Strinati}}, \bibinfo {author} {\bibfnamefont {H.~J.}\ \bibnamefont
  {Mattausch}},\ and\ \bibinfo {author} {\bibfnamefont {W.}~\bibnamefont
  {Hanke}},\ }\bibfield  {title} {\bibinfo {title} {Dynamical aspects of
  correlation corrections in a covalent crystal},\ }\href
  {https://doi.org/10.1103/PhysRevB.25.2867} {\bibfield  {journal} {\bibinfo
  {journal} {Phys. Rev. B}\ }\textbf {\bibinfo {volume} {25}},\ \bibinfo
  {pages} {2867} (\bibinfo {year} {1982}{\natexlab{a}})}\BibitemShut {NoStop}%
\bibitem [{\citenamefont {Hybertsen}\ and\ \citenamefont
  {Louie}(1986{\natexlab{a}})}]{Hybertsen_1986}%
  \BibitemOpen
  \bibfield  {author} {\bibinfo {author} {\bibfnamefont {M.~S.}\ \bibnamefont
  {Hybertsen}}\ and\ \bibinfo {author} {\bibfnamefont {S.~G.}\ \bibnamefont
  {Louie}},\ }\bibfield  {title} {\bibinfo {title} {Electron correlation in
  semiconductors and insulators: Band gaps and quasiparticle energies},\ }\href
  {https://doi.org/10.1103/PhysRevB.34.5390} {\bibfield  {journal} {\bibinfo
  {journal} {Phys. Rev. B}\ }\textbf {\bibinfo {volume} {34}},\ \bibinfo
  {pages} {5390} (\bibinfo {year} {1986}{\natexlab{a}})}\BibitemShut {NoStop}%
\bibitem [{\citenamefont {Godby}\ \emph
  {et~al.}(1988{\natexlab{a}})\citenamefont {Godby}, \citenamefont
  {Schl\"uter},\ and\ \citenamefont {Sham}}]{Godby_1988}%
  \BibitemOpen
  \bibfield  {author} {\bibinfo {author} {\bibfnamefont {R.~W.}\ \bibnamefont
  {Godby}}, \bibinfo {author} {\bibfnamefont {M.}~\bibnamefont {Schl\"uter}},\
  and\ \bibinfo {author} {\bibfnamefont {L.~J.}\ \bibnamefont {Sham}},\
  }\bibfield  {title} {\bibinfo {title} {Self-energy operators and
  exchange-correlation potentials in semiconductors},\ }\href
  {https://doi.org/10.1103/PhysRevB.37.10159} {\bibfield  {journal} {\bibinfo
  {journal} {Phys. Rev. B}\ }\textbf {\bibinfo {volume} {37}},\ \bibinfo
  {pages} {10159} (\bibinfo {year} {1988}{\natexlab{a}})}\BibitemShut {NoStop}%
\bibitem [{\citenamefont {Strinati}(1988{\natexlab{a}})}]{Strinati_1988}%
  \BibitemOpen
  \bibfield  {author} {\bibinfo {author} {\bibfnamefont {G.}~\bibnamefont
  {Strinati}},\ }\bibfield  {title} {\bibinfo {title} {Application of the
  green’s functions method to the study of the optical properties of
  semiconductors},\ }\href {https://doi.org/10.1007/BF02725962} {\bibfield
  {journal} {\bibinfo  {journal} {La Rivista del Nuovo Cimento (1978-1999)}\
  }\textbf {\bibinfo {volume} {11}},\ \bibinfo {pages} {1} (\bibinfo {year}
  {1988}{\natexlab{a}})}\BibitemShut {NoStop}%
\bibitem [{\citenamefont {Onida}\ \emph
  {et~al.}(2002{\natexlab{a}})\citenamefont {Onida}, \citenamefont {Reining},\
  and\ \citenamefont {Rubio}}]{Onida_2002}%
  \BibitemOpen
  \bibfield  {author} {\bibinfo {author} {\bibfnamefont {G.}~\bibnamefont
  {Onida}}, \bibinfo {author} {\bibfnamefont {L.}~\bibnamefont {Reining}},\
  and\ \bibinfo {author} {\bibfnamefont {A.}~\bibnamefont {Rubio}},\ }\bibfield
   {title} {\bibinfo {title} {Electronic excitations: density-functional versus
  many-body green's-function approaches},\ }\href
  {https://doi.org/10.1103/RevModPhys.74.601} {\bibfield  {journal} {\bibinfo
  {journal} {Rev. Mod. Phys.}\ }\textbf {\bibinfo {volume} {74}},\ \bibinfo
  {pages} {601} (\bibinfo {year} {2002}{\natexlab{a}})}\BibitemShut {NoStop}%
\bibitem [{\citenamefont {Trevisanutto}\ \emph {et~al.}(2008)\citenamefont
  {Trevisanutto}, \citenamefont {Giorgetti}, \citenamefont {Reining},
  \citenamefont {Ladisa},\ and\ \citenamefont {Olevano}}]{Trevisanutto_2008}%
  \BibitemOpen
  \bibfield  {author} {\bibinfo {author} {\bibfnamefont {P.~E.}\ \bibnamefont
  {Trevisanutto}}, \bibinfo {author} {\bibfnamefont {C.}~\bibnamefont
  {Giorgetti}}, \bibinfo {author} {\bibfnamefont {L.}~\bibnamefont {Reining}},
  \bibinfo {author} {\bibfnamefont {M.}~\bibnamefont {Ladisa}},\ and\ \bibinfo
  {author} {\bibfnamefont {V.}~\bibnamefont {Olevano}},\ }\bibfield  {title}
  {\bibinfo {title} {Ab initio $gw$ many-body effects in graphene},\ }\href
  {https://doi.org/10.1103/PhysRevLett.101.226405} {\bibfield  {journal}
  {\bibinfo  {journal} {Phys. Rev. Lett.}\ }\textbf {\bibinfo {volume} {101}},\
  \bibinfo {pages} {226405} (\bibinfo {year} {2008})}\BibitemShut {NoStop}%
\bibitem [{\citenamefont {Attaccalite}\ and\ \citenamefont
  {Rubio}(2009)}]{Attaccalite_2009}%
  \BibitemOpen
  \bibfield  {author} {\bibinfo {author} {\bibfnamefont {C.}~\bibnamefont
  {Attaccalite}}\ and\ \bibinfo {author} {\bibfnamefont {A.}~\bibnamefont
  {Rubio}},\ }\bibfield  {title} {\bibinfo {title} {Fermi velocity
  renormalization in doped graphene},\ }\href
  {https://doi.org/https://doi.org/10.1002/pssb.200982335} {\bibfield
  {journal} {\bibinfo  {journal} {Physica Status Solidi (b)}\ }\textbf
  {\bibinfo {volume} {246}},\ \bibinfo {pages} {2523} (\bibinfo {year}
  {2009})}\BibitemShut {NoStop}%
\bibitem [{\citenamefont {Guandalini}\ \emph {et~al.}(2024)\citenamefont
  {Guandalini}, \citenamefont {Leon}, \citenamefont {D'Amico}, \citenamefont
  {Cardoso}, \citenamefont {Ferretti}, \citenamefont {Rontani},\ and\
  \citenamefont {Varsano}}]{Guandalini_2024}%
  \BibitemOpen
  \bibfield  {author} {\bibinfo {author} {\bibfnamefont {A.}~\bibnamefont
  {Guandalini}}, \bibinfo {author} {\bibfnamefont {D.~A.}\ \bibnamefont
  {Leon}}, \bibinfo {author} {\bibfnamefont {P.}~\bibnamefont {D'Amico}},
  \bibinfo {author} {\bibfnamefont {C.}~\bibnamefont {Cardoso}}, \bibinfo
  {author} {\bibfnamefont {A.}~\bibnamefont {Ferretti}}, \bibinfo {author}
  {\bibfnamefont {M.}~\bibnamefont {Rontani}},\ and\ \bibinfo {author}
  {\bibfnamefont {D.}~\bibnamefont {Varsano}},\ }\bibfield  {title} {\bibinfo
  {title} {Efficient $gw$ calculations via interpolation of the screened
  interaction in momentum and frequency space: The case of graphene},\ }\href
  {https://doi.org/10.1103/PhysRevB.109.075120} {\bibfield  {journal} {\bibinfo
   {journal} {Phys. Rev. B}\ }\textbf {\bibinfo {volume} {109}},\ \bibinfo
  {pages} {075120} (\bibinfo {year} {2024})}\BibitemShut {NoStop}%
\bibitem [{\citenamefont {Elias}\ \emph {et~al.}(2011)\citenamefont {Elias},
  \citenamefont {Gorbachev}, \citenamefont {Mayorov}, \citenamefont {Morozov},
  \citenamefont {Zhukov}, \citenamefont {Blake}, \citenamefont {Ponomarenko},
  \citenamefont {Grigorieva}, \citenamefont {Novoselov}, \citenamefont
  {Guinea},\ and\ \citenamefont {Geim}}]{Elias_11}%
  \BibitemOpen
  \bibfield  {author} {\bibinfo {author} {\bibfnamefont {D.~C.}\ \bibnamefont
  {Elias}}, \bibinfo {author} {\bibfnamefont {R.~V.}\ \bibnamefont
  {Gorbachev}}, \bibinfo {author} {\bibfnamefont {A.~S.}\ \bibnamefont
  {Mayorov}}, \bibinfo {author} {\bibfnamefont {S.~V.}\ \bibnamefont
  {Morozov}}, \bibinfo {author} {\bibfnamefont {A.~A.}\ \bibnamefont {Zhukov}},
  \bibinfo {author} {\bibfnamefont {P.}~\bibnamefont {Blake}}, \bibinfo
  {author} {\bibfnamefont {L.~A.}\ \bibnamefont {Ponomarenko}}, \bibinfo
  {author} {\bibfnamefont {I.~V.}\ \bibnamefont {Grigorieva}}, \bibinfo
  {author} {\bibfnamefont {K.~S.}\ \bibnamefont {Novoselov}}, \bibinfo {author}
  {\bibfnamefont {F.}~\bibnamefont {Guinea}},\ and\ \bibinfo {author}
  {\bibfnamefont {A.~K.}\ \bibnamefont {Geim}},\ }\bibfield  {title} {\bibinfo
  {title} {Dirac cones reshaped by interaction effects in suspended graphene},\
  }\href {https://doi.org/10.1038/nphys2049} {\bibfield  {journal} {\bibinfo
  {journal} {Nat. Phys.}\ }\textbf {\bibinfo {volume} {7}},\ \bibinfo {pages}
  {701} (\bibinfo {year} {2011})}\BibitemShut {NoStop}%
\bibitem [{\citenamefont {Sonntag}\ \emph {et~al.}(2018)\citenamefont
  {Sonntag}, \citenamefont {Reichardt}, \citenamefont {Wirtz}, \citenamefont
  {Beschoten}, \citenamefont {Katsnelson}, \citenamefont {Libisch},\ and\
  \citenamefont {Stampfer}}]{Sonntag_2018}%
  \BibitemOpen
  \bibfield  {author} {\bibinfo {author} {\bibfnamefont {J.}~\bibnamefont
  {Sonntag}}, \bibinfo {author} {\bibfnamefont {S.}~\bibnamefont {Reichardt}},
  \bibinfo {author} {\bibfnamefont {L.}~\bibnamefont {Wirtz}}, \bibinfo
  {author} {\bibfnamefont {B.}~\bibnamefont {Beschoten}}, \bibinfo {author}
  {\bibfnamefont {M.~I.}\ \bibnamefont {Katsnelson}}, \bibinfo {author}
  {\bibfnamefont {F.}~\bibnamefont {Libisch}},\ and\ \bibinfo {author}
  {\bibfnamefont {C.}~\bibnamefont {Stampfer}},\ }\bibfield  {title} {\bibinfo
  {title} {Impact of many-body effects on landau levels in graphene},\ }\href
  {https://doi.org/10.1103/PhysRevLett.120.187701} {\bibfield  {journal}
  {\bibinfo  {journal} {Phys. Rev. Lett.}\ }\textbf {\bibinfo {volume} {120}},\
  \bibinfo {pages} {187701} (\bibinfo {year} {2018})}\BibitemShut {NoStop}%
\bibitem [{\citenamefont {Castro~Neto}\ \emph {et~al.}(2009)\citenamefont
  {Castro~Neto}, \citenamefont {Guinea}, \citenamefont {Peres}, \citenamefont
  {Novoselov},\ and\ \citenamefont {Geim}}]{CastroNeto_rev2009}%
  \BibitemOpen
  \bibfield  {author} {\bibinfo {author} {\bibfnamefont {A.~H.}\ \bibnamefont
  {Castro~Neto}}, \bibinfo {author} {\bibfnamefont {F.}~\bibnamefont {Guinea}},
  \bibinfo {author} {\bibfnamefont {N.~M.~R.}\ \bibnamefont {Peres}}, \bibinfo
  {author} {\bibfnamefont {K.~S.}\ \bibnamefont {Novoselov}},\ and\ \bibinfo
  {author} {\bibfnamefont {A.~K.}\ \bibnamefont {Geim}},\ }\bibfield  {title}
  {\bibinfo {title} {The electronic properties of graphene},\ }\href
  {https://doi.org/10.1103/RevModPhys.81.109} {\bibfield  {journal} {\bibinfo
  {journal} {Rev. Mod. Phys.}\ }\textbf {\bibinfo {volume} {81}},\ \bibinfo
  {pages} {109} (\bibinfo {year} {2009})}\BibitemShut {NoStop}%
\bibitem [{\citenamefont {González}\ \emph {et~al.}(1994)\citenamefont
  {González}, \citenamefont {Guinea},\ and\ \citenamefont
  {Vozmediano}}]{Gonzales_1994}%
  \BibitemOpen
  \bibfield  {author} {\bibinfo {author} {\bibfnamefont {J.}~\bibnamefont
  {González}}, \bibinfo {author} {\bibfnamefont {F.}~\bibnamefont {Guinea}},\
  and\ \bibinfo {author} {\bibfnamefont {M.}~\bibnamefont {Vozmediano}},\
  }\bibfield  {title} {\bibinfo {title} {Non-fermi liquid behavior of electrons
  in the half-filled honeycomb lattice (a renormalization group approach)},\
  }\href {https://doi.org/https://doi.org/10.1016/0550-3213(94)90410-3}
  {\bibfield  {journal} {\bibinfo  {journal} {Nuclear Physics B}\ }\textbf
  {\bibinfo {volume} {424}},\ \bibinfo {pages} {595} (\bibinfo {year}
  {1994})}\BibitemShut {NoStop}%
\bibitem [{\citenamefont {Gonz\'alez}\ \emph {et~al.}(1999)\citenamefont
  {Gonz\'alez}, \citenamefont {Guinea},\ and\ \citenamefont
  {Vozmediano}}]{Gonzales_1999}%
  \BibitemOpen
  \bibfield  {author} {\bibinfo {author} {\bibfnamefont {J.}~\bibnamefont
  {Gonz\'alez}}, \bibinfo {author} {\bibfnamefont {F.}~\bibnamefont {Guinea}},\
  and\ \bibinfo {author} {\bibfnamefont {M.~A.~H.}\ \bibnamefont
  {Vozmediano}},\ }\bibfield  {title} {\bibinfo {title} {Marginal-fermi-liquid
  behavior from two-dimensional coulomb interaction},\ }\href
  {https://doi.org/10.1103/PhysRevB.59.R2474} {\bibfield  {journal} {\bibinfo
  {journal} {Phys. Rev. B}\ }\textbf {\bibinfo {volume} {59}},\ \bibinfo
  {pages} {R2474} (\bibinfo {year} {1999})}\BibitemShut {NoStop}%
\bibitem [{\citenamefont {Stauber}\ \emph {et~al.}(2017)\citenamefont
  {Stauber}, \citenamefont {Parida}, \citenamefont {Trushin}, \citenamefont
  {Ulybyshev}, \citenamefont {Boyda},\ and\ \citenamefont
  {Schliemann}}]{Stauber_17}%
  \BibitemOpen
  \bibfield  {author} {\bibinfo {author} {\bibfnamefont {T.}~\bibnamefont
  {Stauber}}, \bibinfo {author} {\bibfnamefont {P.}~\bibnamefont {Parida}},
  \bibinfo {author} {\bibfnamefont {M.}~\bibnamefont {Trushin}}, \bibinfo
  {author} {\bibfnamefont {M.~V.}\ \bibnamefont {Ulybyshev}}, \bibinfo {author}
  {\bibfnamefont {D.~L.}\ \bibnamefont {Boyda}},\ and\ \bibinfo {author}
  {\bibfnamefont {J.}~\bibnamefont {Schliemann}},\ }\bibfield  {title}
  {\bibinfo {title} {Interacting electrons in graphene: Fermi velocity
  renormalization and optical response},\ }\href
  {https://doi.org/10.1103/PhysRevLett.118.266801} {\bibfield  {journal}
  {\bibinfo  {journal} {Phys. Rev. Lett.}\ }\textbf {\bibinfo {volume} {118}},\
  \bibinfo {pages} {266801} (\bibinfo {year} {2017})}\BibitemShut {NoStop}%
\bibitem [{\citenamefont {Martin}\ \emph {et~al.}(2016)\citenamefont {Martin},
  \citenamefont {Reining},\ and\ \citenamefont {Ceperley}}]{Reining_16}%
  \BibitemOpen
  \bibfield  {author} {\bibinfo {author} {\bibfnamefont {R.~M.}\ \bibnamefont
  {Martin}}, \bibinfo {author} {\bibfnamefont {L.}~\bibnamefont {Reining}},\
  and\ \bibinfo {author} {\bibfnamefont {D.~M.}\ \bibnamefont {Ceperley}},\
  }\href@noop {} {\emph {\bibinfo {title} {Interacting Electrons: Theory and
  Computational Approaches}}}\ (\bibinfo  {publisher} {Cambridge University
  Press},\ \bibinfo {year} {2016})\BibitemShut {NoStop}%
\bibitem [{\citenamefont {Hedin}(1965{\natexlab{b}})}]{Hedin_65}%
  \BibitemOpen
  \bibfield  {author} {\bibinfo {author} {\bibfnamefont {L.}~\bibnamefont
  {Hedin}},\ }\bibfield  {title} {\bibinfo {title} {New method for calculating
  the one-particle green's function with application to the electron-gas
  problem},\ }\href {https://doi.org/10.1103/PhysRev.139.A796} {\bibfield
  {journal} {\bibinfo  {journal} {Phys. Rev.}\ }\textbf {\bibinfo {volume}
  {139}},\ \bibinfo {pages} {A796} (\bibinfo {year}
  {1965}{\natexlab{b}})}\BibitemShut {NoStop}%
\bibitem [{\citenamefont {Strinati}\ \emph
  {et~al.}(1982{\natexlab{b}})\citenamefont {Strinati}, \citenamefont
  {Mattausch},\ and\ \citenamefont {Hanke}}]{Strinati_82}%
  \BibitemOpen
  \bibfield  {author} {\bibinfo {author} {\bibfnamefont {G.}~\bibnamefont
  {Strinati}}, \bibinfo {author} {\bibfnamefont {H.~J.}\ \bibnamefont
  {Mattausch}},\ and\ \bibinfo {author} {\bibfnamefont {W.}~\bibnamefont
  {Hanke}},\ }\bibfield  {title} {\bibinfo {title} {Dynamical aspects of
  correlation corrections in a covalent crystal},\ }\href
  {https://doi.org/10.1103/PhysRevB.25.2867} {\bibfield  {journal} {\bibinfo
  {journal} {Phys. Rev. B}\ }\textbf {\bibinfo {volume} {25}},\ \bibinfo
  {pages} {2867} (\bibinfo {year} {1982}{\natexlab{b}})}\BibitemShut {NoStop}%
\bibitem [{\citenamefont {Hybertsen}\ and\ \citenamefont
  {Louie}(1986{\natexlab{b}})}]{Hybertsen_86}%
  \BibitemOpen
  \bibfield  {author} {\bibinfo {author} {\bibfnamefont {M.~S.}\ \bibnamefont
  {Hybertsen}}\ and\ \bibinfo {author} {\bibfnamefont {S.~G.}\ \bibnamefont
  {Louie}},\ }\bibfield  {title} {\bibinfo {title} {Electron correlation in
  semiconductors and insulators: Band gaps and quasiparticle energies},\ }\href
  {https://doi.org/10.1103/PhysRevB.34.5390} {\bibfield  {journal} {\bibinfo
  {journal} {Phys. Rev. B}\ }\textbf {\bibinfo {volume} {34}},\ \bibinfo
  {pages} {5390} (\bibinfo {year} {1986}{\natexlab{b}})}\BibitemShut {NoStop}%
\bibitem [{\citenamefont {Godby}\ \emph
  {et~al.}(1988{\natexlab{b}})\citenamefont {Godby}, \citenamefont
  {Schl\"uter},\ and\ \citenamefont {Sham}}]{Godby_88}%
  \BibitemOpen
  \bibfield  {author} {\bibinfo {author} {\bibfnamefont {R.~W.}\ \bibnamefont
  {Godby}}, \bibinfo {author} {\bibfnamefont {M.}~\bibnamefont {Schl\"uter}},\
  and\ \bibinfo {author} {\bibfnamefont {L.~J.}\ \bibnamefont {Sham}},\
  }\bibfield  {title} {\bibinfo {title} {Self-energy operators and
  exchange-correlation potentials in semiconductors},\ }\href
  {https://doi.org/10.1103/PhysRevB.37.10159} {\bibfield  {journal} {\bibinfo
  {journal} {Phys. Rev. B}\ }\textbf {\bibinfo {volume} {37}},\ \bibinfo
  {pages} {10159} (\bibinfo {year} {1988}{\natexlab{b}})}\BibitemShut {NoStop}%
\bibitem [{\citenamefont {Strinati}(1988{\natexlab{b}})}]{Strinati_88}%
  \BibitemOpen
  \bibfield  {author} {\bibinfo {author} {\bibfnamefont {G.}~\bibnamefont
  {Strinati}},\ }\bibfield  {title} {\bibinfo {title} {Application of the
  green’s functions method to the study of the optical properties of
  semiconductors},\ }\href@noop {} {\bibfield  {journal} {\bibinfo  {journal}
  {La Rivista del Nuovo Cimento (1978-1999)}\ }\textbf {\bibinfo {volume}
  {11}},\ \bibinfo {pages} {1} (\bibinfo {year}
  {1988}{\natexlab{b}})}\BibitemShut {NoStop}%
\bibitem [{\citenamefont {Onida}\ \emph
  {et~al.}(2002{\natexlab{b}})\citenamefont {Onida}, \citenamefont {Reining},\
  and\ \citenamefont {Rubio}}]{Onida_02}%
  \BibitemOpen
  \bibfield  {author} {\bibinfo {author} {\bibfnamefont {G.}~\bibnamefont
  {Onida}}, \bibinfo {author} {\bibfnamefont {L.}~\bibnamefont {Reining}},\
  and\ \bibinfo {author} {\bibfnamefont {A.}~\bibnamefont {Rubio}},\ }\bibfield
   {title} {\bibinfo {title} {Electronic excitations: density-functional versus
  many-body green's-function approaches},\ }\href
  {https://doi.org/10.1103/RevModPhys.74.601} {\bibfield  {journal} {\bibinfo
  {journal} {Rev. Mod. Phys.}\ }\textbf {\bibinfo {volume} {74}},\ \bibinfo
  {pages} {601} (\bibinfo {year} {2002}{\natexlab{b}})}\BibitemShut {NoStop}%
\bibitem [{\citenamefont {Gatti}\ and\ \citenamefont
  {Sottile}(2013)}]{Gatti_13}%
  \BibitemOpen
  \bibfield  {author} {\bibinfo {author} {\bibfnamefont {M.}~\bibnamefont
  {Gatti}}\ and\ \bibinfo {author} {\bibfnamefont {F.}~\bibnamefont
  {Sottile}},\ }\bibfield  {title} {\bibinfo {title} {Exciton dispersion from
  first principles},\ }\href {https://doi.org/10.1103/PhysRevB.88.155113}
  {\bibfield  {journal} {\bibinfo  {journal} {Phys. Rev. B}\ }\textbf {\bibinfo
  {volume} {88}},\ \bibinfo {pages} {155113} (\bibinfo {year}
  {2013})}\BibitemShut {NoStop}%
\bibitem [{\citenamefont {H\"user}\ \emph {et~al.}(2013)\citenamefont
  {H\"user}, \citenamefont {Olsen},\ and\ \citenamefont
  {Thygesen}}]{Huser_2013}%
  \BibitemOpen
  \bibfield  {author} {\bibinfo {author} {\bibfnamefont {F.}~\bibnamefont
  {H\"user}}, \bibinfo {author} {\bibfnamefont {T.}~\bibnamefont {Olsen}},\
  and\ \bibinfo {author} {\bibfnamefont {K.~S.}\ \bibnamefont {Thygesen}},\
  }\bibfield  {title} {\bibinfo {title} {How dielectric screening in
  two-dimensional crystals affects the convergence of excited-state
  calculations: Monolayer mos${}_{2}$},\ }\href
  {https://doi.org/10.1103/PhysRevB.88.245309} {\bibfield  {journal} {\bibinfo
  {journal} {Phys. Rev. B}\ }\textbf {\bibinfo {volume} {88}},\ \bibinfo
  {pages} {245309} (\bibinfo {year} {2013})}\BibitemShut {NoStop}%
\bibitem [{\citenamefont {Latini}\ \emph {et~al.}(2015)\citenamefont {Latini},
  \citenamefont {Olsen},\ and\ \citenamefont {Thygesen}}]{Latini_15}%
  \BibitemOpen
  \bibfield  {author} {\bibinfo {author} {\bibfnamefont {S.}~\bibnamefont
  {Latini}}, \bibinfo {author} {\bibfnamefont {T.}~\bibnamefont {Olsen}},\ and\
  \bibinfo {author} {\bibfnamefont {K.~S.}\ \bibnamefont {Thygesen}},\
  }\bibfield  {title} {\bibinfo {title} {Excitons in van der waals
  heterostructures: The important role of dielectric screening},\ }\href
  {https://doi.org/10.1103/PhysRevB.92.245123} {\bibfield  {journal} {\bibinfo
  {journal} {Phys. Rev. B}\ }\textbf {\bibinfo {volume} {92}},\ \bibinfo
  {pages} {245123} (\bibinfo {year} {2015})}\BibitemShut {NoStop}%
\bibitem [{\citenamefont {Qiu}\ \emph {et~al.}(2016)\citenamefont {Qiu},
  \citenamefont {Felipe},\ and\ \citenamefont {Louie}}]{qiu2016screening}%
  \BibitemOpen
  \bibfield  {author} {\bibinfo {author} {\bibfnamefont {D.~Y.}\ \bibnamefont
  {Qiu}}, \bibinfo {author} {\bibfnamefont {H.}~\bibnamefont {Felipe}},\ and\
  \bibinfo {author} {\bibfnamefont {S.~G.}\ \bibnamefont {Louie}},\ }\bibfield
  {title} {\bibinfo {title} {Screening and many-body effects in two-dimensional
  crystals: Monolayer mos 2},\ }\href
  {https://doi.org/10.1103/PhysRevB.93.235435} {\bibfield  {journal} {\bibinfo
  {journal} {Phys. Rev. B}\ }\textbf {\bibinfo {volume} {93}},\ \bibinfo
  {pages} {235435} (\bibinfo {year} {2016})}\BibitemShut {NoStop}%
\bibitem [{\citenamefont {Guandalini}\ \emph
  {et~al.}(2023{\natexlab{b}})\citenamefont {Guandalini}, \citenamefont
  {D'Amico}, \citenamefont {Ferretti},\ and\ \citenamefont
  {Varsano}}]{Guandalini2023npjCM}%
  \BibitemOpen
  \bibfield  {author} {\bibinfo {author} {\bibfnamefont {A.}~\bibnamefont
  {Guandalini}}, \bibinfo {author} {\bibfnamefont {P.}~\bibnamefont {D'Amico}},
  \bibinfo {author} {\bibfnamefont {A.}~\bibnamefont {Ferretti}},\ and\
  \bibinfo {author} {\bibfnamefont {D.}~\bibnamefont {Varsano}},\ }\bibfield
  {title} {\bibinfo {title} {Efficient gw calculations in two dimensional
  materials through a stochastic integration of the screened potential},\
  }\href {https://doi.org/10.1038/s41524-023-00989-7} {\bibfield  {journal}
  {\bibinfo  {journal} {npj Computational Materials}\ }\textbf {\bibinfo
  {volume} {9}},\ \bibinfo {pages} {44} (\bibinfo {year}
  {2023}{\natexlab{b}})}\BibitemShut {NoStop}%
\bibitem [{\citenamefont {Ismail-Beigi}(2006)}]{Beigi_2006}%
  \BibitemOpen
  \bibfield  {author} {\bibinfo {author} {\bibfnamefont {S.}~\bibnamefont
  {Ismail-Beigi}},\ }\bibfield  {title} {\bibinfo {title} {Truncation of
  periodic image interactions for confined systems},\ }\href
  {https://doi.org/10.1103/PhysRevB.73.233103} {\bibfield  {journal} {\bibinfo
  {journal} {Phys. Rev. B}\ }\textbf {\bibinfo {volume} {73}},\ \bibinfo
  {pages} {233103} (\bibinfo {year} {2006})}\BibitemShut {NoStop}%
\bibitem [{\citenamefont {Rozzi}\ \emph {et~al.}(2006)\citenamefont {Rozzi},
  \citenamefont {Varsano}, \citenamefont {Marini}, \citenamefont {Gross},\ and\
  \citenamefont {Rubio}}]{Rozzi_2006}%
  \BibitemOpen
  \bibfield  {author} {\bibinfo {author} {\bibfnamefont {C.~A.}\ \bibnamefont
  {Rozzi}}, \bibinfo {author} {\bibfnamefont {D.}~\bibnamefont {Varsano}},
  \bibinfo {author} {\bibfnamefont {A.}~\bibnamefont {Marini}}, \bibinfo
  {author} {\bibfnamefont {E.~K.~U.}\ \bibnamefont {Gross}},\ and\ \bibinfo
  {author} {\bibfnamefont {A.}~\bibnamefont {Rubio}},\ }\bibfield  {title}
  {\bibinfo {title} {Exact coulomb cutoff technique for supercell
  calculations},\ }\href {https://doi.org/10.1103/PhysRevB.73.205119}
  {\bibfield  {journal} {\bibinfo  {journal} {Phys. Rev. B}\ }\textbf {\bibinfo
  {volume} {73}},\ \bibinfo {pages} {205119} (\bibinfo {year}
  {2006})}\BibitemShut {NoStop}%
\bibitem [{\citenamefont {Macheda}\ \emph {et~al.}(2023)\citenamefont
  {Macheda}, \citenamefont {Sohier}, \citenamefont {Barone},\ and\
  \citenamefont {Mauri}}]{PhysRevB.107.094308}%
  \BibitemOpen
  \bibfield  {author} {\bibinfo {author} {\bibfnamefont {F.}~\bibnamefont
  {Macheda}}, \bibinfo {author} {\bibfnamefont {T.}~\bibnamefont {Sohier}},
  \bibinfo {author} {\bibfnamefont {P.}~\bibnamefont {Barone}},\ and\ \bibinfo
  {author} {\bibfnamefont {F.}~\bibnamefont {Mauri}},\ }\bibfield  {title}
  {\bibinfo {title} {Electron-phonon interaction and phonon frequencies in
  two-dimensional doped semiconductors},\ }\href
  {https://doi.org/10.1103/PhysRevB.107.094308} {\bibfield  {journal} {\bibinfo
   {journal} {Phys. Rev. B}\ }\textbf {\bibinfo {volume} {107}},\ \bibinfo
  {pages} {094308} (\bibinfo {year} {2023})}\BibitemShut {NoStop}%
\bibitem [{\citenamefont {Sohier}\ \emph {et~al.}(2015)\citenamefont {Sohier},
  \citenamefont {Calandra},\ and\ \citenamefont {Mauri}}]{Sohier2015}%
  \BibitemOpen
  \bibfield  {author} {\bibinfo {author} {\bibfnamefont {T.}~\bibnamefont
  {Sohier}}, \bibinfo {author} {\bibfnamefont {M.}~\bibnamefont {Calandra}},\
  and\ \bibinfo {author} {\bibfnamefont {F.}~\bibnamefont {Mauri}},\ }\bibfield
   {title} {\bibinfo {title} {Density-functional calculation of static
  screening in two-dimensional materials: The long-wavelength dielectric
  function of graphene},\ }\href {https://doi.org/10.1103/PhysRevB.91.165428}
  {\bibfield  {journal} {\bibinfo  {journal} {Phys. Rev. B}\ }\textbf {\bibinfo
  {volume} {91}},\ \bibinfo {pages} {165428} (\bibinfo {year}
  {2015})}\BibitemShut {NoStop}%
\bibitem [{\citenamefont {Nazarov}(2015)}]{Nazarov_15}%
  \BibitemOpen
  \bibfield  {author} {\bibinfo {author} {\bibfnamefont {V.~U.}\ \bibnamefont
  {Nazarov}},\ }\bibfield  {title} {\bibinfo {title} {Electronic excitations in
  quasi-2d crystals: what theoretical quantities are relevant to experiment?},\
  }\href {https://doi.org/10.1088/1367-2630/17/7/073018} {\bibfield  {journal}
  {\bibinfo  {journal} {New Journal of Physics}\ }\textbf {\bibinfo {volume}
  {17}},\ \bibinfo {pages} {073018} (\bibinfo {year} {2015})}\BibitemShut
  {NoStop}%
\bibitem [{\citenamefont {Cudazzo}\ \emph {et~al.}(2011)\citenamefont
  {Cudazzo}, \citenamefont {Tokatly},\ and\ \citenamefont
  {Rubio}}]{Cudazzo_2011}%
  \BibitemOpen
  \bibfield  {author} {\bibinfo {author} {\bibfnamefont {P.}~\bibnamefont
  {Cudazzo}}, \bibinfo {author} {\bibfnamefont {I.~V.}\ \bibnamefont
  {Tokatly}},\ and\ \bibinfo {author} {\bibfnamefont {A.}~\bibnamefont
  {Rubio}},\ }\bibfield  {title} {\bibinfo {title} {Dielectric screening in
  two-dimensional insulators: Implications for excitonic and impurity states in
  graphane},\ }\href {https://doi.org/10.1103/PhysRevB.84.085406} {\bibfield
  {journal} {\bibinfo  {journal} {Phys. Rev. B}\ }\textbf {\bibinfo {volume}
  {84}},\ \bibinfo {pages} {085406} (\bibinfo {year} {2011})}\BibitemShut
  {NoStop}%
\bibitem [{\citenamefont {van Schilfgaarde}\ \emph {et~al.}(2006)\citenamefont
  {van Schilfgaarde}, \citenamefont {Kotani},\ and\ \citenamefont
  {Faleev}}]{Schilfgaarde_2006}%
  \BibitemOpen
  \bibfield  {author} {\bibinfo {author} {\bibfnamefont {M.}~\bibnamefont {van
  Schilfgaarde}}, \bibinfo {author} {\bibfnamefont {T.}~\bibnamefont
  {Kotani}},\ and\ \bibinfo {author} {\bibfnamefont {S.}~\bibnamefont
  {Faleev}},\ }\bibfield  {title} {\bibinfo {title} {Quasiparticle
  self-consistent $gw$ theory},\ }\href
  {https://doi.org/10.1103/PhysRevLett.96.226402} {\bibfield  {journal}
  {\bibinfo  {journal} {Phys. Rev. Lett.}\ }\textbf {\bibinfo {volume} {96}},\
  \bibinfo {pages} {226402} (\bibinfo {year} {2006})}\BibitemShut {NoStop}%
\bibitem [{\citenamefont {Hwang}\ and\ \citenamefont
  {Das~Sarma}(2007{\natexlab{b}})}]{Hwang_07}%
  \BibitemOpen
  \bibfield  {author} {\bibinfo {author} {\bibfnamefont {E.~H.}\ \bibnamefont
  {Hwang}}\ and\ \bibinfo {author} {\bibfnamefont {S.}~\bibnamefont
  {Das~Sarma}},\ }\bibfield  {title} {\bibinfo {title} {Dielectric function,
  screening, and plasmons in two-dimensional graphene},\ }\href
  {https://doi.org/10.1103/PhysRevB.75.205418} {\bibfield  {journal} {\bibinfo
  {journal} {Phys. Rev. B}\ }\textbf {\bibinfo {volume} {75}},\ \bibinfo
  {pages} {205418} (\bibinfo {year} {2007}{\natexlab{b}})}\BibitemShut
  {NoStop}%
\bibitem [{\citenamefont {Giannozzi}\ \emph {et~al.}(2020)\citenamefont
  {Giannozzi}, \citenamefont {Baseggio}, \citenamefont {Bonfà}, \citenamefont
  {Brunato}, \citenamefont {Car}, \citenamefont {Carnimeo}, \citenamefont
  {Cavazzoni}, \citenamefont {de~Gironcoli}, \citenamefont {Delugas},
  \citenamefont {Ferrari~Ruffino}, \citenamefont {Ferretti}, \citenamefont
  {Marzari}, \citenamefont {Timrov}, \citenamefont {Urru},\ and\ \citenamefont
  {Baroni}}]{QE_2020}%
  \BibitemOpen
  \bibfield  {author} {\bibinfo {author} {\bibfnamefont {P.}~\bibnamefont
  {Giannozzi}}, \bibinfo {author} {\bibfnamefont {O.}~\bibnamefont {Baseggio}},
  \bibinfo {author} {\bibfnamefont {P.}~\bibnamefont {Bonfà}}, \bibinfo
  {author} {\bibfnamefont {D.}~\bibnamefont {Brunato}}, \bibinfo {author}
  {\bibfnamefont {R.}~\bibnamefont {Car}}, \bibinfo {author} {\bibfnamefont
  {I.}~\bibnamefont {Carnimeo}}, \bibinfo {author} {\bibfnamefont
  {C.}~\bibnamefont {Cavazzoni}}, \bibinfo {author} {\bibfnamefont
  {S.}~\bibnamefont {de~Gironcoli}}, \bibinfo {author} {\bibfnamefont
  {P.}~\bibnamefont {Delugas}}, \bibinfo {author} {\bibfnamefont
  {F.}~\bibnamefont {Ferrari~Ruffino}}, \bibinfo {author} {\bibfnamefont
  {A.}~\bibnamefont {Ferretti}}, \bibinfo {author} {\bibfnamefont
  {N.}~\bibnamefont {Marzari}}, \bibinfo {author} {\bibfnamefont
  {I.}~\bibnamefont {Timrov}}, \bibinfo {author} {\bibfnamefont
  {A.}~\bibnamefont {Urru}},\ and\ \bibinfo {author} {\bibfnamefont
  {S.}~\bibnamefont {Baroni}},\ }\bibfield  {title} {\bibinfo {title} {Quantum
  espresso toward the exascale},\ }\href {https://doi.org/10.1063/5.0005082}
  {\bibfield  {journal} {\bibinfo  {journal} {J. Chem. Phys.}\ }\textbf
  {\bibinfo {volume} {152}},\ \bibinfo {pages} {154105} (\bibinfo {year}
  {2020})}\BibitemShut {NoStop}%
\bibitem [{\citenamefont {Perdew}\ and\ \citenamefont {Zunger}(1981)}]{LDA}%
  \BibitemOpen
  \bibfield  {author} {\bibinfo {author} {\bibfnamefont {J.~P.}\ \bibnamefont
  {Perdew}}\ and\ \bibinfo {author} {\bibfnamefont {A.}~\bibnamefont
  {Zunger}},\ }\bibfield  {title} {\bibinfo {title} {Self-interaction
  correction to density-functional approximations for many-electron systems},\
  }\href {https://doi.org/10.1103/PhysRevB.23.5048} {\bibfield  {journal}
  {\bibinfo  {journal} {Phys. Rev. B}\ }\textbf {\bibinfo {volume} {23}},\
  \bibinfo {pages} {5048} (\bibinfo {year} {1981})}\BibitemShut {NoStop}%
\bibitem [{\citenamefont {Marini}\ \emph {et~al.}(2009)\citenamefont {Marini},
  \citenamefont {Hogan}, \citenamefont {Grüning},\ and\ \citenamefont
  {Varsano}}]{yambo_2009}%
  \BibitemOpen
  \bibfield  {author} {\bibinfo {author} {\bibfnamefont {A.}~\bibnamefont
  {Marini}}, \bibinfo {author} {\bibfnamefont {C.}~\bibnamefont {Hogan}},
  \bibinfo {author} {\bibfnamefont {M.}~\bibnamefont {Grüning}},\ and\
  \bibinfo {author} {\bibfnamefont {D.}~\bibnamefont {Varsano}},\ }\bibfield
  {title} {\bibinfo {title} {yambo: An ab initio tool for excited state
  calculations},\ }\href
  {https://doi.org/https://doi.org/10.1016/j.cpc.2009.02.003} {\bibfield
  {journal} {\bibinfo  {journal} {Comput. Phys. Commun.}\ }\textbf {\bibinfo
  {volume} {180}},\ \bibinfo {pages} {1392 } (\bibinfo {year}
  {2009})}\BibitemShut {NoStop}%
\bibitem [{\citenamefont {Sangalli}\ \emph {et~al.}(2019)\citenamefont
  {Sangalli}, \citenamefont {Ferretti}, \citenamefont {Miranda}, \citenamefont
  {Attaccalite}, \citenamefont {Marri}, \citenamefont {Cannuccia},
  \citenamefont {Melo}, \citenamefont {Marsili}, \citenamefont {Paleari},
  \citenamefont {Marrazzo}, \citenamefont {Prandini}, \citenamefont
  {Bonf{\`{a}}}, \citenamefont {Atambo}, \citenamefont {Affinito},
  \citenamefont {Palummo}, \citenamefont {Molina-S{\'{a}}nchez}, \citenamefont
  {Hogan}, \citenamefont {Grüning}, \citenamefont {Varsano},\ and\
  \citenamefont {Marini}}]{yambo_2019}%
  \BibitemOpen
  \bibfield  {author} {\bibinfo {author} {\bibfnamefont {D.}~\bibnamefont
  {Sangalli}}, \bibinfo {author} {\bibfnamefont {A.}~\bibnamefont {Ferretti}},
  \bibinfo {author} {\bibfnamefont {H.}~\bibnamefont {Miranda}}, \bibinfo
  {author} {\bibfnamefont {C.}~\bibnamefont {Attaccalite}}, \bibinfo {author}
  {\bibfnamefont {I.}~\bibnamefont {Marri}}, \bibinfo {author} {\bibfnamefont
  {E.}~\bibnamefont {Cannuccia}}, \bibinfo {author} {\bibfnamefont
  {P.}~\bibnamefont {Melo}}, \bibinfo {author} {\bibfnamefont {M.}~\bibnamefont
  {Marsili}}, \bibinfo {author} {\bibfnamefont {F.}~\bibnamefont {Paleari}},
  \bibinfo {author} {\bibfnamefont {A.}~\bibnamefont {Marrazzo}}, \bibinfo
  {author} {\bibfnamefont {G.}~\bibnamefont {Prandini}}, \bibinfo {author}
  {\bibfnamefont {P.}~\bibnamefont {Bonf{\`{a}}}}, \bibinfo {author}
  {\bibfnamefont {M.~O.}\ \bibnamefont {Atambo}}, \bibinfo {author}
  {\bibfnamefont {F.}~\bibnamefont {Affinito}}, \bibinfo {author}
  {\bibfnamefont {M.}~\bibnamefont {Palummo}}, \bibinfo {author} {\bibfnamefont
  {A.}~\bibnamefont {Molina-S{\'{a}}nchez}}, \bibinfo {author} {\bibfnamefont
  {C.}~\bibnamefont {Hogan}}, \bibinfo {author} {\bibfnamefont
  {M.}~\bibnamefont {Grüning}}, \bibinfo {author} {\bibfnamefont
  {D.}~\bibnamefont {Varsano}},\ and\ \bibinfo {author} {\bibfnamefont
  {A.}~\bibnamefont {Marini}},\ }\bibfield  {title} {\bibinfo {title}
  {Many-body perturbation theory calculations using the yambo code},\ }\href
  {https://doi.org/10.1088/1361-648x/ab15d0} {\bibfield  {journal} {\bibinfo
  {journal} {J. Phys.: Condens. Matter}\ }\textbf {\bibinfo {volume} {31}},\
  \bibinfo {pages} {325902} (\bibinfo {year} {2019})}\BibitemShut {NoStop}%
\bibitem [{\citenamefont {Sohier}\ \emph {et~al.}(2017)\citenamefont {Sohier},
  \citenamefont {Calandra},\ and\ \citenamefont {Mauri}}]{PhysRevB.96.075448}%
  \BibitemOpen
  \bibfield  {author} {\bibinfo {author} {\bibfnamefont {T.}~\bibnamefont
  {Sohier}}, \bibinfo {author} {\bibfnamefont {M.}~\bibnamefont {Calandra}},\
  and\ \bibinfo {author} {\bibfnamefont {F.}~\bibnamefont {Mauri}},\ }\bibfield
   {title} {\bibinfo {title} {Density functional perturbation theory for gated
  two-dimensional heterostructures: Theoretical developments and application to
  flexural phonons in graphene},\ }\href
  {https://doi.org/10.1103/PhysRevB.96.075448} {\bibfield  {journal} {\bibinfo
  {journal} {Phys. Rev. B}\ }\textbf {\bibinfo {volume} {96}},\ \bibinfo
  {pages} {075448} (\bibinfo {year} {2017})}\BibitemShut {NoStop}%
\bibitem [{\citenamefont {Godby}\ and\ \citenamefont
  {Needs}(1989)}]{Godby_1989}%
  \BibitemOpen
  \bibfield  {author} {\bibinfo {author} {\bibfnamefont {R.~W.}\ \bibnamefont
  {Godby}}\ and\ \bibinfo {author} {\bibfnamefont {R.~J.}\ \bibnamefont
  {Needs}},\ }\bibfield  {title} {\bibinfo {title} {Metal-insulator transition
  in kohn-sham theory and quasiparticle theory},\ }\href
  {https://doi.org/10.1103/PhysRevLett.62.1169} {\bibfield  {journal} {\bibinfo
   {journal} {Phys. Rev. Lett.}\ }\textbf {\bibinfo {volume} {62}},\ \bibinfo
  {pages} {1169} (\bibinfo {year} {1989})}\BibitemShut {NoStop}%
\bibitem [{\citenamefont {Monkhorst}\ and\ \citenamefont
  {Pack}(1976)}]{Monkhorst_1976}%
  \BibitemOpen
  \bibfield  {author} {\bibinfo {author} {\bibfnamefont {H.~J.}\ \bibnamefont
  {Monkhorst}}\ and\ \bibinfo {author} {\bibfnamefont {J.~D.}\ \bibnamefont
  {Pack}},\ }\bibfield  {title} {\bibinfo {title} {Special points for
  brillouin-zone integrations},\ }\href
  {https://doi.org/10.1103/PhysRevB.13.5188} {\bibfield  {journal} {\bibinfo
  {journal} {Phys. Rev. B}\ }\textbf {\bibinfo {volume} {13}},\ \bibinfo
  {pages} {5188} (\bibinfo {year} {1976})}\BibitemShut {NoStop}%
\bibitem [{\citenamefont {Binci}\ \emph {et~al.}(2021)\citenamefont {Binci},
  \citenamefont {Barone},\ and\ \citenamefont {Mauri}}]{Binci_2021}%
  \BibitemOpen
  \bibfield  {author} {\bibinfo {author} {\bibfnamefont {L.}~\bibnamefont
  {Binci}}, \bibinfo {author} {\bibfnamefont {P.}~\bibnamefont {Barone}},\ and\
  \bibinfo {author} {\bibfnamefont {F.}~\bibnamefont {Mauri}},\ }\bibfield
  {title} {\bibinfo {title} {First-principles theory of infrared vibrational
  spectroscopy of metals and semimetals: Application to graphite},\ }\href
  {https://doi.org/10.1103/PhysRevB.103.134304} {\bibfield  {journal} {\bibinfo
   {journal} {Phys. Rev. B}\ }\textbf {\bibinfo {volume} {103}},\ \bibinfo
  {pages} {134304} (\bibinfo {year} {2021})}\BibitemShut {NoStop}%
\bibitem [{\citenamefont {Benedict}\ \emph {et~al.}(1998)\citenamefont
  {Benedict}, \citenamefont {Shirley},\ and\ \citenamefont
  {Bohn}}]{Benedict_1998}%
  \BibitemOpen
  \bibfield  {author} {\bibinfo {author} {\bibfnamefont {L.~X.}\ \bibnamefont
  {Benedict}}, \bibinfo {author} {\bibfnamefont {E.~L.}\ \bibnamefont
  {Shirley}},\ and\ \bibinfo {author} {\bibfnamefont {R.~B.}\ \bibnamefont
  {Bohn}},\ }\bibfield  {title} {\bibinfo {title} {Optical absorption of
  insulators and the electron-hole interaction: An ab initio calculation},\
  }\href {https://doi.org/10.1103/PhysRevLett.80.4514} {\bibfield  {journal}
  {\bibinfo  {journal} {Phys. Rev. Lett.}\ }\textbf {\bibinfo {volume} {80}},\
  \bibinfo {pages} {4514} (\bibinfo {year} {1998})}\BibitemShut {NoStop}%
\bibitem [{\citenamefont {Benedict}\ and\ \citenamefont
  {Shirley}(1999)}]{Benedict_1999}%
  \BibitemOpen
  \bibfield  {author} {\bibinfo {author} {\bibfnamefont {L.~X.}\ \bibnamefont
  {Benedict}}\ and\ \bibinfo {author} {\bibfnamefont {E.~L.}\ \bibnamefont
  {Shirley}},\ }\bibfield  {title} {\bibinfo {title} {Ab initio calculation of
  ${\ensuremath{\epsilon}}_{2}(\ensuremath{\omega})$ including the
  electron-hole interaction: Application to gan and ${\mathrm{caf}}_{2}$},\
  }\href {https://doi.org/10.1103/PhysRevB.59.5441} {\bibfield  {journal}
  {\bibinfo  {journal} {Phys. Rev. B}\ }\textbf {\bibinfo {volume} {59}},\
  \bibinfo {pages} {5441} (\bibinfo {year} {1999})}\BibitemShut {NoStop}%
\bibitem [{\citenamefont {Grüning}\ \emph {et~al.}(2011)\citenamefont
  {Grüning}, \citenamefont {Marini},\ and\ \citenamefont
  {Gonze}}]{Gruning_2011}%
  \BibitemOpen
  \bibfield  {author} {\bibinfo {author} {\bibfnamefont {M.}~\bibnamefont
  {Grüning}}, \bibinfo {author} {\bibfnamefont {A.}~\bibnamefont {Marini}},\
  and\ \bibinfo {author} {\bibfnamefont {X.}~\bibnamefont {Gonze}},\ }\bibfield
   {title} {\bibinfo {title} {Implementation and testing of lanczos-based
  algorithms for random-phase approximation eigenproblems},\ }\href
  {https://doi.org/https://doi.org/10.1016/j.commatsci.2011.02.021} {\bibfield
  {journal} {\bibinfo  {journal} {Computational Materials Science}\ }\textbf
  {\bibinfo {volume} {50}},\ \bibinfo {pages} {2148} (\bibinfo {year}
  {2011})}\BibitemShut {NoStop}%
\bibitem [{\citenamefont {Crossno}\ \emph {et~al.}(2016)\citenamefont
  {Crossno}, \citenamefont {Shi}, \citenamefont {Wang}, \citenamefont {Liu},
  \citenamefont {Harzheim}, \citenamefont {Lucas}, \citenamefont {Sachdev},
  \citenamefont {Kim}, \citenamefont {Taniguchi}, \citenamefont {Watanabe},
  \citenamefont {Ohki},\ and\ \citenamefont {Fong}}]{Crossno_2016}%
  \BibitemOpen
  \bibfield  {author} {\bibinfo {author} {\bibfnamefont {J.}~\bibnamefont
  {Crossno}}, \bibinfo {author} {\bibfnamefont {J.~K.}\ \bibnamefont {Shi}},
  \bibinfo {author} {\bibfnamefont {K.}~\bibnamefont {Wang}}, \bibinfo {author}
  {\bibfnamefont {X.}~\bibnamefont {Liu}}, \bibinfo {author} {\bibfnamefont
  {A.}~\bibnamefont {Harzheim}}, \bibinfo {author} {\bibfnamefont
  {A.}~\bibnamefont {Lucas}}, \bibinfo {author} {\bibfnamefont
  {S.}~\bibnamefont {Sachdev}}, \bibinfo {author} {\bibfnamefont
  {P.}~\bibnamefont {Kim}}, \bibinfo {author} {\bibfnamefont {T.}~\bibnamefont
  {Taniguchi}}, \bibinfo {author} {\bibfnamefont {K.}~\bibnamefont {Watanabe}},
  \bibinfo {author} {\bibfnamefont {T.~A.}\ \bibnamefont {Ohki}},\ and\
  \bibinfo {author} {\bibfnamefont {K.~C.}\ \bibnamefont {Fong}},\ }\bibfield
  {title} {\bibinfo {title} {Observation of the dirac fluid and the breakdown
  of the wiedemann-franz law in graphene},\ }\href
  {https://doi.org/10.1126/science.aad0343} {\bibfield  {journal} {\bibinfo
  {journal} {Science}\ }\textbf {\bibinfo {volume} {351}},\ \bibinfo {pages}
  {1058} (\bibinfo {year} {2016})}\BibitemShut {NoStop}%
\bibitem [{\citenamefont {Lazzeri}\ and\ \citenamefont
  {Mauri}(2006)}]{Lazzeri_2006}%
  \BibitemOpen
  \bibfield  {author} {\bibinfo {author} {\bibfnamefont {M.}~\bibnamefont
  {Lazzeri}}\ and\ \bibinfo {author} {\bibfnamefont {F.}~\bibnamefont
  {Mauri}},\ }\bibfield  {title} {\bibinfo {title} {Nonadiabatic kohn anomaly
  in a doped graphene monolayer},\ }\href
  {https://doi.org/10.1103/PhysRevLett.97.266407} {\bibfield  {journal}
  {\bibinfo  {journal} {Phys. Rev. Lett.}\ }\textbf {\bibinfo {volume} {97}},\
  \bibinfo {pages} {266407} (\bibinfo {year} {2006})}\BibitemShut {NoStop}%
\bibitem [{\citenamefont {Mak}\ \emph {et~al.}(2012)\citenamefont {Mak},
  \citenamefont {Ju}, \citenamefont {Wang},\ and\ \citenamefont
  {Heinz}}]{mak2012optical}%
  \BibitemOpen
  \bibfield  {author} {\bibinfo {author} {\bibfnamefont {K.~F.}\ \bibnamefont
  {Mak}}, \bibinfo {author} {\bibfnamefont {L.}~\bibnamefont {Ju}}, \bibinfo
  {author} {\bibfnamefont {F.}~\bibnamefont {Wang}},\ and\ \bibinfo {author}
  {\bibfnamefont {T.~F.}\ \bibnamefont {Heinz}},\ }\bibfield  {title} {\bibinfo
  {title} {Optical spectroscopy of graphene: From the far infrared to the
  ultraviolet},\ }\href
  {https://doi.org/https://doi.org/10.1016/j.ssc.2012.04.064} {\bibfield
  {journal} {\bibinfo  {journal} {Solid state communications}\ }\textbf
  {\bibinfo {volume} {152}},\ \bibinfo {pages} {1341} (\bibinfo {year}
  {2012})}\BibitemShut {NoStop}%
\bibitem [{\citenamefont {Chang}\ \emph {et~al.}(2014)\citenamefont {Chang},
  \citenamefont {Liu}, \citenamefont {Liu}, \citenamefont {Zhong},\ and\
  \citenamefont {Norris}}]{chang2014extracting}%
  \BibitemOpen
  \bibfield  {author} {\bibinfo {author} {\bibfnamefont {Y.-C.}\ \bibnamefont
  {Chang}}, \bibinfo {author} {\bibfnamefont {C.-H.}\ \bibnamefont {Liu}},
  \bibinfo {author} {\bibfnamefont {C.-H.}\ \bibnamefont {Liu}}, \bibinfo
  {author} {\bibfnamefont {Z.}~\bibnamefont {Zhong}},\ and\ \bibinfo {author}
  {\bibfnamefont {T.~B.}\ \bibnamefont {Norris}},\ }\bibfield  {title}
  {\bibinfo {title} {Extracting the complex optical conductivity of mono-and
  bilayer graphene by ellipsometry},\ }\bibfield  {journal} {\bibinfo
  {journal} {Applied Physics Letters}\ }\textbf {\bibinfo {volume} {104}},\
  \href {https://doi.org/10.1063/1.4887364} {10.1063/1.4887364} (\bibinfo
  {year} {2014})\BibitemShut {NoStop}%
\bibitem [{\citenamefont {Li}\ \emph {et~al.}(2016)\citenamefont {Li},
  \citenamefont {Cheng}, \citenamefont {Liang}, \citenamefont {Tian},
  \citenamefont {Liang}, \citenamefont {Peng}, \citenamefont {Walker},
  \citenamefont {Gundlach},\ and\ \citenamefont {Nguyen}}]{li2016broadband}%
  \BibitemOpen
  \bibfield  {author} {\bibinfo {author} {\bibfnamefont {W.}~\bibnamefont
  {Li}}, \bibinfo {author} {\bibfnamefont {G.}~\bibnamefont {Cheng}}, \bibinfo
  {author} {\bibfnamefont {Y.}~\bibnamefont {Liang}}, \bibinfo {author}
  {\bibfnamefont {B.}~\bibnamefont {Tian}}, \bibinfo {author} {\bibfnamefont
  {X.}~\bibnamefont {Liang}}, \bibinfo {author} {\bibfnamefont
  {L.}~\bibnamefont {Peng}}, \bibinfo {author} {\bibfnamefont {A.~H.}\
  \bibnamefont {Walker}}, \bibinfo {author} {\bibfnamefont {D.~J.}\
  \bibnamefont {Gundlach}},\ and\ \bibinfo {author} {\bibfnamefont {N.~V.}\
  \bibnamefont {Nguyen}},\ }\bibfield  {title} {\bibinfo {title} {Broadband
  optical properties of graphene by spectroscopic ellipsometry},\ }\href
  {https://doi.org/10.1016/j.carbon.2015.12.007} {\bibfield  {journal}
  {\bibinfo  {journal} {Carbon}\ }\textbf {\bibinfo {volume} {99}},\ \bibinfo
  {pages} {348} (\bibinfo {year} {2016})}\BibitemShut {NoStop}%
\bibitem [{\citenamefont {Liang}\ and\ \citenamefont
  {Yang}(2015)}]{Liang_2015}%
  \BibitemOpen
  \bibfield  {author} {\bibinfo {author} {\bibfnamefont {Y.}~\bibnamefont
  {Liang}}\ and\ \bibinfo {author} {\bibfnamefont {L.}~\bibnamefont {Yang}},\
  }\bibfield  {title} {\bibinfo {title} {Carrier plasmon induced nonlinear band
  gap renormalization in two-dimensional semiconductors},\ }\href
  {https://doi.org/10.1103/PhysRevLett.114.063001} {\bibfield  {journal}
  {\bibinfo  {journal} {Phys. Rev. Lett.}\ }\textbf {\bibinfo {volume} {114}},\
  \bibinfo {pages} {063001} (\bibinfo {year} {2015})}\BibitemShut {NoStop}%
\bibitem [{\citenamefont {Venezuela}\ \emph {et~al.}(2011)\citenamefont
  {Venezuela}, \citenamefont {Lazzeri},\ and\ \citenamefont
  {Mauri}}]{Venezuela_11}%
  \BibitemOpen
  \bibfield  {author} {\bibinfo {author} {\bibfnamefont {P.}~\bibnamefont
  {Venezuela}}, \bibinfo {author} {\bibfnamefont {M.}~\bibnamefont {Lazzeri}},\
  and\ \bibinfo {author} {\bibfnamefont {F.}~\bibnamefont {Mauri}},\ }\bibfield
   {title} {\bibinfo {title} {Theory of double-resonant raman spectra in
  graphene: Intensity and line shape of defect-induced and two-phonon bands},\
  }\href {https://doi.org/10.1103/PhysRevB.84.035433} {\bibfield  {journal}
  {\bibinfo  {journal} {Phys. Rev. B}\ }\textbf {\bibinfo {volume} {84}},\
  \bibinfo {pages} {035433} (\bibinfo {year} {2011})}\BibitemShut {NoStop}%
\end{thebibliography}
\end{document}